\documentclass{aa}

\usepackage{verbatim}
\usepackage{graphicx}
\usepackage{epstopdf}
\usepackage{subfigure}
\usepackage{hyperref}
\usepackage{mathtools}
\usepackage{amsmath}
\usepackage{textcomp}
\usepackage{natbib}

\title{The active lives of stars: a complete description of rotation and XUV evolution of F, G, K, and M dwarfs\thanks{Our code for calculating stellar rotation and XUV evolution tracks using our model can be obtained from \url{https://github.com/ColinPhilipJohnstone/Mors} and a full set of evolutionary tracks will be available with the fully published version of this paper.}}

\titlerunning{The active lives of stars}

\author{C. P. Johnstone\inst{\ref{nhm},\ref{vienna}}, M. Bartel\inst{\ref{vienna}}, M. G\"udel\inst{\ref{vienna}}}

\institute{
Natural History Museum, Burgring 7, A-1010 Vienna, Austria \label{nhm} \\
\email{colin.johnstone@univie.ac.at}
\and
University of Vienna, Department of Astrophysics, T\"{u}rkenschanzstrasse 17, 1180 Vienna, Austria \label{vienna}
}

\abstract{}{
We study the evolution of rotation and high energy X-ray, extreme ultraviolet (EUV), {and Ly-$\alpha$ emission} for F, G, K, and M dwarfs, with masses between 0.1 and 1.2~M$_\odot$, and provide our evolutionary code and a freely available set of evolutionary tracks for use in planetary atmosphere studies.
}{
{We develop a physical rotational evolution model constrained by observed rotation distributions in young stellar clusters.
Using rotation, X-ray, EUV, and Ly-$\alpha$ measurements, we derive empirical relations for the dependences of high energy emission on stellar parameters.
Our description of X-ray evolution is validated using measurements of X-ray distributions in young clusters.}
}{
A star's X-ray, EUV, and Ly-$\alpha$ evolution is determined by its mass and initial rotation rate, with initial rotation being less important for lower mass stars. 
At all ages, solar mass stars are significantly more X-ray luminous than lower mass stars and stars that are born as rapid rotators remain highly active longer than those born as slow rotators.
At all evolutionary stages, habitable zone planets receive higher X-ray and EUV fluxes when orbiting lower mass stars due to their longer evolutionary timescales.
{The rates of flares follow similar evolutionary trends with higher mass stars flaring more often than lower mass stars at all ages, though habitable zone planets are likely influenced by flares more when orbiting lower mass stars.}
}{
Our results show that single decay-laws are insufficient to describe stellar activity evolution and highlight the need for a more comprehensive description based on the evolution of rotation, including also the effects of short-term variability.
Planets at similar orbital distances from their host stars receive significantly more X-ray and EUV energy over their lifetimes when orbiting higher mass stars.
The common belief that M dwarfs are more X-ray and EUV active than G dwarfs is justified only when considering the fluxes received by planets with similar effective temperatures, such as those in the habitable zone. 
}

\begin{document}

\maketitle


\section{Introduction}

Stellar magnetic activity is crucially important for the formation and evolution of planetary atmospheres and surface conditions.
Magnetic fields cause the outer atmospheres of stars to be heated to million degree temperatures, leading to the emission of X-ray and ultraviolet (XUV) radiation and the acceleration of magnetised stellar winds which can have a diverse range of effects on planets and their atmospheres. 
This high-energy radiation is absorbed in the upper atmospheres of planets leading to dissociation, ionisation, and heating (\citealt{Roble88}; \citealt{MurrayClay09}), which enhances and drives atmospheric loss processes, including rapid hydrodynamic losses if the star's activity is high enough (\citealt{Tian05}).
The details of how the activities of stars evolve are important for determining the eventual state of a planet's atmosphere (\citealt{Luger15}; \citealt{Johnstone15c}; \citealt{Kubyshkina19}).

When stars are first born, they have very strong magnetic fields and high XUV luminosities (\citealt{Yang11}) which both decay rapidly during the first few million years (\citealt{Gregory12}; \citealt{Gregory16}).
At later ages, rotational spin-down causes a significant decay in magnetic fields and high energy emission (\citealt{Guedel97}; \citealt{Vidotto14}).
For the XUV emission, this decay is more rapid at shorter wavelengths causing the shape of the XUV spectrum to evolve (\citealt{Ribas05}; \citealt{Claire12}).
The timescale for activity evolution is a strong function of mass with lower mass stars remaining active for much longer (\citealt{West08}; \citealt{ReinersBasri08}).
It is known empirically that a star's X-ray emission is determined by its mass, age, and rotation rate (\citealt{Pizzolato03}) meaning that its activity evolution is linked to its rotational evolution.
Stars born as fast rotators remain active longer than those born as slow rotators (\citealt{Tu15}).
Stars with different initial rotation rates can have very different influences on how the atmospheres of planets evolve (\citealt{Johnstone15c}; \citealt{Johnstone20}).

A star's long term rotational evolution depends primarily on its mass and initial rotation rate (\citealt{GalletBouvier15}; \citealt{Matt15}), though other factors play a role, such as the lifetime of the early circumstellar gas disk during the classical T Tauri phase (\citealt{Herbst02}) and the star's metallicity (\citealt{Amard20}).
At ages of $\sim$1~Myr, the rotation rates of stars are distributed between approximately a few to a few tens of times the rate of the current Sun (\citealt{Affer13}).
In the first few Myr, this rotation distribution appears to remain approximately constant with no clear evolution (\citealt{Rebull04}) except possibly for stars with masses below 0.4~M$_\odot$ (\citealt{HendersonStassun12}).
After this phase, pre-main-sequence contraction causes them to spin up until the zero-age main-sequence (ZAMS), and during this time the initial wide distribution of rotation rates gets wider. 
After the ZAMS, angular momentum removal by stellar winds causes stars to spin down and the distribution converges until most stars follow a single mass and age dependent value.
How long this convergence takes depends on stellar mass, with solar mass stars converging in the first billion years and lower mass stars taking longer, while for fully convective M~dwarfs, it is possible that no convergence takes place (\citealt{Irwin11}).
At ages of a few Gyr, lower mass stars tend to be slower rotators than higher mass stars (\citealt{Nielsen13}).
For a detailed review, see \mbox{\citet{Bouvier14}}.

In this paper, we study the evolution of stellar rotation and XUV emission on the pre-main-sequence and the main-sequence for stars with masses between 0.1 and 1.2~M$_\odot$ and provide a comprehensive empirical description of these phenomena.
We explore how long stars with different masses and initial rotation rates remain active and discuss the influences on the evolution of planetary atmospheres. 
In Section~\ref{sect:rotevo} we develop a description of rotational evolution, in Section~\ref{sect:xray} study the evolution of X-ray emission, in Section~\ref{sect:xuvplanet} we study EUV and Ly-$\alpha$ emission, in Section~\ref{sect:hz} we discuss XUV fluxes in the habitable zone, in Section~\ref{sect:flares} we discuss the contributions of flares, and in Section~\ref{sect:discuss} we summarise our results.
In Appendix~\ref{appendix:package}, we describe our freely available grid of rotation and XUV evolution models. 


\section{Rotational evolution} \label{sect:rotevo}

\subsection{Rotational evolution model} \label{sect:rotmodel}

\begin{table}
\centering
\begin{tabular}{ccc}
Year & Cluster name & Reference \\
\hline
12~Myr & h~Per &  \citet{Moraux13} \\
40~Myr & NGC~2547 & \citet{Irwin08} \\
150~Myr & Pleiades & \citet{Hartman10} \\
&  &  \citet{Rebull16} \\
& M50 & \citet{Irwin09} \\
& M35 & \citet{Meibom09} \\
& NGC~2516 & \citet{Irwin07} \\
550~Myr & M37 & \citet{Hartman09} \\
650~Myr & Praesepe & \citet{Douglas17} \\
& Hyades & \citet{Douglas19} \\
1000~Myr & NGC~6811 & \citet{Curtis19} \\
& NGC~752 & \citet{Agueros18} \\
2500~Myr & NGC~6819 & \citet{Meibom15} \\
\hline
\end{tabular}
\caption{
Sources of the observational constraints on rotational evolution used in this paper. 
}
\label{table:rotcluster}
\end{table}

In this paper, we extend our rotation model developed in previous studies (\citealt{Johnstone15b}; \citealt{Tu15};  \citealt{Johnstone19b}) to describe the full rotational evolution between 1~Myr and the end of the main-sequence for stars with masses between 0.1 and 1.2~M$_\odot$.
Our model describes the star as being composed of an envelope and a core, with the envelope corresponding to the outer convective zone and the core corresponding to everything else, and these two components are assumed to rotate as solid bodies and at separate rates. 
The rotation rates of the two components are influenced by core-envelope angular momentum exchanges and changes in the internal structure of the star, and the rotation of the envelope is additionally influenced by angular momentum loss by a magnetised stellar wind.
We use the stellar evolution models of \mbox{\citet{Spada13}} to get all parameters related to the star's internal structure, including convective turnover times, and use their set of models with initial metallicities, compositions, and mixing length parameters that best match the case of the Sun.
Throughout this paper, we define the Sun's rotation rate, $\Omega_\odot$, as \mbox{$2.67 \times 10^{-6}$~rad~s$^{-1}$}.

To understand how a star's rotation evolves, we must consider the both the changes in the angular momentum values and the moments of interia of the core and the envelope.
In our model, this is given by
\begin{equation} \label{eqn:dOmegaCoredt}
\frac{d\Omega_\mathrm{core}}{dt} = \frac{1}{I_\mathrm{core}} \left( -\tau_\mathrm{ce} - \tau_\mathrm{cg} - \Omega_\mathrm{core} \frac{d I_\mathrm{core} }{dt} \right),
\end{equation}
\begin{equation} \label{eqn:dOmegaEnvdt}
\frac{d\Omega_\mathrm{env}}{dt} = \frac{1}{I_\mathrm{env}} \left( \tau_\mathrm{w} + \tau_\mathrm{ce} + \tau_\mathrm{cg} + \tau_\mathrm{dl} - \Omega_\mathrm{env} \frac{d I_\mathrm{env} }{dt} \right),
\end{equation}
where $\Omega_\mathrm{core}$ and $\Omega_\mathrm{env}$ are the core and envelope angular velocities, $I_\mathrm{core}$ and $I_\mathrm{env}$ are the core and envelope moments of inertia, $\tau_\mathrm{w}$ is the stellar wind spin-down torque, $\tau_\mathrm{ce}$ is the core-envelope coupling torque, $\tau_\mathrm{cg}$ the core-growth torque, and $\tau_\mathrm{dl}$ the disk-locking torque.
For fully convective stars (\mbox{$M_\star \lesssim 0.35 M_\odot$}), the distinction between core and envelope is not considered and the stars are assumed to rotate entirely as solid bodies.
For these stars, we replace the above equations with
\begin{equation} \label{eqn:dOmegadt}
\frac{d\Omega_\star}{dt} = \frac{1}{I_\star} \left( \tau_\mathrm{w} + \tau_\mathrm{dl} - \Omega_\star \frac{d I_\star }{dt} \right),
\end{equation}
where $\Omega_\star$ and $I_\star$ are the star's rotation rate and moment of inertia.

\begin{figure}
\centering
\includegraphics[width=0.4\textwidth]{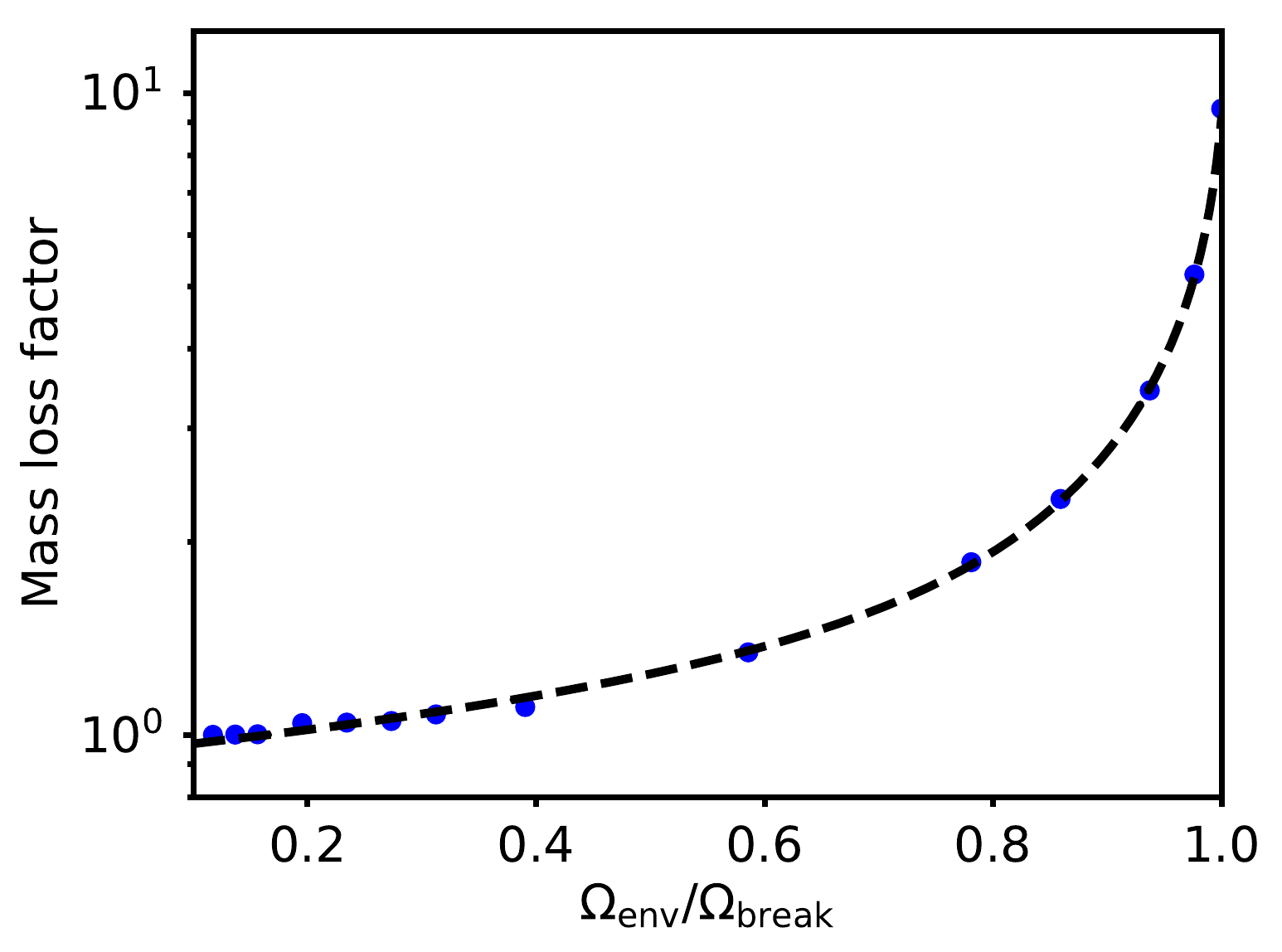}
\includegraphics[width=0.4\textwidth]{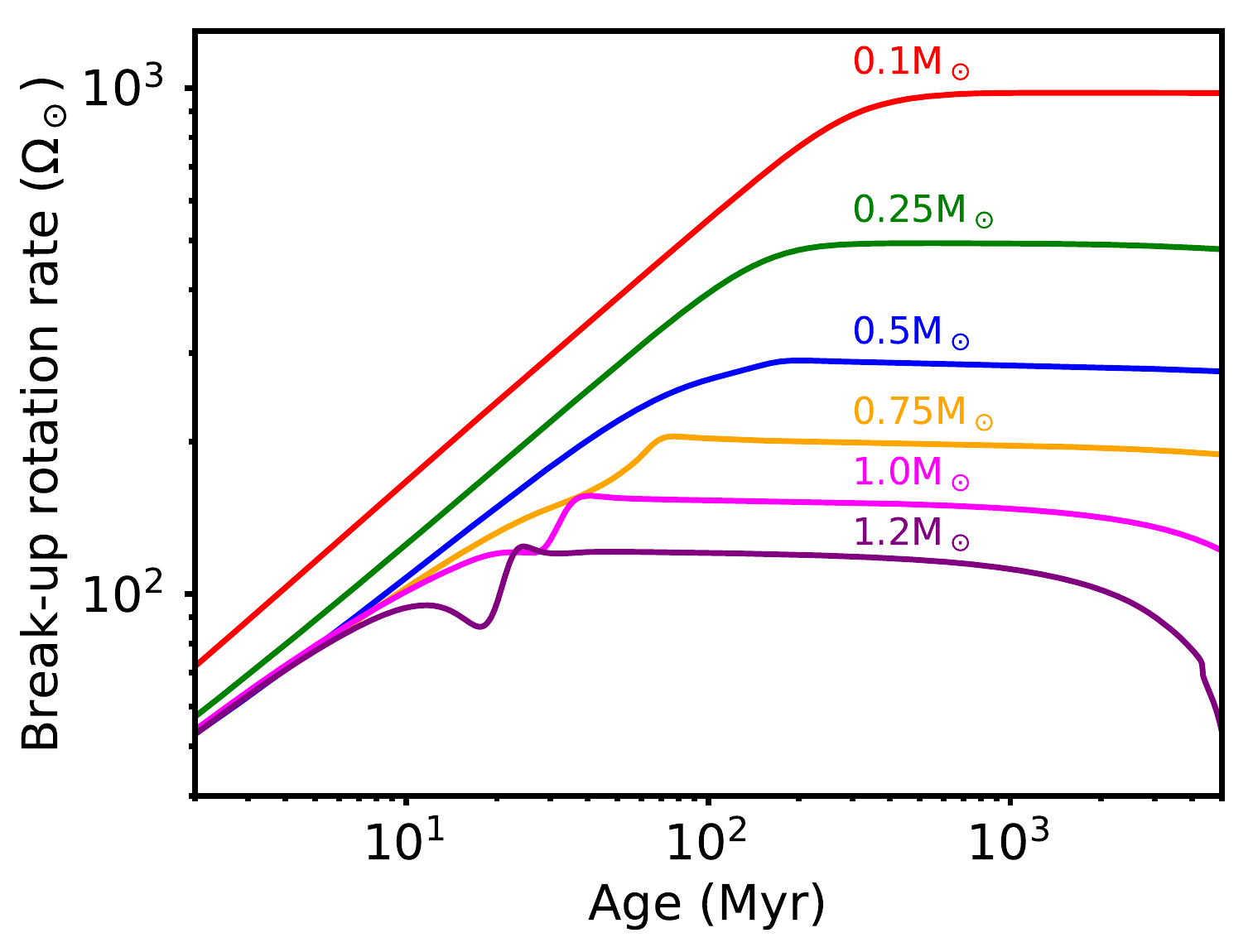}
\caption{
\emph{Upper-panel:}
Factor $f$ in Eqn.~\ref{eqn:MdotRossby}, as a function of rotation with the blue circles showing the results of MHD simulations for rapidly rotating stars by \citet{Johnstone17} and the dashed black line showing our fit formula given by Eqn.~\ref{eqn:magcentrifugal}.
\emph{Lower-panel:}
The evolution of the break-up rotation rate for different stellar masses. 
}
\label{fig:breakup}
\end{figure}

As in previous work, we calculate the stellar wind torque using
\begin{equation} \label{eqn:modifyMatttorque}
\tau_\mathrm{w} = - K_\tau \tau',
\end{equation}
where \mbox{$K_\tau$} is a parameter that we use to better reproduce the Skumanich spin-down of the modern Sun and we use \mbox{$K_\tau=11$} as derived in Section~4.1 of \mbox{\citet{Johnstone15b}}.
This value depends on the specific wind torque formula used and our assumption of the solar wind mass loss rate and the Sun's dipole field strength.
We calculate $\tau'$ using 
\begin{equation} \label{eqn:matttorque}
\tau' = K_1^2 B_\mathrm{dip}^{4m} \dot{M}_\star^{1-2m} R_\star^{4m+2} \frac{\Omega_\mathrm{env}}{(K_2^2 v_\mathrm{esc}^2 + \Omega_\mathrm{env}^2 R_\star^2)^m},
\end{equation}
where $B_\mathrm{dip}$ is the strength of the dipole component of the star's magnetic field, $\dot{M}_\star$ is the wind mass loss rate, $R_\star$ and $M_\star$ are the star's radius and mass, and $v_\mathrm{esc}$ is the surface escape velocity (\mbox{$=\sqrt{2 G M_\star / R_\star}$}).
The other parameters in this equation are given by $K_1 = 1.3$, $K_2 = 0.0506$, and $m = 0.2177$.
\mbox{\citet{Matt12}} derived this relation from a large number of magnetohydrodynamic wind simulations assuming axis-symmetry, a dipole magnetic field, and a polytropic equation of state for the heating with approximately constant wind temperatures.
We do not consider the influence of non-dipolar magnetic field geometries, although this can have an influence on the wind torque (\mbox{\citealt{HolzwarthJardine05}}; \mbox{\citealt{Reville15}}; \mbox{\citealt{Garraffo16}}).
Stellar magnetic fields are mostly composed of a combination of dipole and higher order components and since the dipole component rapidly becomes dominant further from the star's surface, it tends to be the dominant component for angular momentum loss even for stars with very non-dipolar fields (\citealt{Finley18}).
However, since the thermal structure of the wind influenced the angular momentum loss (\citealt{Cohen17}; \citealt{Pantolmos17}), we should expect some deviation from Eqn.~\ref{eqn:matttorque} due to the importance of the magnetic field geometry on the wind heating and acceleration and due to the observed dependence of stellar coronal temperatures on activity level (\citealt{JohnstoneGuedel15}).
Also, \citet{See19} argued that non-dipolar geometries could be more important for stars with very high mass loss rates. 
For our purposes, it is not necessary to consider these details. 

Both the dipole field strength and the mass loss rate in Eqn.~\ref{eqn:matttorque} are manifestations of the star's magnetic activity, which depends on the star's mass, age, and rotation rate. 
Many activity parameters are tightly correlated with the Rossby number, $Ro$, which is a dimensionless parameter defined by \mbox{$Ro = P_\mathrm{rot} / \tau_\mathrm{c}$}, where $P_\mathrm{rot}$ is the rotation period and $\tau_\mathrm{c}$ is the convective turnover time. 
Based on the observational correlation between global magnetic field and rotation given by \citet{Vidotto14}, we use
\begin{equation} \label{eqn:dipoleRossby}
B_\mathrm{dip} = \left \{
\begin{array}{ll}
B_{\mathrm{dip},\odot} \left( \frac{Ro_\mathrm{sat}}{Ro_\odot} \right)^{-1.32}, & \text{if }  Ro \le Ro_\mathrm{sat},\\
B_{\mathrm{dip},\odot} \left( \frac{Ro}{Ro_\odot} \right)^{-1.32}, & \text{otherwise},\\
\end{array} \right.
\end{equation}
where $Ro_\odot$ and $B_{\mathrm{dip},\odot}$ are the Rossby number and dipole field strength of the modern Sun {and $Ro_\mathrm{sat}$ is the Rossby number of the saturation threshold}.
For $B_{\mathrm{dip},\odot}$, we take the value 1.35~G (\mbox{\citealt{Johnstone15b}}).  
Very little is known about the mass loss rates of low-mass stars other than the modern Sun and our best option is to assume a plausible scaling law for the mass loss rate, $\dot{M}_\star$, given by
\begin{equation} \label{eqn:MdotRossby}
\dot{M}_\star = \left \{
\begin{array}{ll}
f \dot{M}_\odot \left( \frac{R_\star}{R_\odot} \right)^2 \left( \frac{Ro_\mathrm{sat}}{Ro_\odot} \right)^{a_\mathrm{w}} \left( \frac{M_\star}{M_\odot} \right)^{b_\mathrm{w}}, & \text{if }  Ro \le Ro_\mathrm{sat},\\
f \dot{M}_\odot \left( \frac{R_\star}{R_\odot} \right)^2 \left( \frac{Ro}{Ro_\odot} \right)^{a_\mathrm{w}} \left( \frac{M_\star}{M_\odot} \right)^{b_\mathrm{w}}, & \text{otherwise}  ,\\
\end{array} \right.
\end{equation}
where \mbox{$\dot{M}_\odot = 1.4 \times 10^{-14}$~M$_\odot$~yr$^{-1}$} is the current Sun's mass loss rate, $a_\mathrm{w}$ and $b_\mathrm{w}$ are fit parameters, and $f$ is the magneto-centrifugal factor discussed below.
As described in Appendix~\ref{appendix:fitting}, we find \mbox{$a_\mathrm{w}=-1.76$} and \mbox{$b_\mathrm{w}=0.649$}. 
We use the Rossby number instead of the rotation rate in the above relations since this captures the effect of young pre-main-sequence stars being saturated even at low rotation rates observed for activity related phenomena such as X-ray emission and magnetic field strength (\citealt{Briggs07}; \citealt{Johnstone14}).
Based on activity indicators such as X-ray luminosity, we assume that saturation occurs at a constant Rossby number for all stellar masses and ages and is the same for all activity related phenomena, allowing us to derive $Ro_\mathrm{sat}$ from X-ray observations.
The value of $Ro_\mathrm{sat}$ depends on the convective turnover times used, and in Section~\ref{sect:Xrayrelation} we find  \mbox{$Ro_\mathrm{sat} = 0.0605$} for the convective turnover times given by \citet{Spada13}.

For very rapidly rotating stars, we include also magneto-centrifugal effects. 
As winds propagate away from their stars, azimuthal components form for both the propagation velocity and the magnetic field, causing additional radial acceleration by centrifugal and Lorentz forces (\citealt{WeberDavis67}).
While these forces are negligible for the modern solar wind, they can be significant for the winds of rapidly rotating stars with strong magnetic fields (\citealt{BelcherMacGregor76}; \citealt{Johnstone17}).
When a star's rotation is close to the break-up rotation rate, its mass loss is enhanced by these effects, likely causing additional spin-down which should stop pre-main-sequence stas from spinning up to or past break-up.
We therefore include in Eqn.~\ref{eqn:MdotRossby} an additional multiplicative factor given by
\begin{equation} \label{eqn:magcentrifugal}
f \left( \Omega_\mathrm{env} \right) = \left \{
\begin{array}{ll}
1, & \text{if }  x \le 0.1,\\
0.93 \left( 1.01 - x \right)^{-0.43} e^{0.31 x^{7.5}}, & \text{otherwise},\\
\end{array} \right.
\end{equation}
where \mbox{$x = \Omega_\mathrm{env} / \Omega_\mathrm{break}$}.
This equation is based on the 1D MHD simulations of \citet{Johnstone17} and is shown in Fig.~\ref{fig:breakup}.
As they spin up, stars reach break-up when the Keplerian co-rotation radius becomes equal to the star's radius at the equator, which happens at a rotation rate of \mbox{$\Omega_\mathrm{break} = \left( G M_\star / R_\mathrm{e}^3 \right)^{1/2}$}, where $R_\mathrm{e}$ is the star's equatorial radius at the break-up threshold. 
Defining $R_\mathrm{p}$ as the star's polar radius at the break-up threshold, bulging at the equators of very rapidly rotating stars means that \mbox{$R_\mathrm{e} = 1.5 R_\mathrm{p}$} (\citealt{Maeder09}), which leads to
\begin{equation} \label{eqn:breakup}
\Omega_\mathrm{break} = \left( \frac{2}{3} \right)^{\frac{3}{2}} \left( \frac{G M_\star}{R_\mathrm{p}^3} \right)^{\frac{1}{2}}.
\end{equation}
We assume here that \mbox{$R_\mathrm{p} = R_\star$} where $R_\star$ is the star's radius in the absence of rotation and therefore do not consider the influence of rapid rotation on the polar radius, which is anyway likely to change $R_\mathrm{p}$ by only a few percent (\citealt{Ekstroem08}).
The evolution of $\Omega_\mathrm{break}$ is shown in Fig.~\ref{fig:breakup}.

For the exchanges of angular momentum between the core and the envelope, we adopt the approach used in \citet{MacGregorBrenner91}, \citet{Spada11}, and \mbox{\citet{GalletBouvier15}}, where the torque is given by
\begin{equation} \label{eqn:CEtorque}
\tau_\mathrm{ce} = \frac{\Delta J}{ t_\mathrm{ce} },
\end{equation}
where $t_\mathrm{ce}$ is the core-envelope coupling timescale and \mbox{$\Delta J$} is the angular momentum that must be transferred between the two components in order to make them rotate with the same speed.
We define this torque such that positive values imply angular momentum transfer from the core to the envelope, meaning that $\Delta J$ is given by
\begin{equation} \label{eqn:deltaJ}
\Delta J 
= 
\frac{ I_\mathrm{env} I_\mathrm{core} }{ I_\mathrm{env} + I_\mathrm{core} } 
\left( \Omega_\mathrm{core} - \Omega_\mathrm{env} \right),
\end{equation}
which implies that \mbox{$\Delta J=0$} when \mbox{$\Omega_\mathrm{core}=\Omega_\mathrm{env}$}.
For $t_\mathrm{ce}$, we assume 
\begin{equation} \label{eqn:tCE}
t_\mathrm{ce} = a_\mathrm{ce} \left( |\Omega_\mathrm{env} - \Omega_\mathrm{core}| \right)^{b_\mathrm{ce}} \left( \frac{M_\star}{M_\odot} \right)^{c_\mathrm{ce}},
\end{equation}
where $a_\mathrm{ce}$, $b_\mathrm{ce}$, and $c_\mathrm{ce}$ are fit parameters. 
This is similar to the assumption made by \citet{Spada11} and \citet{Johnstone19b} with the additional mass dependence.
As described in Appendix~\ref{appendix:fitting}, we find \mbox{$a_\mathrm{ce}=25.6$}, \mbox{$b_\mathrm{ce}=-3.25 \times 10^{-2}$}, and \mbox{$c_\mathrm{ce}=-0.448$} when $\Omega_\mathrm{env}$ and $\Omega_\mathrm{core}$ have units of $\Omega_\odot$ and $t_\mathrm{ce}$ has units of Myr.

\begin{figure*}
\centering
\includegraphics[width=0.95\textwidth]{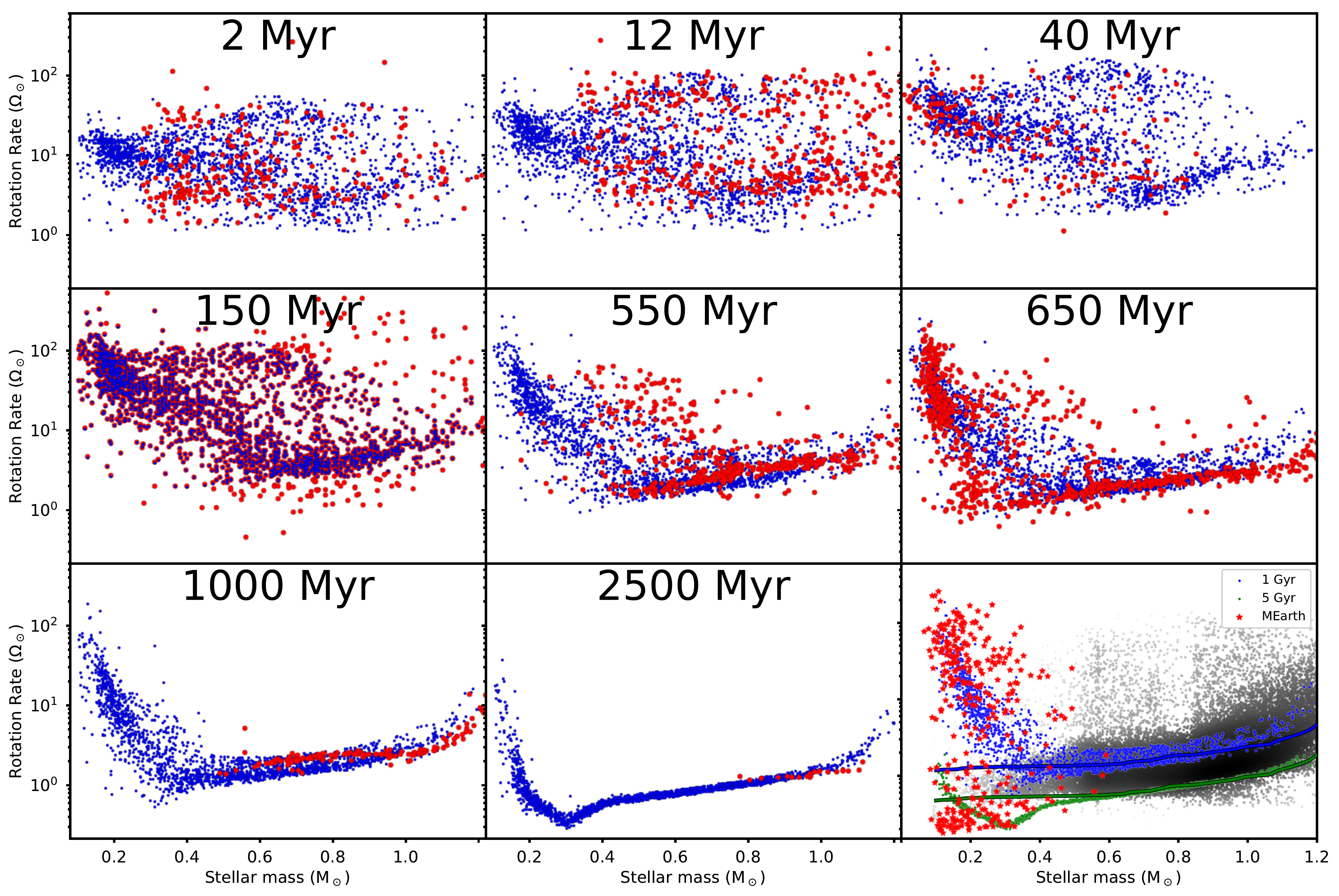}
\caption{
Stellar rotation distribution at each observed age bin listed in Table~\ref{table:rotcluster}.
Red and blue circles show observed and modelled distributions, with the modelled distributions calculated by evolving the observed 150~Myr distribution between 1 and 5000~Myr.
In the 150~Myr panel, red circles show stars that we are unable to fit with our rotation model using realistic initial rotation rates. 
In the lower right panel, black show the distribution measured by \emph{Kepler} (\citealt{McQuillan14}; \citealt{Santos19}), red stars show the distribution determined by the MEarth Project (\citealt{Newton16}), blue and green circles show the 1 and 5~Gyr modelled distributions, and solid lines show predictions from the gyrochrological relation of \citet{MamajekHillenbrand08}.
}
\label{fig:rotdist}
\end{figure*}

The core-growth torque, $\tau_\mathrm{cg}$, in Eqns.~\ref{eqn:dOmegaCoredt} and \ref{eqn:dOmegaEnvdt} represents a different type of angular momentum exchange between the core and the envelope.
The growth of the core at young ages means that material that is part of the envelope becomes part of the core, and this material possesses angular momentum that is therefore transported from the envelope to the core. 
Assuming a positive value corresponds to angular momentum transport from the envelope to the core, this is given by
\begin{equation} \label{eqn:CGtorque}
\tau_\mathrm{cg} = - \frac{2}{3} R_\mathrm{core}^2 \Omega_\mathrm{env} \frac{dM_\mathrm{core}}{dt},
\end{equation}
where $M_\mathrm{core}$ and $R_\mathrm{core}$ are the core mass and radius. 
During the pre-main-sequence phase, both $I_\mathrm{core}$ and $I_\mathrm{env}$ change rapidly due to two effects: firstly, the growth of the core increases $I_\mathrm{core}$ and decreases $I_\mathrm{env}$, and secondly, the radial distribution of mass changes.
This torque balances the contribution of the former effect to both $dI_\mathrm{env}/dt$ and $dI_\mathrm{core}/dt$ in Eqns.~\ref{eqn:dOmegaCoredt} and \ref{eqn:dOmegaEnvdt}, meaning that during the core growth phase, no unphysically rapid spin-down of the core takes place as $I_\mathrm{core}$ increases.
The above is valid when \mbox{$dM_\mathrm{core}/dt > 0$}, and when \mbox{$dM_\mathrm{core}/dt < 0$}, $\Omega_\mathrm{env}$ should be replaced by $\Omega_\mathrm{core}$.

The final ingredient in this model is disk-locking, which is an \mbox{\emph{ad hoc}} assumption that is commonly used to reproduce the lack of observed spin-up during the early pre-main-sequence when stars still possess circumstellar gas disks (e.g. \citealt{Allain98}; \citealt{GalletBouvier13}).
This lack of spin-up is surprising since these stars are contracting and accreting high angular momentum material from their disks, meaning that something must be removing angular momentum from the star during this phase. 
This could be enhanced stellar winds driven by the accretion of material from the disk onto the stellar surface (\citealt{MattPudritz08}; \citealt{Cranmer09}).
It is normal in rotation models to simply keep the star's surface rotation rate constant for the first few million years of the star's life. 
To do this, we assume a disk-locking torque acting on the envelope that cancels all other terms in Eqn.~\ref{eqn:dOmegaEnvdt}, given by
\begin{equation} \label{eqn:disklocking}
\tau_\mathrm{dl} = \left \{
\begin{array}{ll}
-\tau_\mathrm{w} - \tau_\mathrm{ce} - \tau_\mathrm{cg} + \Omega_\mathrm{env} \frac{d I_\mathrm{env} }{dt}, & \text{if } t \le t_\mathrm{disk},\\
0, & \text{otherwise}.\\
\end{array} \right.
\end{equation}
where $t_\mathrm{disk}$ is the disk-locking time.
As in \mbox{\citet{Tu15}} and \citet{Johnstone19b}, we assume
\begin{equation} \label{eqn:timeCE}
t_\mathrm{disk} = 13.5 \left( \frac{ \Omega_0 }{ \Omega_\odot } \right)^{-0.5},
\end{equation}
where $\Omega_0$ is the initial (1~Myr) rotation rate of the star in units of $\Omega_\odot$ and $t_\mathrm{disk}$ is in Myr. 
The inverse dependence means that the envelopes of fast rotators start spinning up earlier, which is consistent with the observed distribution of fast rotators in the young $\sim$12~Myr old cluster h~Per (\mbox{\citealt{Moraux13}}).
For $\Omega_0$ below $\Omega_\odot$, this equation can predict unreasonably large values of $t_\mathrm{disk}$ and to avoid this, we suggest setting a maximum value of 15~Myr, though we do not consider any cases with \mbox{ $\Omega_0 < \Omega_\odot$} in this paper.

The five unconstrained parameters in our models are $a_\mathrm{w}$ and $b_\mathrm{w}$ from Eqn.~\ref{eqn:MdotRossby}, and $a_\mathrm{ce}$, $b_\mathrm{ce}$, and $c_\mathrm{ce}$ from Eqn.~\ref{eqn:tCE}.
We determine these parameters using a large number of observational constraints, described in the next section, and a Markov Chain Monte Carlo method for parameter optimisation, described in Appendix~\ref{appendix:fitting}.
We obtain \mbox{$a_\mathrm{w}=-1.76$}, \mbox{$b_\mathrm{w}=0.649$}, \mbox{$a_\mathrm{ce}=25.6$}, \mbox{$b_\mathrm{ce}=-3.25 \times 10^{-2}$}, and \mbox{$c_\mathrm{ce}=-0.448$}. 
While these results could yield important information about stellar wind properties and angular momentum transport within stellar interiors, this is not the aim of this paper and we do not attempt physical interpretations of our parameter fitting results.

\subsection{Observational constraints} \label{sect:rotobs}

To constrain the parameters in our model, we use measured rotation distributions in young clusters.
We assume that the sequence of clusters provides a good representation of the rotational evolution of individual groups of stars, which is reasonable given that multiple clusters with similar ages have similar distributions (Fig.~14 of \citealt{Hartman10}).
A large number of the rotation rates that we use were presented in Section~3 of \mbox{\citet{Johnstone15b}} who collected data for seven clusters with ages between 100 and 1000~Myr: these were the Pleiades ($\sim$125~Myr), M50 ($\sim$130~Myr), M35 ($\sim$150~Myr), NGC~2516 ($\sim$150~Myr), M37 ($\sim$550~Myr), Praesepe ($\sim$580~Myr), and NGC~6811 ($\sim$1000~Myr).
We include additional newer rotation measurements for Pleiades, Praesepe, and NGC~6811.
We also include an additional five clusters, with rotation rates mostly measured by \emph{K2}, with ages between 12 and 2500~Myr.
These are h~Per ($\sim$12~Myr), NGC~2547 ($\sim$40~Myr), Hyades ($\sim$600~Myr), NGC~752 ($\sim$1300~Myr), and NGC~6819 ($\sim$2500~Myr).
We put the twelve clusters into seven age bins, with ages of 12, 40, 150, 550, 650, 1000, and 2500~Myr, as summarised in Table~\ref{table:rotcluster} with references to the original sources for the rotation period measurements.
For most clusters, we use stellar masses reported in the original rotation studies and rederive masses for Pleiades, M35, M37, and NGC~6819 using colour-mass conversions from the data given by \citet{PecautMamajek13}.

It is necessary in our fitting procedure also to have information about the rotation rates of stars with ages later than our oldest cluster, which has an age of $\sim$2.5~Gyr. 
One possibility is to use rotation measurements from \emph{K2} observations of the $\sim$4.0~Gyr cluster M67.
However, inconsistent results for the rotation distribution in this cluster have been found (\citealt{Barnes16}; \citealt{Gonzalez16b}; \citealt{Gonzalez16a}), and \citet{Esselstein18} showed that accurately determining the rotation periods from the available data is challenging, especially given their long 20-35~day rotation periods and the limited 75~day observation window.
Another problem with using M67 is that the age of the cluster is uncertain, and many age determinations are based on the gyrochonology method, which for our purpose is not useful. 
This is also a problem with using the large number of rotation periods measured for field stars.
In this paper, we use the gyrochronological relation given by \citet{MamajekHillenbrand08} to estimate rotation periods at 4.5~Gyr for each mass.
This is given by
\begin{equation}
P_\mathrm{rot} = a \left[ (B-V)_0 - c \right]^b t^n
\end{equation}
where \mbox{$a=0.407$}, \mbox{$b=0.325$}, \mbox{$c=0.495$}, \mbox{$n=0.566$}, $t$ is the age in Myr, and $P_\mathrm{rot}$ is in days.
We convert from $(B-V)_0$ to $M_\star$ using the data from \citet{PecautMamajek13}.

In Fig.~\ref{fig:rotdist}, we show as red circles the observed rotation distribution as a function of mass in each observed age bin. 
We additionally add for comparison the 2~Myr cluster NGC~6530 with rotation rates and masses determined by \citet{HendersonStassun12}.
We also show in the lower right panel the rotation distribution for approximately 40,000 field stars observed by \emph{Kepler} (black circles) using the rotation rates determined by \citet{McQuillan14} and \citet{Santos19} and for approximately 320 mid and late M~dwarfs (red stars) from the MEarth Project (\citealt{Newton16}).
Neither NGC~6530 nor the \emph{Kepler} and MEarth field stars are used in the parameter determination for our rotational evolution model.

\begin{figure}
\centering
\includegraphics[width=0.49\textwidth]{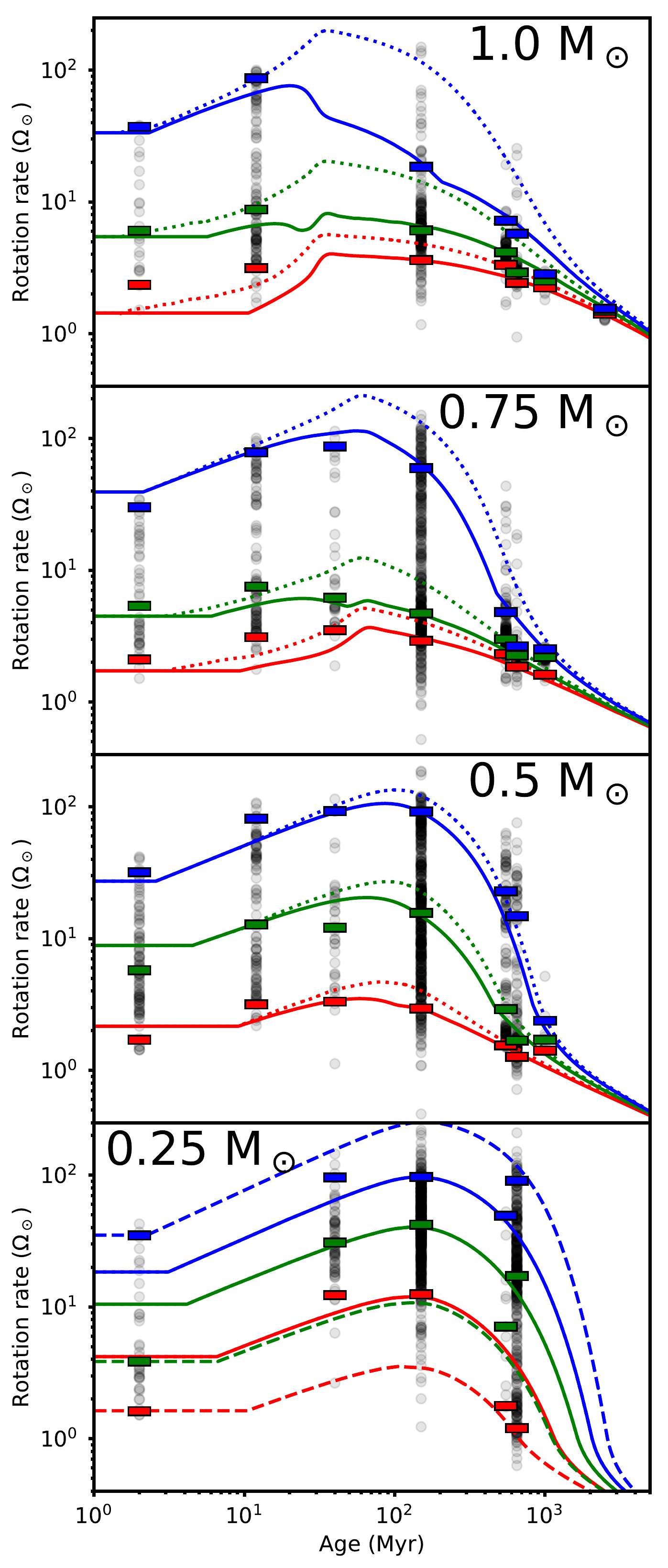}
\caption{
Rotational evolution for stars with masses of 1.0, 0.75, 0.5, and 0.25~M$_\odot$.
The red, green, and blue lines show our slow, medium, and fast rotator tracks with the solid and dotted lines representing the envelope and core rotation rates.
The slow, medium, and fast rotator are defined as the observed 5$^\mathrm{th}$, 50$^\mathrm{th}$, and 95$^\mathrm{th}$ percentiles at 150~Myr.
The grey circles show observed rotation rates and the short horizontal lines show the percentiles from the measurements in each age bin.
In the 0.25~M$_\odot$ mass bin, the dashed lines show the tracks that instead pass through the observed percentiles in the 2~Myr cluster NGC~6530.
}
\label{fig:rottracks}
\end{figure}

\subsection{Results: the evolution of rotation} \label{sect:rotresults}

In Fig.~\ref{fig:rotdist}, we show the observed 150~Myr distribution evolved between 2 and 2500~Myr using our rotation model at each observed age bin.
We use the 150~Myr cluster since it has the most stars and these stars are distributed over the entire mass range. 
The cluster evolution shown in Fig.~\ref{fig:rotdist} is as expected, with an initially wide distribution that becomes wider during the pre-main-sequence spin-up phase and then converges to a mass and age dependent value on the main-sequence as stars spin down. 

The 2~Myr panel in Fig.~\ref{fig:rotdist} shows the starting distribution implied by the 150~Myr cluster and our best fit rotation model.
This fits well the observed distribution in the $\sim$2~Myr cluster NGC~6530, though it appears that between 0.3 and 0.5~M$_\odot$, the observed distribution has more fast rotators. 
Our 2~Myr distribution is approximately mass independent down to 0.3~M$_\odot$, with stars distributed between 1 and 40~$\Omega_\odot$.
At lower masses, the distribution becomes much tighter and shifted towards faster rotation, which has good observational support from the 40~Myr old cluster NGC~2547.
This could have two intepretations: either the initial rotation distribution is mass dependent at such low masses in the way that we show, or the initial distribution is mass independent and the disk-locking times are shorter for such low masses. 
The latter is supported by the results of \citet{HendersonStassun12}, who found evidence for a mass-independent initial rotation distribution at 1--2~Myr followed by mass-dependent spin-up of stars with masses below 0.5~M$_\odot$ in the first 10~Myr (see their Fig.~16).
Since the activities of all of these stars are saturated, both possibilities lead to the same early activity evolution.


A feature clearly visible in Fig.~\ref{fig:rotdist} is the lack of observed spin-down between 650~Myr and 1000~Myr for stars with masses between 0.5 and 0.9~M$_\odot$ despite clear spin-down of solar mass stars, as has been discussed in the recent literature (\citealt{Agueros18}; \citealt{Curtis19}).
This epoch of stalled spin-down must be temporary since rotation measurements of field stars do not show a lack of slowly rotating K~dwarfs.
This lack of spin-down is not reproduced in our model, or any other rotation model to our knowledge, suggesting that additional physical processes are needed to explain the early main-sequence spin-down of K~dwarfs.
This could be a temporary reduction in angular momentum loss, due maybe to changes in the global magnetic field structures and strengths, or potentially a change in core-envelope coupling timescales causing more rapid angular momentum transfer from the core to the envelope.  
Interestingly, the 550~Myr age bin also shows a similar effect.

In the 150~Myr panel, the red circles show observed stars not included in our model distribution. 
While evolutionary tracks can be fit to all slow rotators in the 150~Myr distribution, we do not consider stars that require initial rotation rates below 1$\Omega_\odot$ since there is no observational support for such slow rotation at 1~Myr.
For the rapid rotators, there is a difficulty that we cannot fit realistic evolutionary tracks for some $\sim$1~M$_\odot$ stars, many of which are above the break-up threshold and therefore unrealistic.
While some stars are below break-up, they must have undergone significant spin-down since the zero-age main-sequence and our models suggest that they could not have arrived at their current rotation rates without being above break-up in the past.
{The 550 and 650~Myr clusters also show stars above the upper bound of our model distribution.
A similar difficulty was found in the models of \citet{Matt15}, who suggested that such stars require a modified torque, and this could suggest that at young ages, the torques for stars with the same basic parameters parameter (mass, age, rotation) are not uniform.}
It is also possible that these stars have or had short-period binary or planetary companions or it could be that the measured rotation periods are unrealistically short, which can happen for specific distributions of surface inhomogeneities.

In the lower right panel of Fig.~\ref{fig:rotdist}, we show our model distributions at 1 and 5~Gyr.
At 5~Gyr, the rotation distribution fits very well the expectations from the gyrochonological relation of \citet{MamajekHillenbrand08} at masses above 0.4~M$_\odot$ but shows a dip towards slow rotation for lower masses.
The lower bound of the main cloud of stars in the \emph{Kepler} distribution fits well these 5~Gyr expectations for masses above 0.6~M$_\odot$, but is above our expectations at lower masses.
This could suggest that our description of the later evolution of low mass stars is unrealistic, but it could also be that the \emph{Kepler} sample is missing the slowest rotators since their longer rotation periods and lower photometric variability make period determinations difficult (\citealt{McQuillan14}).
The latter interpretation is supported by the rotation distribution from the MEarth Project which has many M~dwarfs with periods slower than 100~days and is more consistent with our model predictions.
In our model, this dip is due to a peak at this mass in the convective turnover times, $\tau_\mathrm{c}$, that we use which causes more rapid spin down at masses around 0.3~M$_\odot$. 
This feature, realistic or not, does not influence the results of this paper since we use the same $\tau_\mathrm{c}$ values to calculate X-ray emission and the peak in $\tau_\mathrm{c}$ influences those calculations in such a way the feature is not present in our model X-ray distributions.

In Fig.~\ref{fig:rottracks}, we show rotation tracks for stellar masses of 1.0, 0.75, 0.5, and 0.25~M$_\odot$.
In each bin, we show tracks for slow, medium, and fast rotators, defined as the tracks that pass through the 5$^\mathrm{th}$, 50$^\mathrm{th}$, and 95$^\mathrm{th}$ percentiles of the observed 150~Myr rotation distribution. 
These percentiles are calculated using all stars within 0.1~M$_\odot$ of the specified masses and we use the distribution at 150~Myr since that is the most complete and has the largest number of stars. 
For the 1.0, 0.75, and 0.5~M$_\odot$ cases, the tracks clearly give an excellent description of the observed rotational evolution sequence, with the only issue being the slow rotator track being slower than the observed 5$^\mathrm{th}$ percentile in the first two clusters for the 1.0 and 0.75~M$_\odot$ cases.
For the 0.25~M$_\odot$ case, the solid lines are the tracks defined, as above, to go through the corresponding percentiles of the observed distribution at 150~Myr and the dashed lines are defined to go through the percentiles in NGC~6530.
In neither case, do these tracks give a good description at all ages of the evolution implied by the observed percentiles.
Since the saturation threshold for activity is at very low rotation rates for fully convective M dwarfs, this is only an uncertainty for the X-ray evolution after $\sim$2~Gyr.
The rotational evolution of these low mass stars, and the possibility that some fully convective M~dwarfs do not spin down significantly, will be studied in more detail in Bartel et al.~(in prep).


\section{X-ray evolution} \label{sect:xray}

\subsection{X-ray relations} \label{sect:Xrayrelation}

\begin{figure}
\centering
\includegraphics[width=0.49\textwidth]{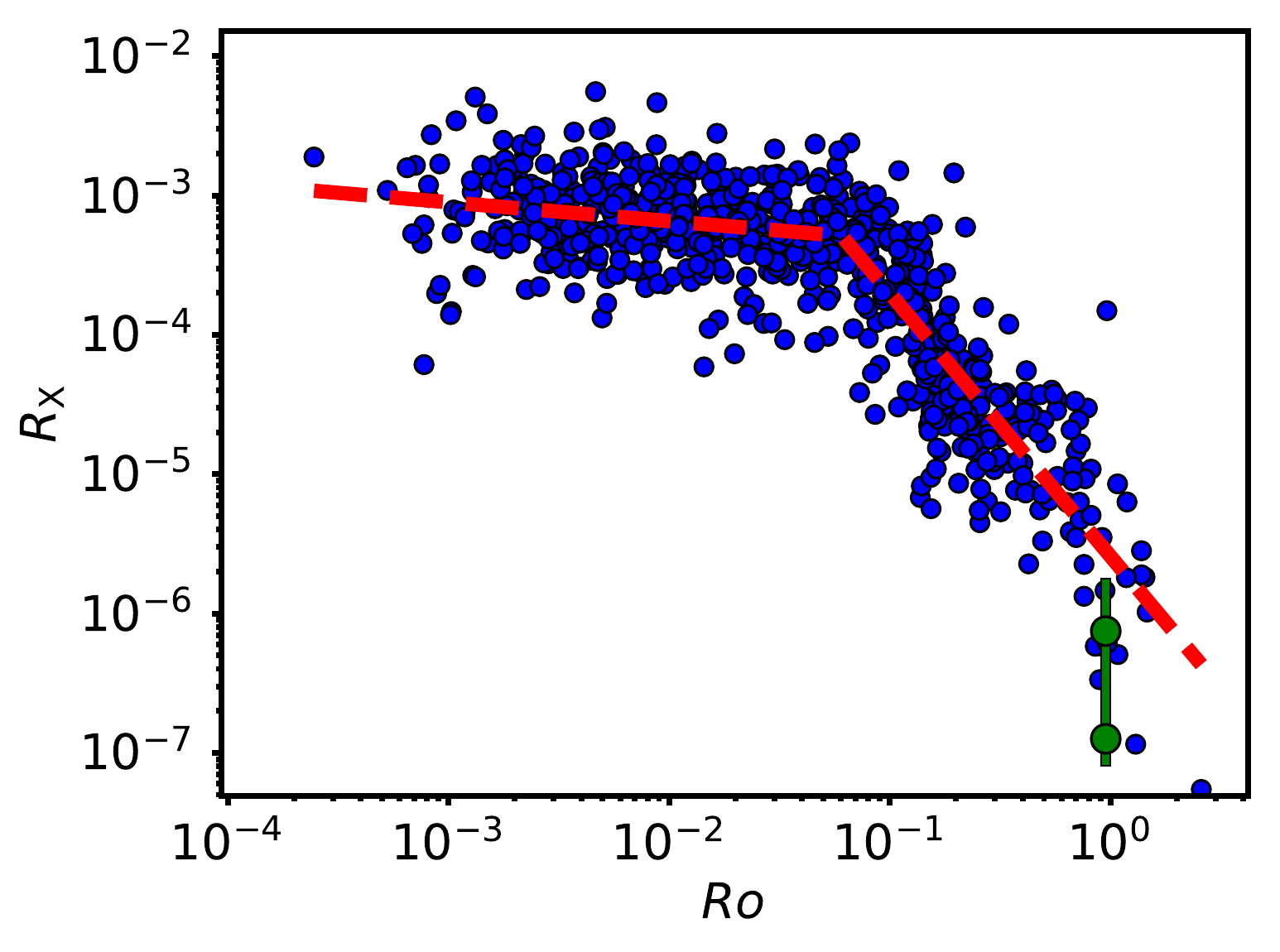}
\caption{
Stellar $R_\mathrm{X}$, defined as \mbox{$L_\mathrm{X}/L_\mathrm{bol}$}, as a function of Rossby number for the sample of main sequence stars collected by \citet{Wright11}.
The Rossby numbers were calculated using the convective turnover times given by \citet{Spada13}.
The dashed red line shows our best fit relation, given by Eqn.~\ref{eqn:RoRx}.
The green line shows the range of values for the non-flaring Sun, with the green circles showing the 10th and 90th percentiles of the Sun's X-ray luminosity. 
}
\label{fig:RoRx}
\end{figure}

\begin{figure}
\centering
\includegraphics[width=0.45\textwidth]{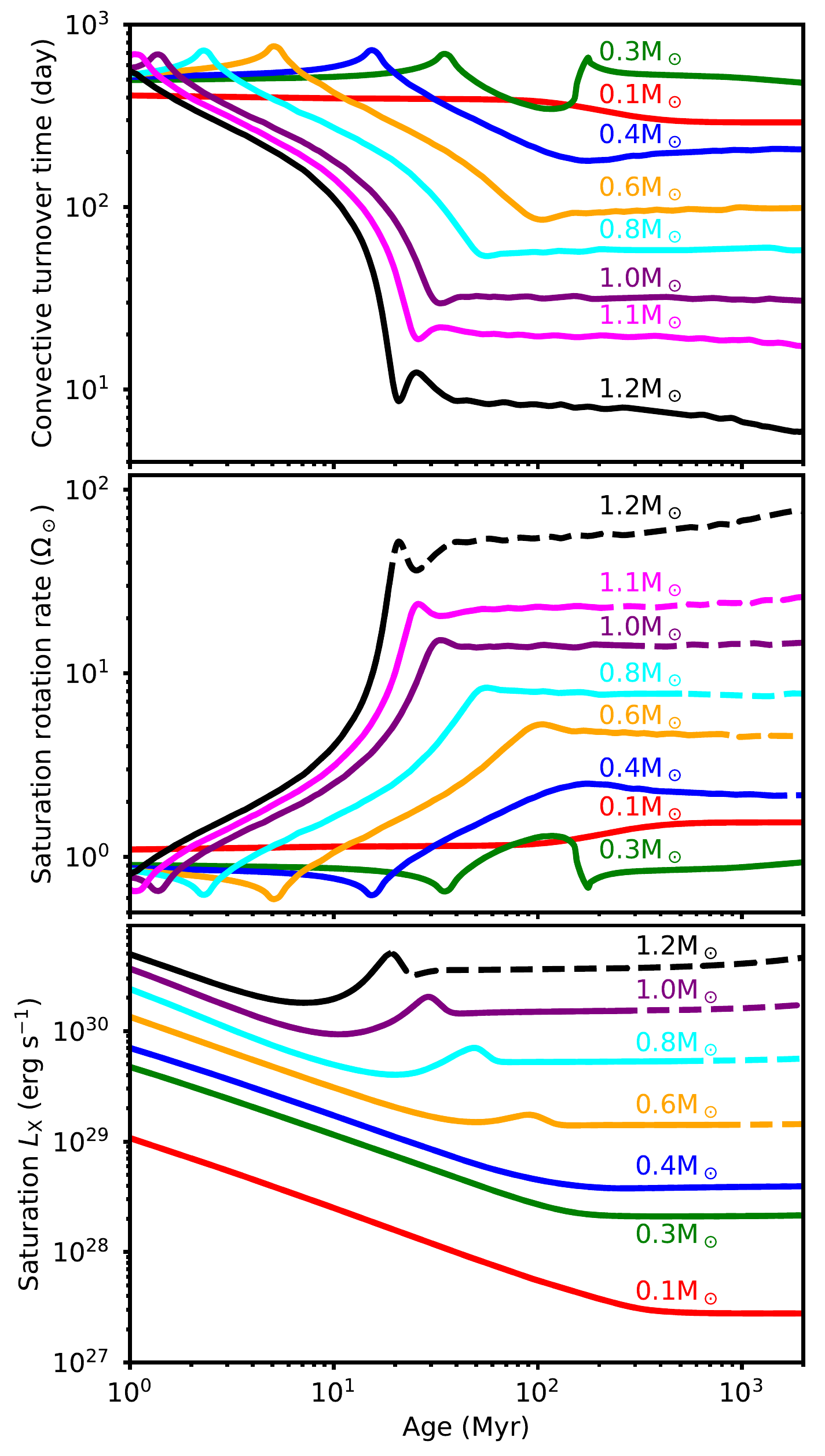}
\caption{
Convective turnover time (\emph{upper-panel}), the rotation rate of the saturation threshold (\emph{middle-panel}), and the X-ray luminosity at the saturation threshold (\emph{lower-panel}) as functions of age for different stellar masses.
In the middle and lower panels, the transition from solid to dashed lines shows the time in the evolution for each mass when 90\% of all stars have dropped below the saturation threshold.  
The convective turnover times are from \citet{Spada13} and the saturation threshold is calculated assuming a saturation Rossby number of 0.0605.
}
\label{fig:satthreshold}
\end{figure}

\begin{figure}
\centering
\includegraphics[width=0.45\textwidth]{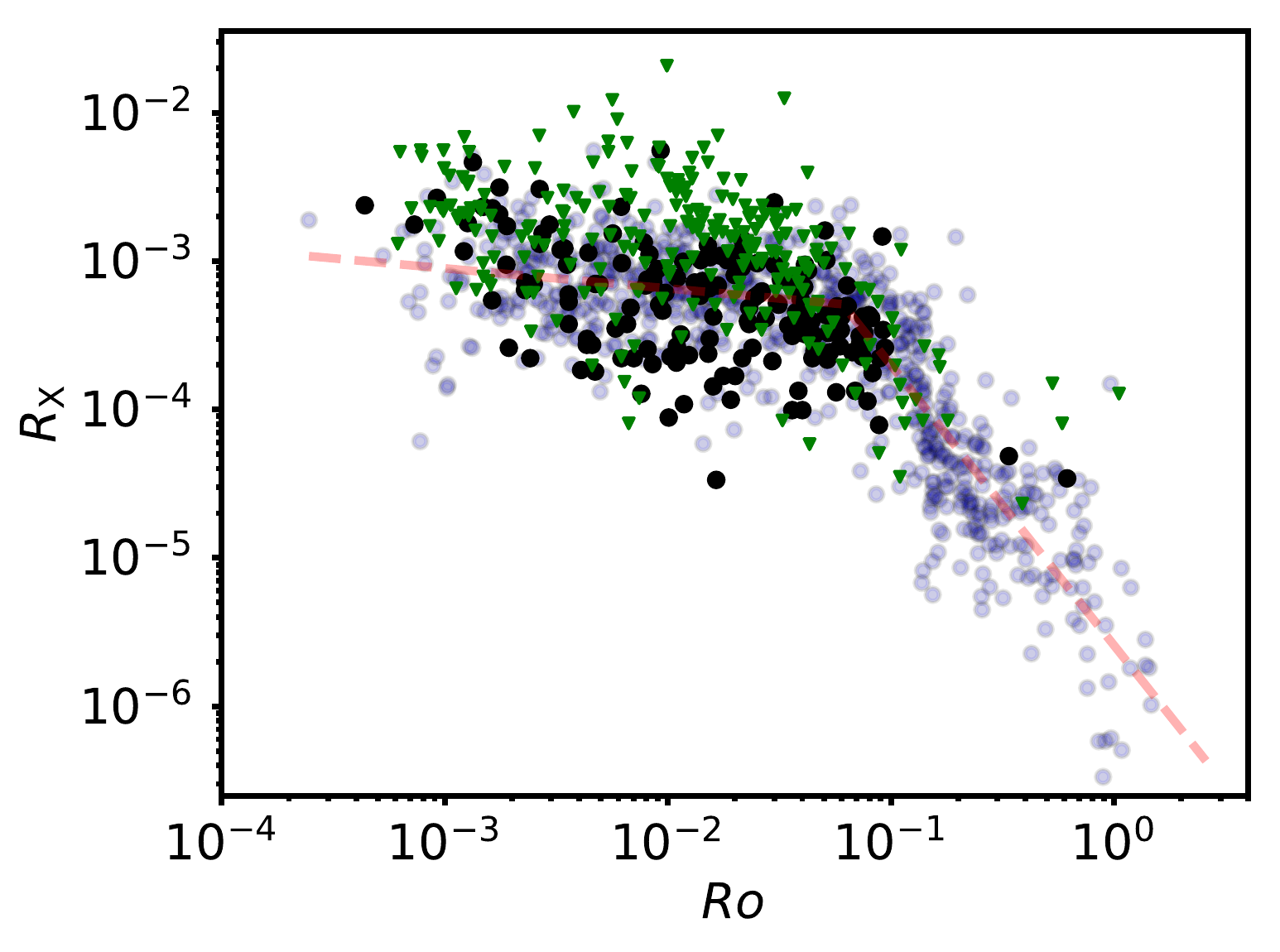}
\includegraphics[width=0.45\textwidth]{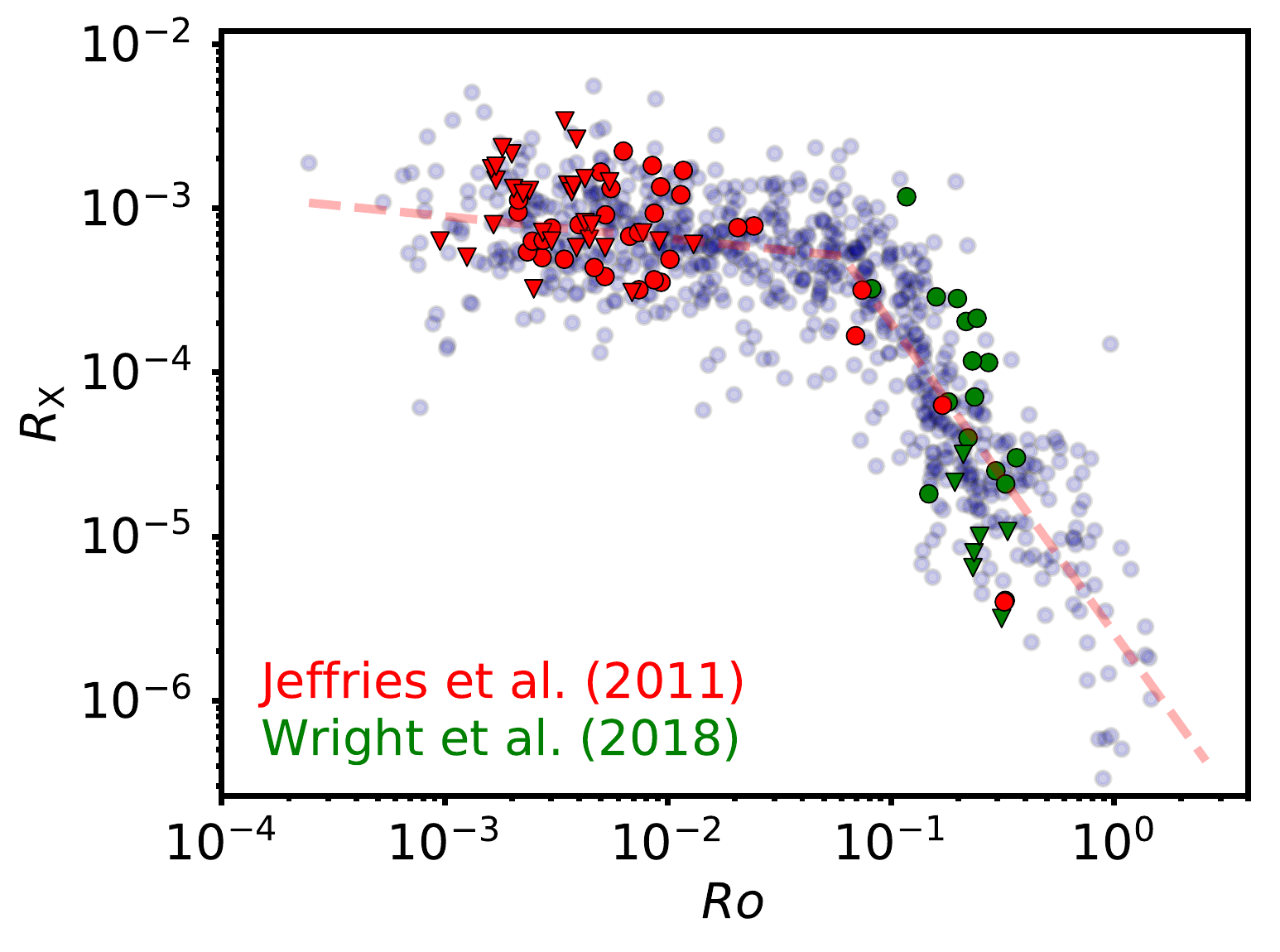}
\caption{
\emph{Upper panel:}
Comparison of the $Ro$--$R_\mathrm{X}$ distribution shown in Fig.~\ref{fig:RoRx} with the stars in the $\sim$12~Myr cluster h~Per with the black circles and green triangles showing X-ray detections and upper limits from \citet{Argiroffi16}.
We recalculate the Rossby numbers using the convective turnover times given by \citet{Spada13} assuming an age of 12~Myr.
\emph{Lower panel:}
Comparison of the $Ro$--$R_\mathrm{X}$ distribution shown in Fig.~\ref{fig:RoRx} with a sample of fully convective M~dwarfs from \citet{Jeffries11} and \citet{Wright18}, with detections and upper limits shown as circles and triangles.
}
\label{fig:RoRx_hPer}
\end{figure}

It is known empirically that a star's X-ray luminosity depends on its rotation rate, mass, and age and that this dependence can be broken down into the unsaturated and saturated regimes.
For main-sequence solar mass stars, the saturation threshold is at a rotation rate of approximately 15$\Omega_\odot$ and for lower mass stars, this threshold is at slower rotation rates, meaning that K and M~dwarfs need to spin down more than G~dwarfs before they enter the unsaturated regime. 
For pre-main-sequence stars, the saturation threshold is at much slower rotation, and in the first 10~Myr, all stars are saturated regardless of their rotation rates. 
The saturation X-ray luminosity is approximately a constant fraction of the bolometric luminosity meaning that among saturated stars, higher mass stars are more X-ray luminous and pre-main-sequence stars tend to be more X-ray luminous than their main-sequence counterparts.
A final factor is a significant spread in measured X-ray luminosities for stars with similar parameters, much or all of which is a result of random and cyclic variability.

The X-ray dependence on stellar parameters can be well described with a power-law dependence between the X-ray luminosity normalised to the bolometric luminosity, \mbox{$R_\mathrm{X} = L_\mathrm{X} / L_\mathrm{bol}$}, and the Rossby number, $Ro$.
In the unsaturated regime, the dependence can also be described by a single mass independence relation between $L_\mathrm{X}$ and rotation rate, and \citet{Reiners14} argued that this description is preferable.
In this paper, we use the $Ro$--$R_\mathrm{X}$ relation since it is able to describe the observed evolution of the saturation threshold for pre-main-sequence stars.
This relation can be described as a broken power-law given by
\begin{equation} \label{eqn:RoRx}
R_\mathrm{X} = \left \{
\begin{array}{ll}
C_1 Ro^{\beta_1}, & \text{if }  Ro \ge Ro_\mathrm{sat},\\
C_2 Ro^{\beta_2}, & \text{if }  Ro \le Ro_\mathrm{sat},\\
\end{array} \right.
\end{equation}
where $C_1$, $C_2$, $\beta_1$, and $\beta_2$ are constants to be determined empirically and $Ro_\mathrm{sat}$ is the saturation Rossby number. 
It is common in the literature to assume that $R_\mathrm{X}$ has a constant value of $R_\mathrm{X,sat}$ in the saturated regime, meaning that \mbox{$\beta_1 = 0$} and \mbox{$C_1 = R_\mathrm{X,sat}$}, though a weak dependence of $R_\mathrm{X}$ on $Ro$ has been pointed out in the literature (\citealt{Reiners14}; \citealt{Magaudda20}).
We make no assumption for $\beta_1$, but fit it to the observed data.
We do not consider the `supersaturation' phenomenon, which is a possible decrease in X-ray emission for the most rapid rotators in the saturated regime (\citealt{Jardine04}; \citealt{Argiroffi16}).

As described in Appendix~\ref{appendix:RoRx}, we constrain the constants in the above equation using the distribution of stars with known X-ray luminosities and rotation periods presented by \citet{Wright11}.
The values of these constants depend on the model or relation used to get the convective turnover times, $\tau_\mathrm{conv}$, and therefore to be consistent with our rotational evolution model, we use the values of $\tau_\mathrm{conv}$ from the stellar evolution models of \citet{Spada13}.
The result for our relation is \mbox{$\beta_1 = -0.135 \pm 0.030$}, \mbox{$\beta_2 = -1.889 \pm 0.079$}, \mbox{$Ro_\mathrm{sat} = 0.0605 \pm 0.00331$}, and \mbox{$R_\mathrm{X,sat} = 5.135 \times 10^{-4} \pm 3.320 \times 10^{-5}$}, where $R_\mathrm{X,sat}$ is the value of $R_\mathrm{X}$ at $Ro_\mathrm{sat}$.
The values of $C_1$ and $C_2$ can be derived from the fact that the two power-laws have equal values of $R_\mathrm{X}$ at the saturation point, meaning that \mbox{$R_\mathrm{X,sat} = C_1 Ro_\mathrm{sat}^{\alpha_1} = C_2 Ro_\mathrm{sat}^{\alpha_2}$}.
This relation is shown in Fig.~\ref{fig:RoRx} and the evolution of the saturation rotation rate and X-ray luminosity implied by our fit is shown in Fig.~\ref{fig:satthreshold}.
Our relation is shallower in the unsaturated regime than many previous estimates and this shallow relation suggests that the Sun is less X-ray active than other stars with similar parameters, which is consistent with results from other activity indicators (\citealt{Reinhold2020}). 

Since the stellar sample of \citet{Wright11} contains only main-sequence stars and does not contain unsaturated fully-convective M dwarfs, it is useful to check if the above scaling law applies on the pre-main-sequence and for all stellar masses, especially when combined with the convective turnover times that we use.
Most very young stellar clusters are not useful since the large convective turnover times put most stars in the saturated regime.
A useful cluster is h~Per since its 12~Myr age means that we might expect some slowly rotating solar mass stars to be unsaturated.
\citet{Argiroffi16} studied X-ray emission in h~Per and found evidence of a relation between $Ro$ and $R_\mathrm{X}$ for slowly rotating solar mass stars.
While their results suggest a much shallower power law relation than the relation found for main-sequence stars, this is based on a very small number of stars with X-ray detections and if we rederive the Rossby numbers using the convective turnover times from \citet{Spada13}, we find that the distribution for h~Per appears consistent with the distribution for main-sequence stars, as we show in the upper panel of Fig.~\ref{fig:RoRx_hPer}.

For fully-convective M dwarfs, \citet{Jeffries11} showed that stars with masses below 0.35~M$_\odot$ follow approximately the same $Ro$--$R_\mathrm{X}$ relation as higher mass stars, and this result has been supported by later studies (\citealt{WrightDrake16}; \citealt{Wright18}; \citealt{Magaudda20}).
In the lower panel of Fig.~\ref{fig:RoRx_hPer}, we show the $Ro$--$R_\mathrm{X}$ distribution for fully convective main-sequence M dwarfs from several studies with $Ro$ recalculated using the convective turnover times from \citet{Spada13}.
The low activity M dwarfs look mostly consistent with the entire sample, though it is notable that so many of the lowest activity M dwarfs are upper limits which could put them below the $Ro$--$R_\mathrm{X}$ distribution.
This could be consistent with the results of \citet{Magaudda20} who found a steeper slope in the unsaturated regime for fully convective stars.
We conclude that it is appropriate with our current knowledge to use the $Ro$--$R_\mathrm{X}$ relation constrained for main-sequence stars on the pre-main-sequence and for all stellar masses. 

The uncertainties given for the fit parameters in Eqn.~\ref{eqn:RoRx} are calculated using the bootstrap method and do not take into account uncertainties in the measured X-ray luminosities and rotation periods, uncertainties in the modelled $\tau_\mathrm{conv}$, biases in our sample of stars, and the method used to fit the relation. 
Several previous studies have estimated $\beta_2$ and a range of values have been derived. 
Considering the entire sample, \citet{Wright11} found \mbox{$\beta_2 = -2.18$} and from a subset of the sample they found \mbox{$\beta_2 = -2.7$}.
The difference between our estimate and their lower estimate is mostly because we use different convective turnover times.
Similarly, \citet{Reiners14} estimated \mbox{$\beta_2 = -2$} and showed that this depends on which stars are used in the fit and which method is used for the fitting, with the OLS($Y|X$) method consistently estimating shallower relations than the OLS Bisector method.
We use the former method for reasons discussed in Appendix~\ref{appendix:RoRx}.

\begin{figure}
\centering
\includegraphics[width=0.45\textwidth]{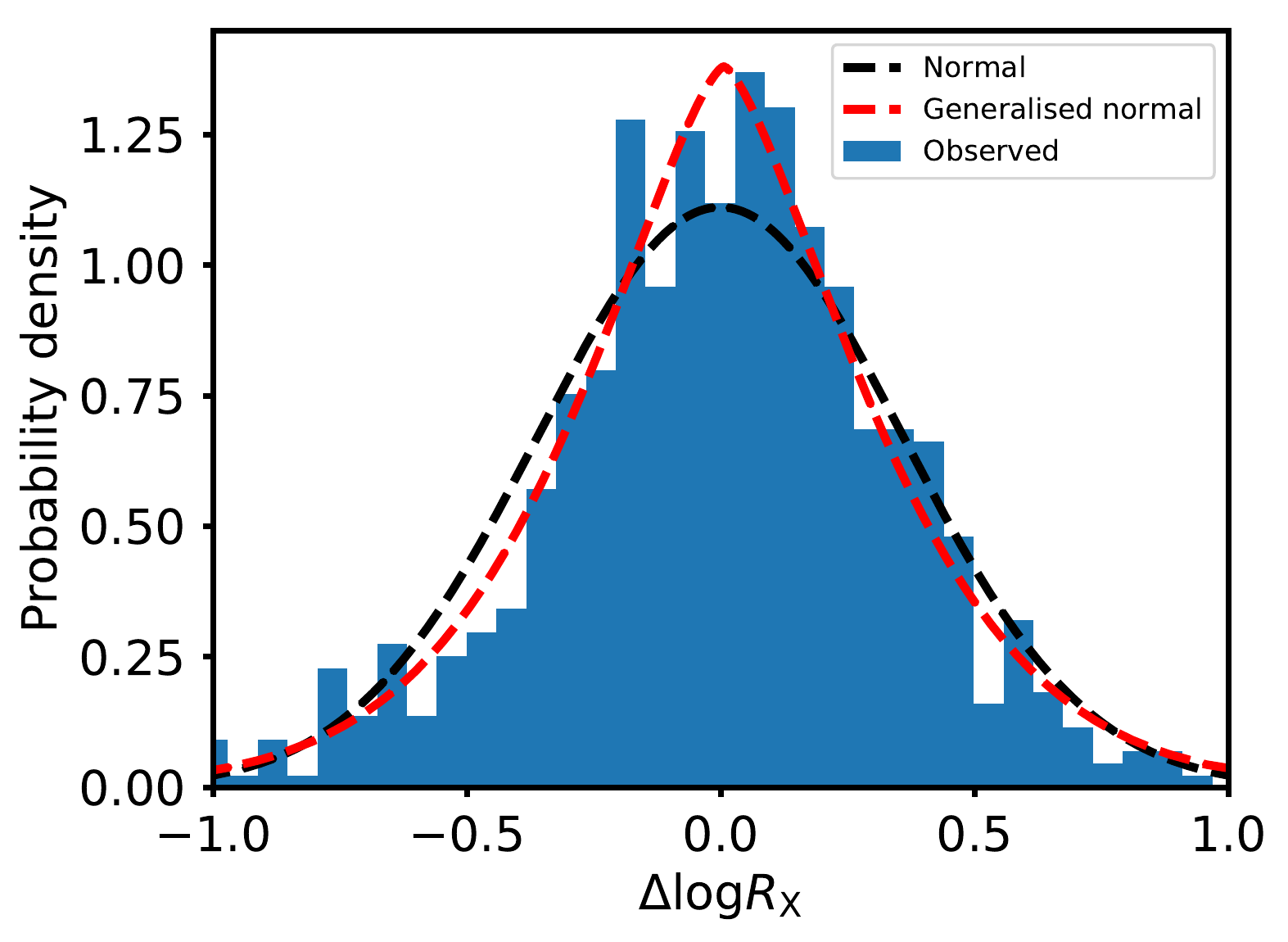}
\includegraphics[width=0.45\textwidth]{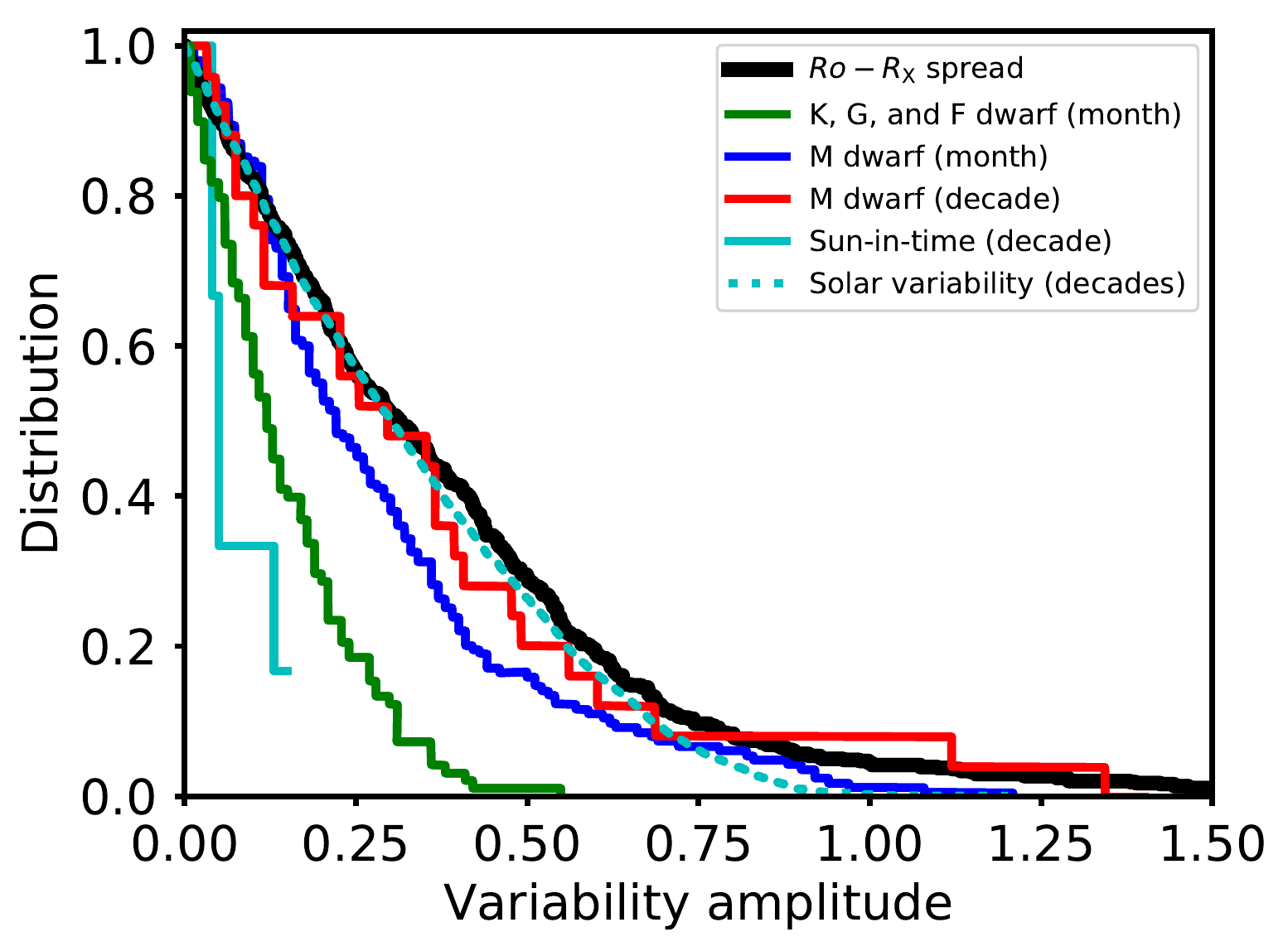}
\caption{ 
\emph{Upper-panel:}
Histogram showing the distribution of \mbox{$\Delta \log R_\mathrm{X}$}, defined as the difference between the $\log$ of the observed $R_\mathrm{X}$ and the $\log$ of the $R_\mathrm{X}$ from our fit formula. 
The black line shows a normal distribution with $\mu=0$ and $\sigma=0.359$ and the red line shows the generalised normal distribution with a probability density function given by \mbox{$\beta \exp \left[ - \left( |\Delta \log R_\mathrm{X} - \mu| / \alpha \right)^\beta \right] / \left[ 2 \alpha \Gamma \left( 1 / \beta \right) \right]$}, where $\beta=1.43$, $\mu=0$, $\alpha=0.4$, and $\Gamma$ is the gamma-function.
\emph{Lower-panel:}
Cumulative distribution functions for the variability amplitudes of several samples of stellar $L_\mathrm{X}$ measurements. 
The black line represents the spread in the $Ro$--$R_\mathrm{X}$ distribution, the dotted line represents variability of the Sun, and the other lines show variability distributions derived from the literature as described in Section~\ref{sect:variability}.
}
\label{fig:RoRxSpread}
\end{figure}

\subsection{Variability and the nature of the spread} \label{sect:variability}

One reason why the parameters in Eqn.~\ref{eqn:RoRx} are difficult to determine is the large spread of approximately two orders of magnitude in the $Ro$--$R_\mathrm{X}$ distribution.
In Fig.~\ref{fig:RoRxSpread}, we show the distribution of \mbox{$\Delta \log R_\mathrm{X}$}, defined as the difference between the measured \mbox{$\log R_\mathrm{X}$} and the value predicted by Eqn.~\ref{eqn:RoRx}.
The distribution can be described as a normal distribution centered around zero with a standard deviation of 0.359, or as a generalised normal distribution as described in Fig.~\ref{fig:RoRxSpread}.
There are three factors that could cause this spread and it is important to understand the contributions of each.
Firstly, inaccuracies in measurements of the X-ray luminosities should cause some spread, and the sample we use is especially susceptible to this since it contains X-ray measurements from several instruments, especially \emph{Einstein}~IPC, ROSAT~PSPC, and \emph{XMM-Newton}.
Each instrument is sensitive to a different photon energy range and the count-rate to flux conversions used to calculate fluxes in the entire X-ray range are often derived using different methods.
Secondly, random and cyclic variability surely play a substantial role in causing the spread in the $Ro$--$R_\mathrm{X}$ relation. 
The magnitude of the variability for the Sun, not including flares, is shown in green in Fig.~ \ref{fig:RoRx} and is similar to but smaller than the overall spread.
Thirdly, the spread could be caused by intrinsic differences between stars, with some stars being intrinsically more active that other stars with similar masses, ages, and rotation rates.

Several studies have estimated the X-ray variability of individual stars.
\citet{Stern95} and \citet{Schmitt95} combined measurements from \emph{Einstein} and ROSAT to study variability in the Hyades and nearby K and M dwarfs on approximately decadal timescales and found that $L_\mathrm{X}$ typically varies by less than a factor of 2.
Some of this variability is cyclic in nature since activity cycles have been seen in other stars (\citealt{BoroSaikia18}) and much of the variability is a result of stellar flares (\citealt{Guedel04}; \citealt{Stelzer07}), though for our purposes, the exact nature of the variability is not important.
\citet{Marino00} and \citet{Marino02} studied variability on timescales of a few months for M, K, G, and F dwarfs with multiple ROSAT measurements and found that on these timescales, M dwarfs are more variable than their higher mass counterparts.
\citet{Marino03} found similar results among a more active sample of stars in the Pleiades.
By comparing with \emph{Einstein} measurements, \citet{Marino00} found larger variability for M dwarf on 8--15~year timescales.
These studies did not find evidence that the variability amplitude depends on activity level or age, consistent with the fact that the spread in the $Ro$--$R_\mathrm{X}$ distribution is also independent of $Ro$. 
They defined the variability amplitude for stars with more than one X-ray luminosity measurement as \mbox{$| \log L_{\mathrm{X},1} - \log L_{\mathrm{X},2} |$} and since they considered only variability of individual stars, this is equivalent to \mbox{$| \log R_{\mathrm{X},1} - \log R_{\mathrm{X},2} |$}.

To compare these results to the spread in the $Ro$--$R_\mathrm{X}$ distribution, we calculate a distribution of variability amplitudes from the sample of stars used to constrain the $Ro$--$R_\mathrm{X}$ relation.
For every star in this sample, we calculate \mbox{$| \log R_{\mathrm{X},1} - \log R_{\mathrm{X},2} |$}, where $\log R_{\mathrm{X},1}$ is the value for that star and $\log R_{\mathrm{X},2}$ is the value for the star with the nearest Rossby number. 
The reason to use $R_\mathrm{X}$ here is that this corrects for different bolometric luminosities. 
We also calculate a variability amplitude distribution for the six stars in the Sun-in-time sample using $L_\mathrm{X}$ measurements by \citet{Guedel97} and \citet{Telleschi05}.
These measurements, summarised in Table~1 of \citet{Telleschi05}, were based on ROSAT and XMM-Newton observations taken 7--10~years apart. 
Finally, we calculate also the solar X-ray variability using daily average $L_\mathrm{X}$ values derived from the <10~nm part of the quiescent (non-flaring) solar spectra provided by the Flare Irradiance Spectrum Model (\citealt{Chamberlin07}) which uses activity proxies to reconstruct the full solar spectrum.
For the solar distribution, we calculate the variability amplitudes for 10,000 pairs of days randomly selected from the time period 1948--2010.

\begin{table}
\centering
\begin{tabular}{ccc}
Year & Cluster name & Reference \\
\hline
2~Myr & Taurus & \citet{Guedel07} \\
12~Myr & h~Per &  \citet{Argiroffi16} \\
40~Myr & NGC~2547 & \citet{Nunez16} \\
50~Myr & $\alpha$~Per & \citet{Prosser96} \\
 & & \citet{Prosser98} \\
 & Blanco~I & \citet{Pillitteri03} \\
150~Myr & NGC~2516 & \citet{Nunez16} \\
 & Pleiades & \citet{Nunez16} \\
300~Myr & NGC~6475 & \citet{Nunez16} \\ 
550~Myr & M37 & \citet{Nunez16} \\
650~Myr & Hyades & \citet{Nunez16} \\
\hline
\end{tabular}
\caption{
Sources of the observational constraints on X-ray evolution used in this paper. 
}
\label{table:xraycluster}
\end{table}

\begin{figure*}
\centering
\includegraphics[width=0.95\textwidth]{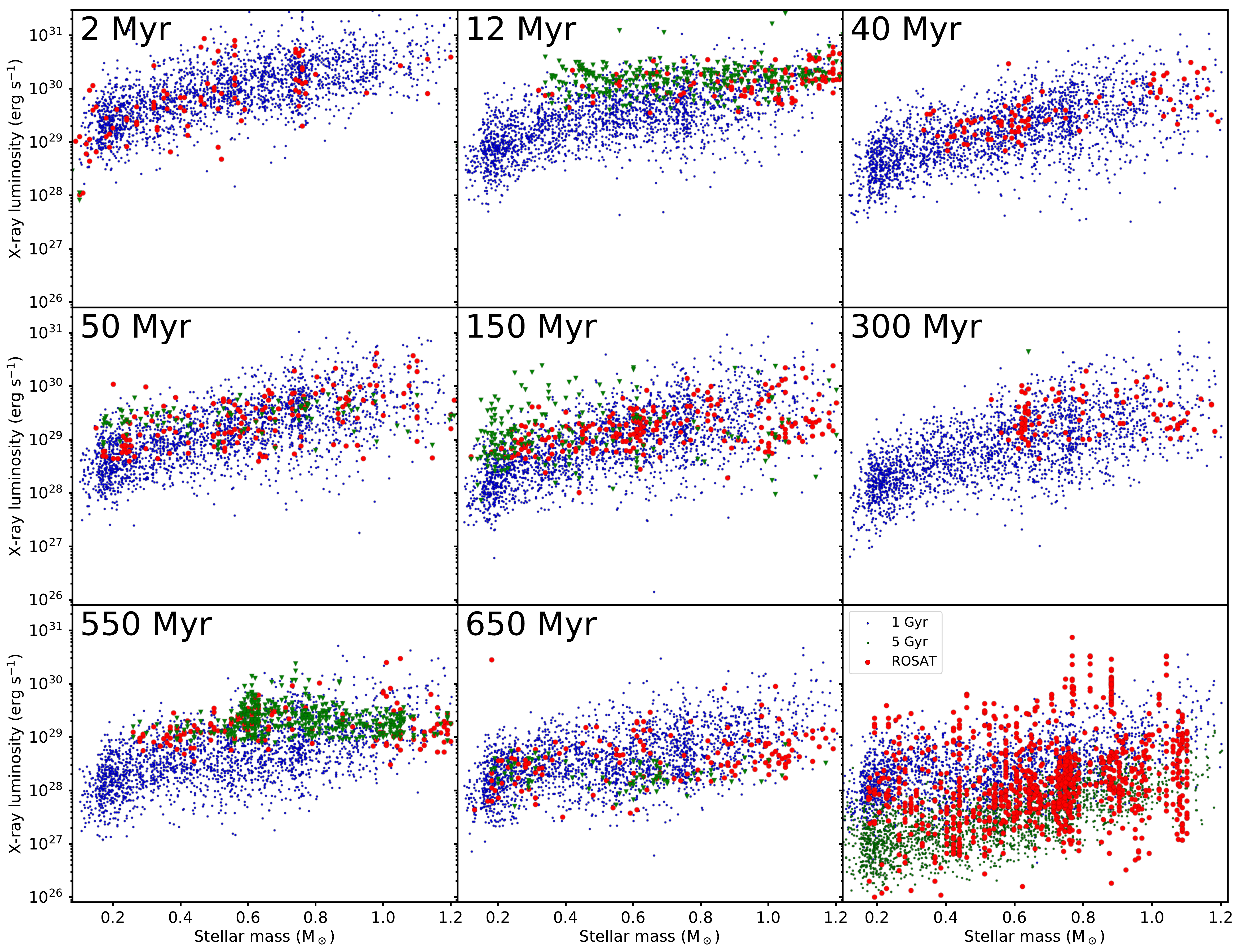}
\caption{
Comparison between stellar X-ray luminosity distributions for several young clusters and our predicted distributions at these ages.
In each panel, the blue circles shows our model distribution, the red circles show measured values, and the green triangles show upper limits.
The final panel shows X-ray luminosities determined by ROSAT for nearby field stars (red circles) compared to our model distributions at ages of 1 and 5~Gyr.
}
\label{fig:XrayClusters}
\end{figure*}

In Fig.~\ref{fig:RoRxSpread}, we show the variability amplitude distribution for the sample of stars used to constrain the $Ro$--$R_\mathrm{X}$ relation and compare it to measured variability distributions from \citet{Marino00} and \citet{Marino02}.
The blue and green lines show variability on monthly timescales for M dwarf and K, G, and F dwarfs respectively, and these distributions are clearly inconsistent with the spread in the \mbox{$Ro$--$R_\mathrm{X}$} distribution.
The red line shows variability for M dwarfs on the longer timescale of 8--15 years and this matches the spread distribution much better, suggesting that the spread in the \mbox{$Ro$--$R_\mathrm{X}$} relation is consistent with variability on longer timescales.
The solar variability is consistent with the spread for small variability amplitudes but is inconsistent with the high variability tail of the spread, possibly because we use non-flaring solar spectra only.
Strangely, the much smaller Sun-in-time sample shows far less variability than all other samples which could be a result of having only six stars in the sample.
Using Kolmogorov–Smirnov tests, we find that the consistency between the spread in the $Ro$--$R_\mathrm{X}$ distribution and the variability in each of the other distributions can be rejected, except for the long-term M dwarf variability which cannot be rejected.

In conclusion, while the nature of the observed spread in the $Ro$--$R_\mathrm{X}$ distribution remains unclear, it appears unnecessary to assume that the spread suggests intrinsic differences between the X-ray activities of stars with similar masses, ages, and rotation rates. 
Despite the uncertainties, it is reasonable to assume that the spread is mostly caused by stellar variability on timescales of decades or longer.
If true, stars have $L_\mathrm{X}$ values more than twice the average 20\% of the time, more than three times the average 9\% of the time, and more than five times the average 3\% of the time, and they spend similar amounts of time with these factors below the average. 
These variations could be important when considering the effects of stellar XUV emission on the long-term evolution of planetary atmospheres.

\begin{figure*}
\centering
\includegraphics[width=0.94\textwidth]{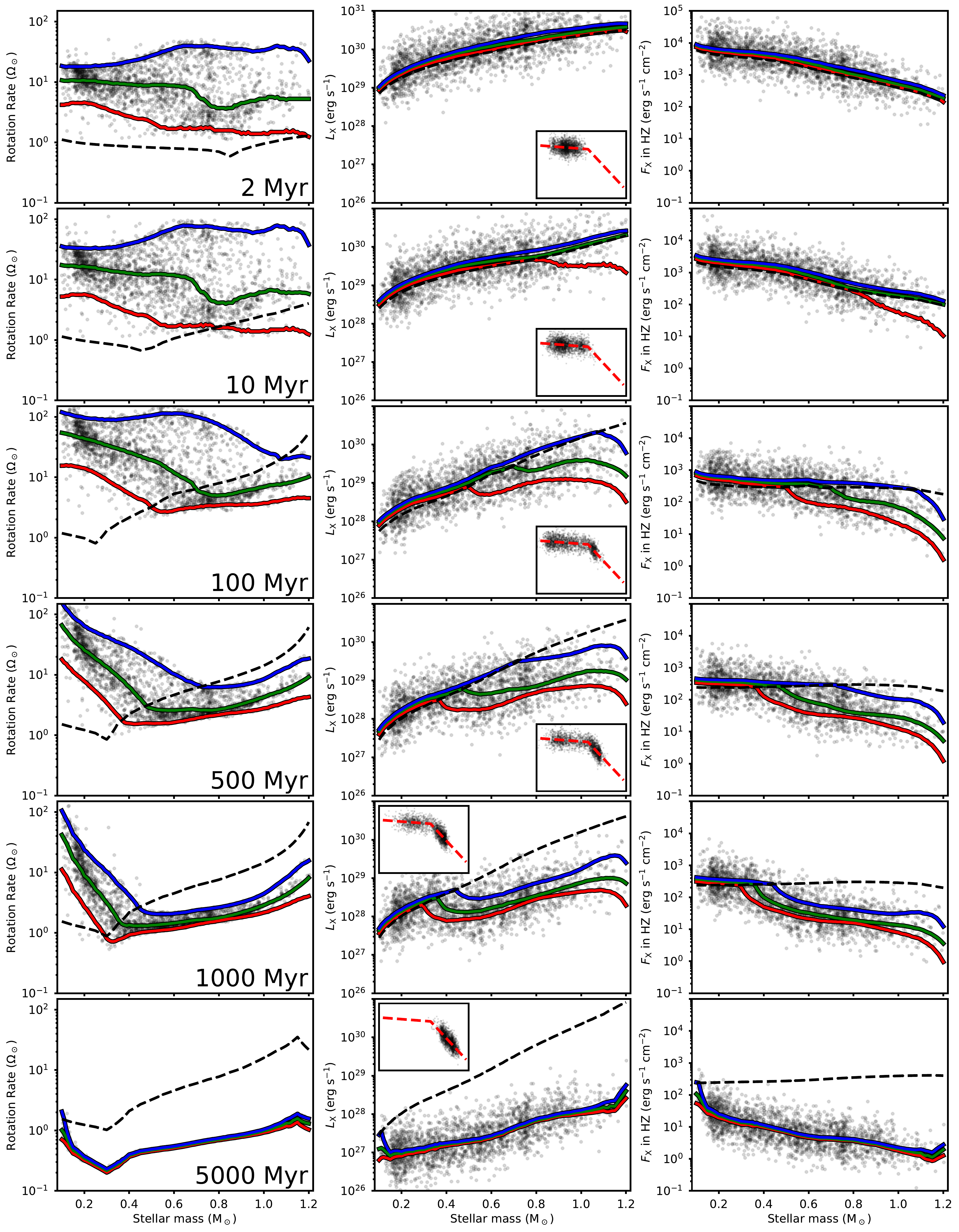}
\caption{
Evolution of the distributions of rotation (\emph{left-column}), X-ray luminosity (\emph{middle-column}), and X-ray flux in the habitable zone (\emph{right-column}) against stellar mass, with each row showing the ages labeled in the left column.
In each panel, the black circles show the model distribution, the red, green, and blue lines show the values for the slow, medium, and fast rotator cases, and the dashed black line shows the saturation threshold values.  
In the middle column, the insets show $R_\mathrm{X}$ against $Ro$ for the model distribution, with the red line showing the empirical relation given by Eqn.~\ref{eqn:RoRx}.
The X-ray flux in the habitable zone is defined as the flux at an orbital distance half-way between the moist and maximum greenhouse limits and these limits are calculated at all ages assuming the stellar properties at 5~Gyr.
}
\label{fig:XrayEvo}
\end{figure*}

\begin{figure}
\centering
\includegraphics[width=0.45\textwidth]{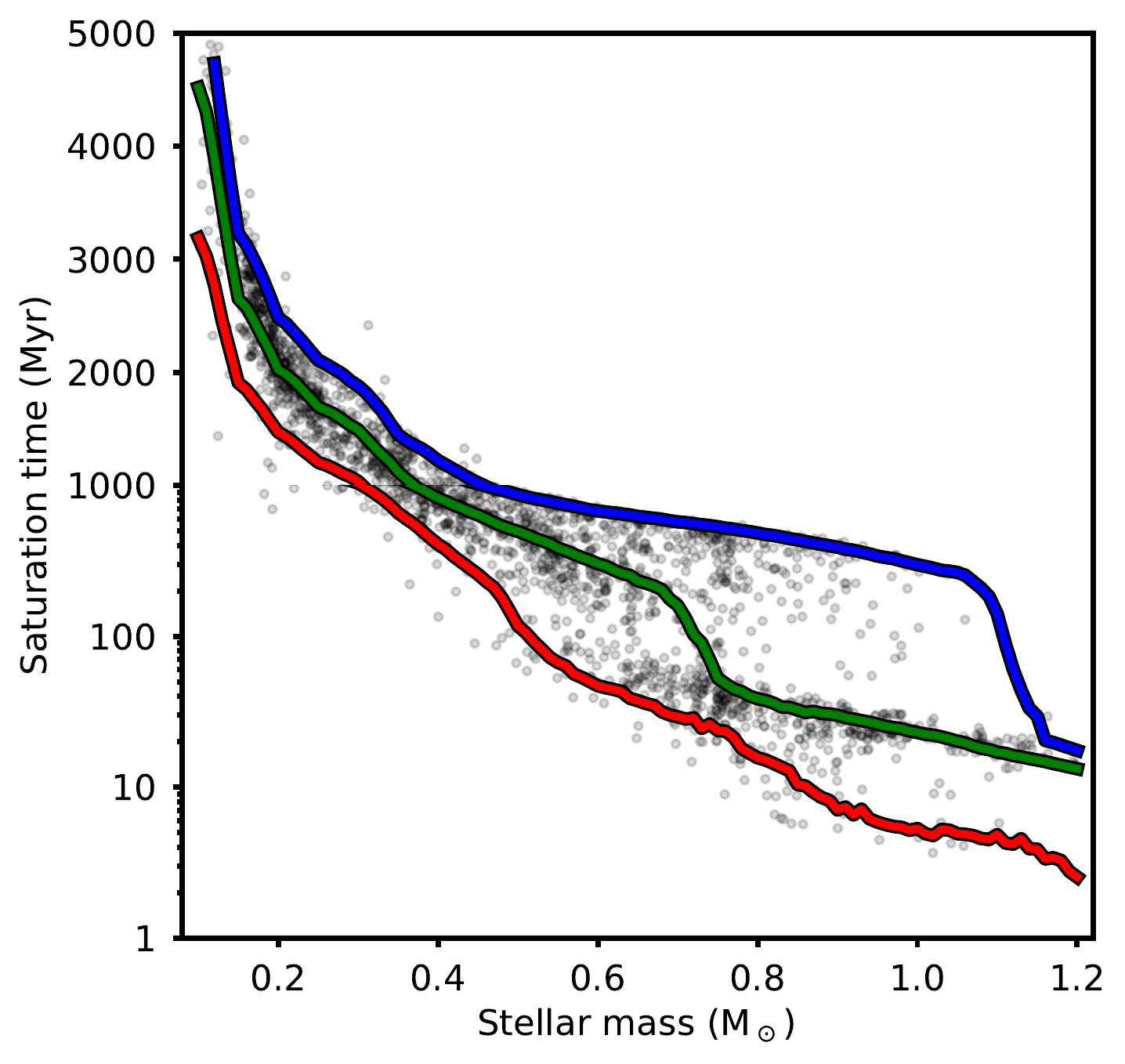}
\caption{ 
The age at which a star falls below the saturation threshold as a function of stellar mass for slow (red), medium (green), and fast (blue) rotators.
The background circles show when each of the stars in our model distribution become unsaturated. 
Note that the y-axis is logarithmic for ages up to 1000~Myr and linear at later ages. 
}
\label{fig:satage}
\end{figure}

\begin{figure}
\centering
\includegraphics[width=0.41\textwidth]{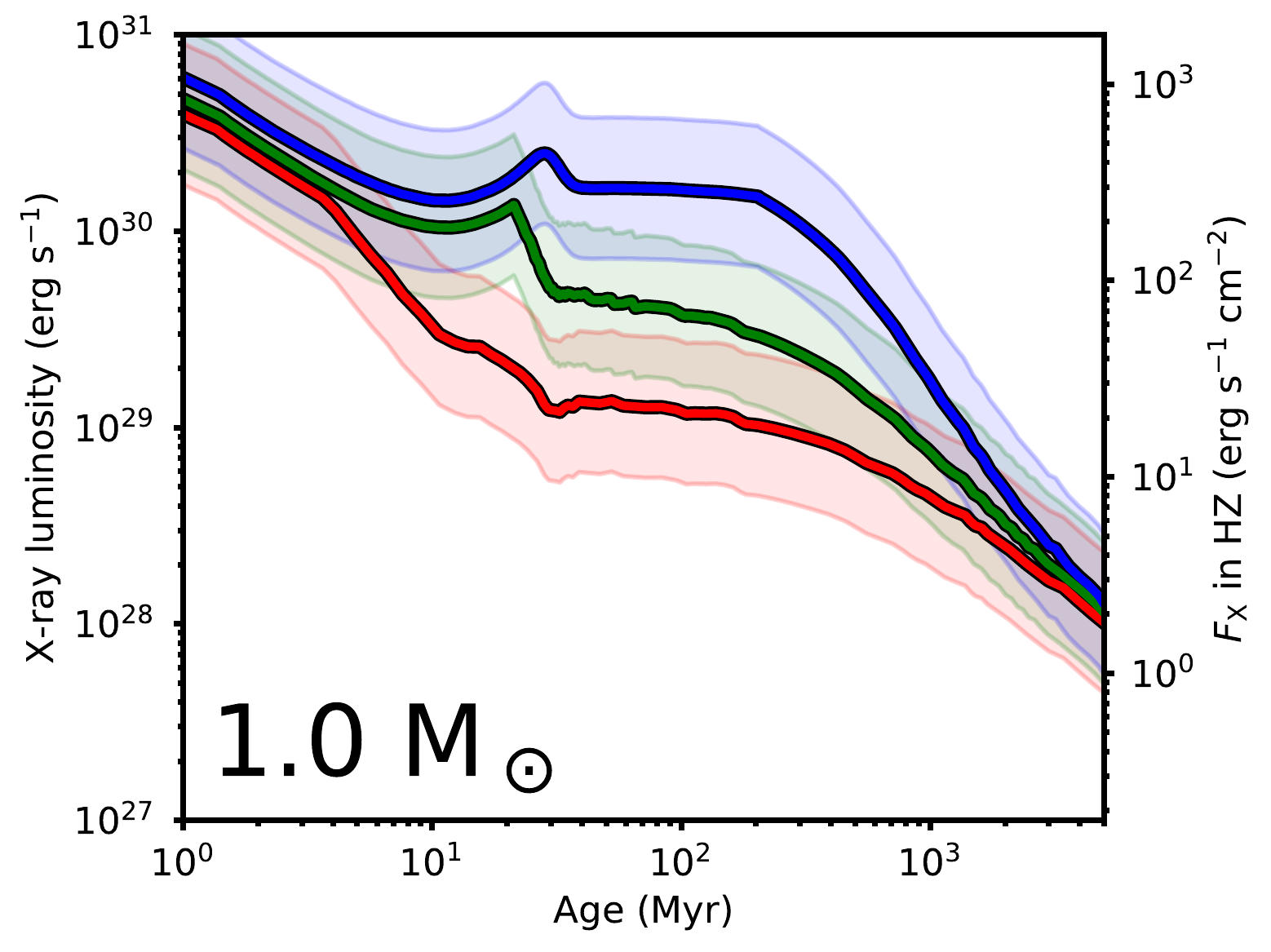}
\includegraphics[width=0.41\textwidth]{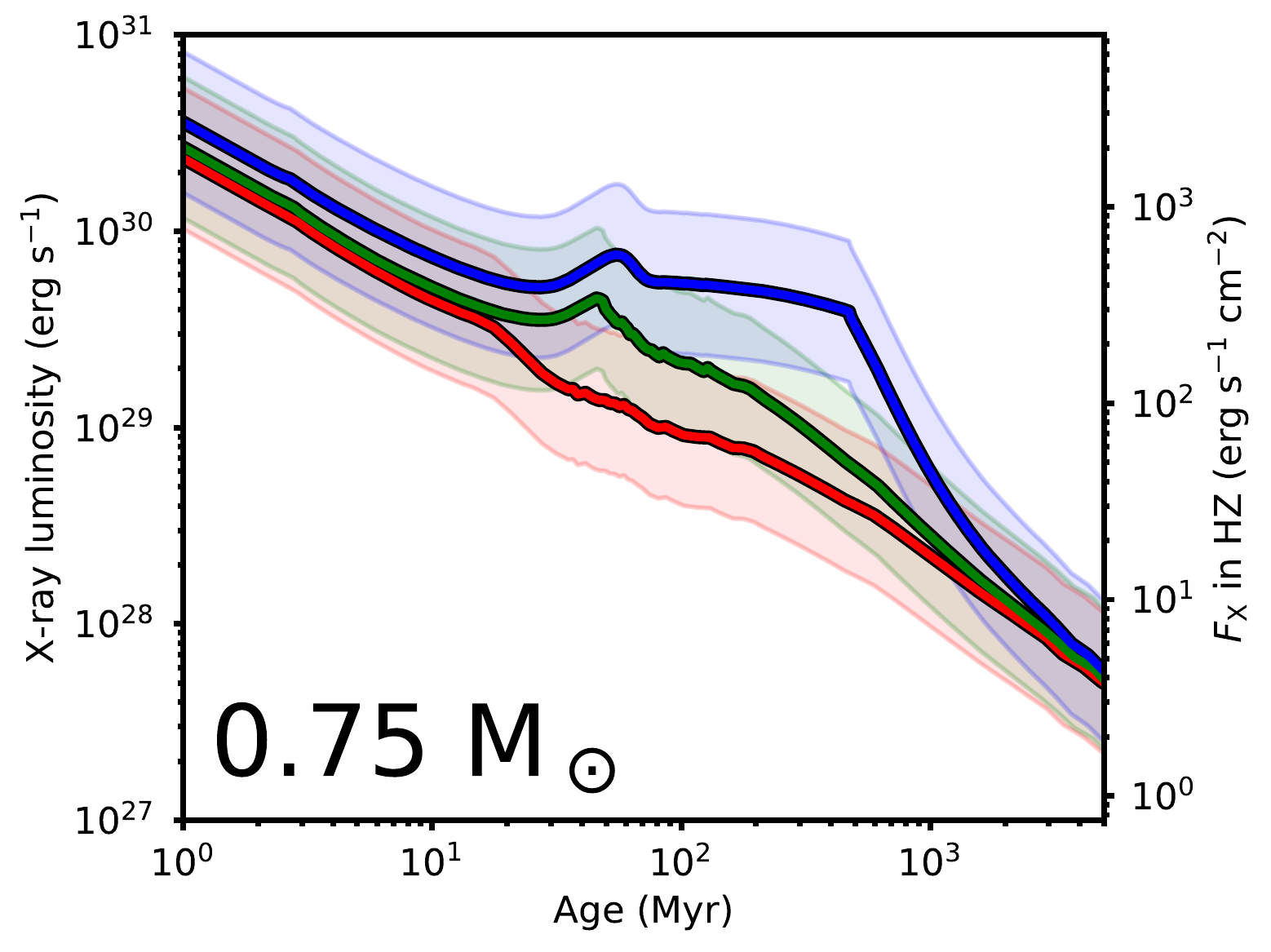}
\includegraphics[width=0.41\textwidth]{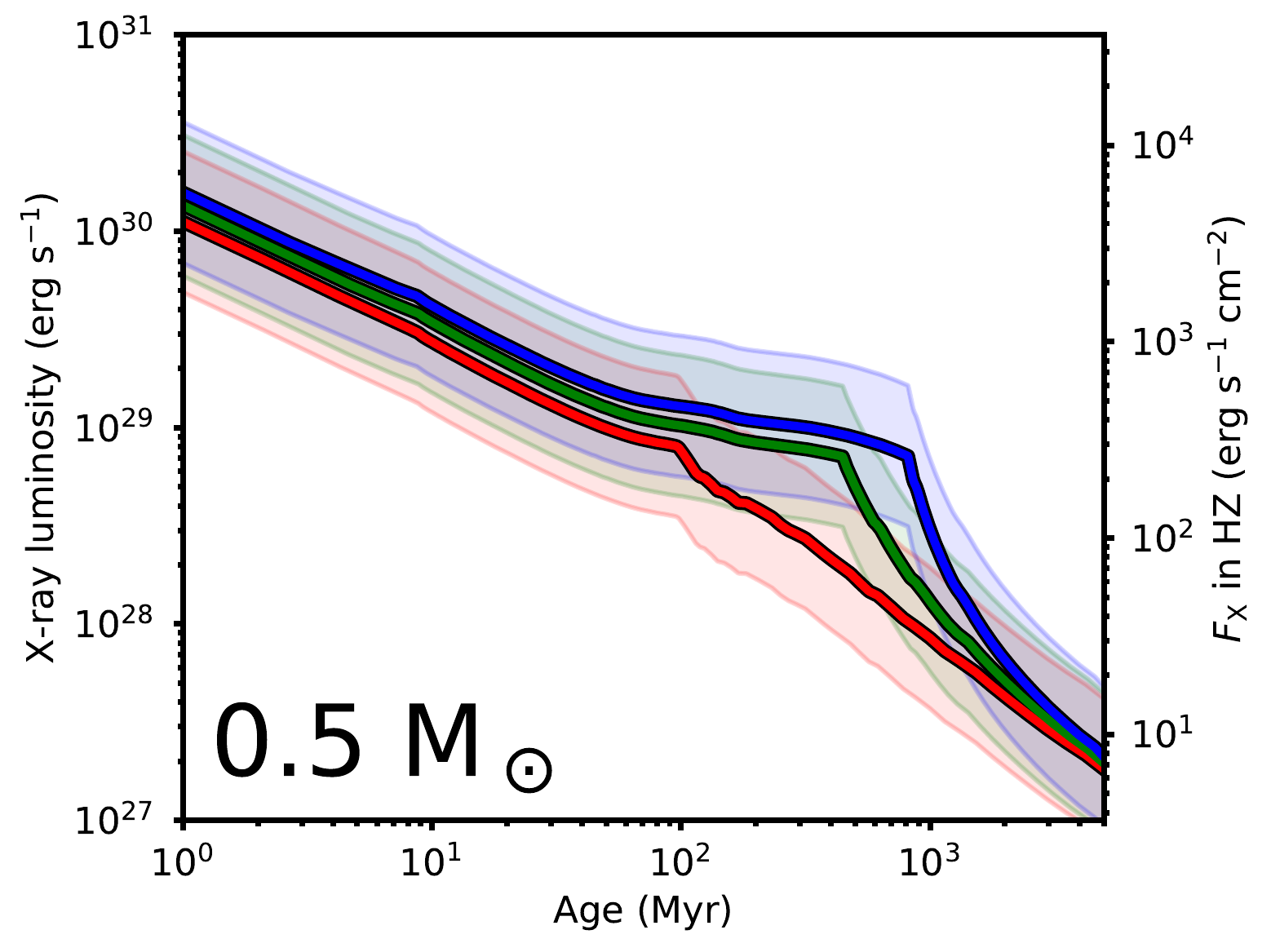}
\includegraphics[width=0.41\textwidth]{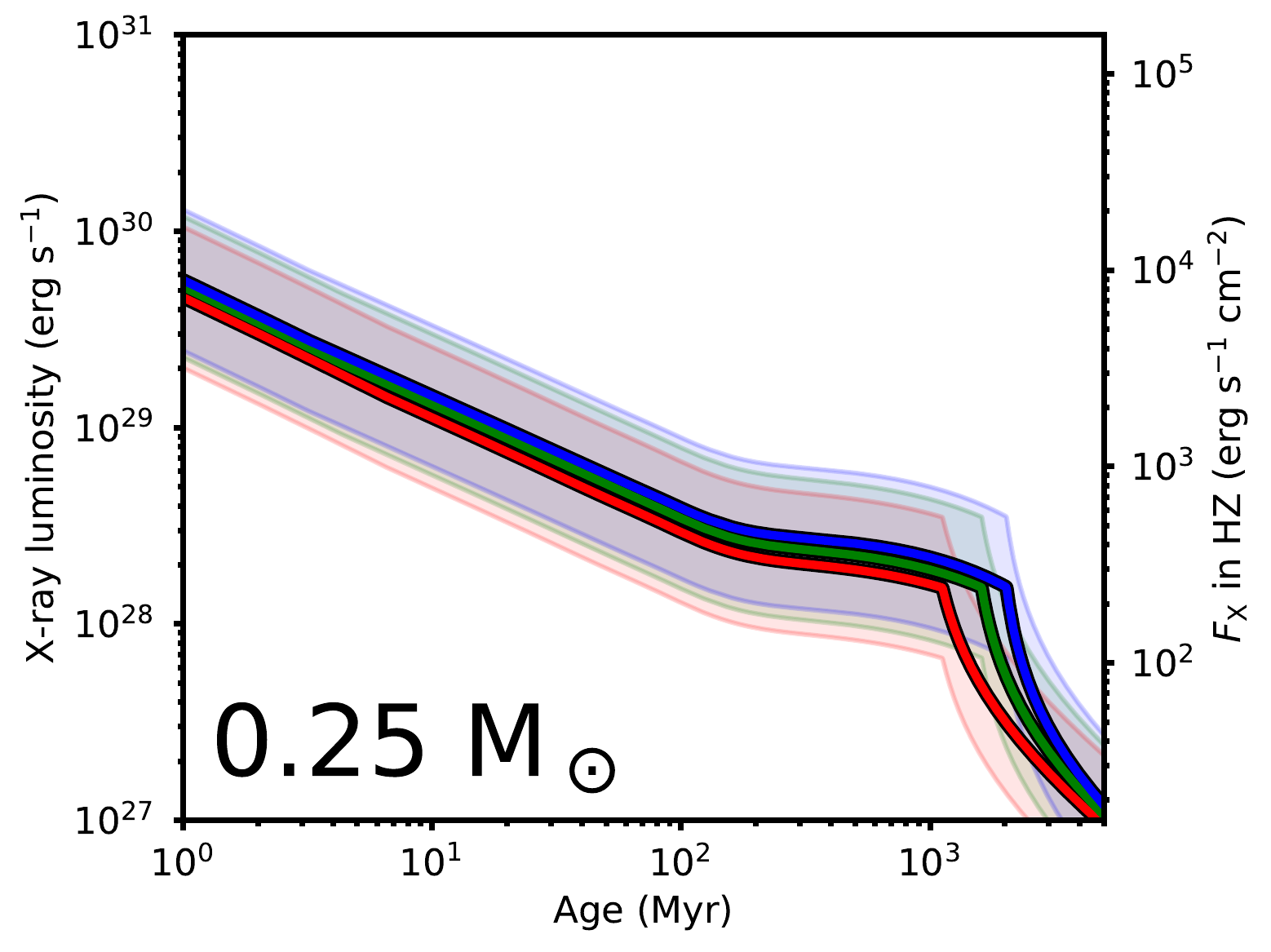}
\caption{ 
Evolutionary tracks for stellar X-ray luminosity for slow, medium, and fast rotators with several masses.
The shaded areas around each line represents the spread around the best fit relation, showing one standard deviation of the spread above and below the average. 
}
\label{fig:xraytracks}
\end{figure}

\begin{figure*}
\centering
\includegraphics[width=0.45\textwidth]{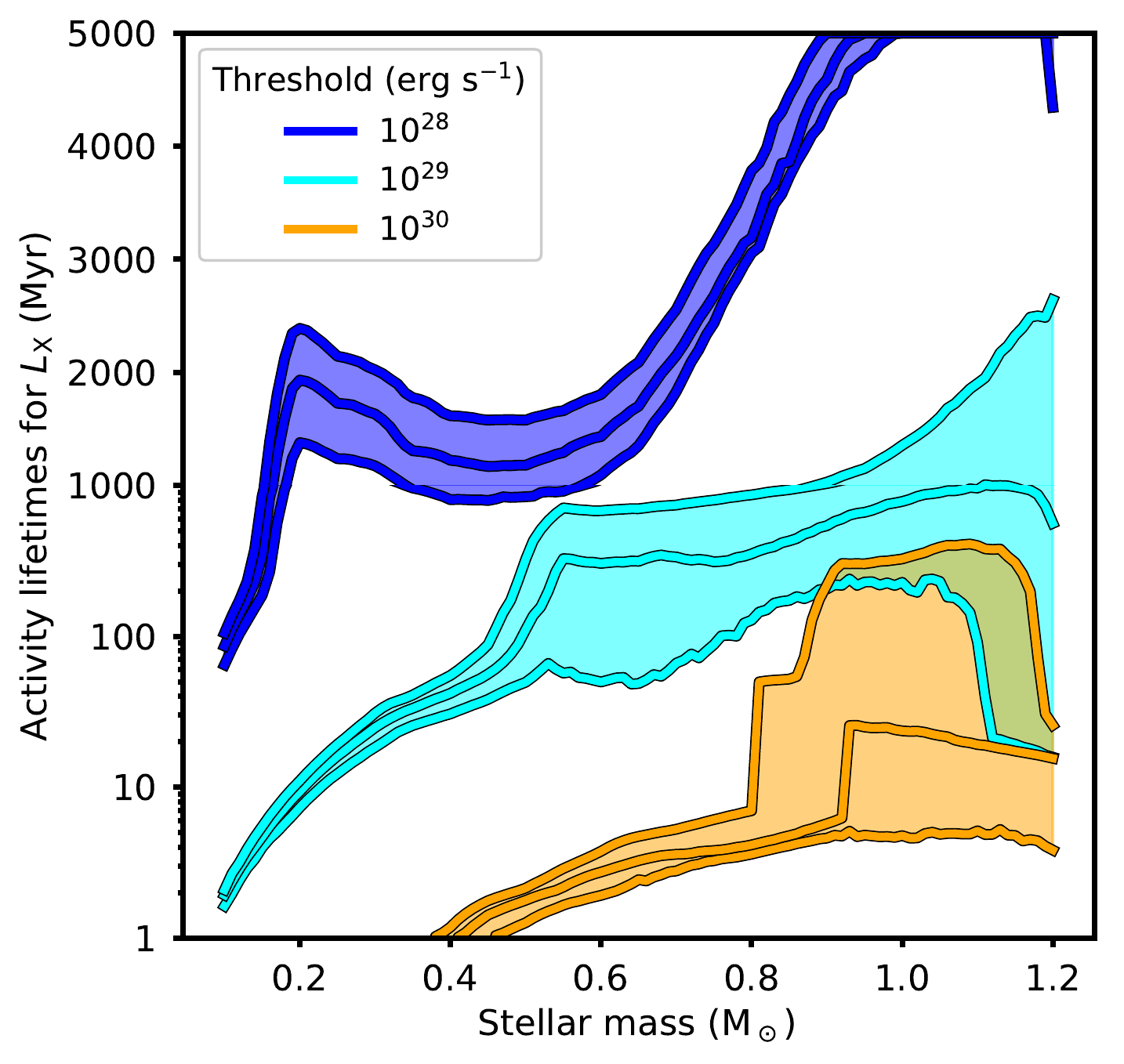}
\includegraphics[width=0.45\textwidth]{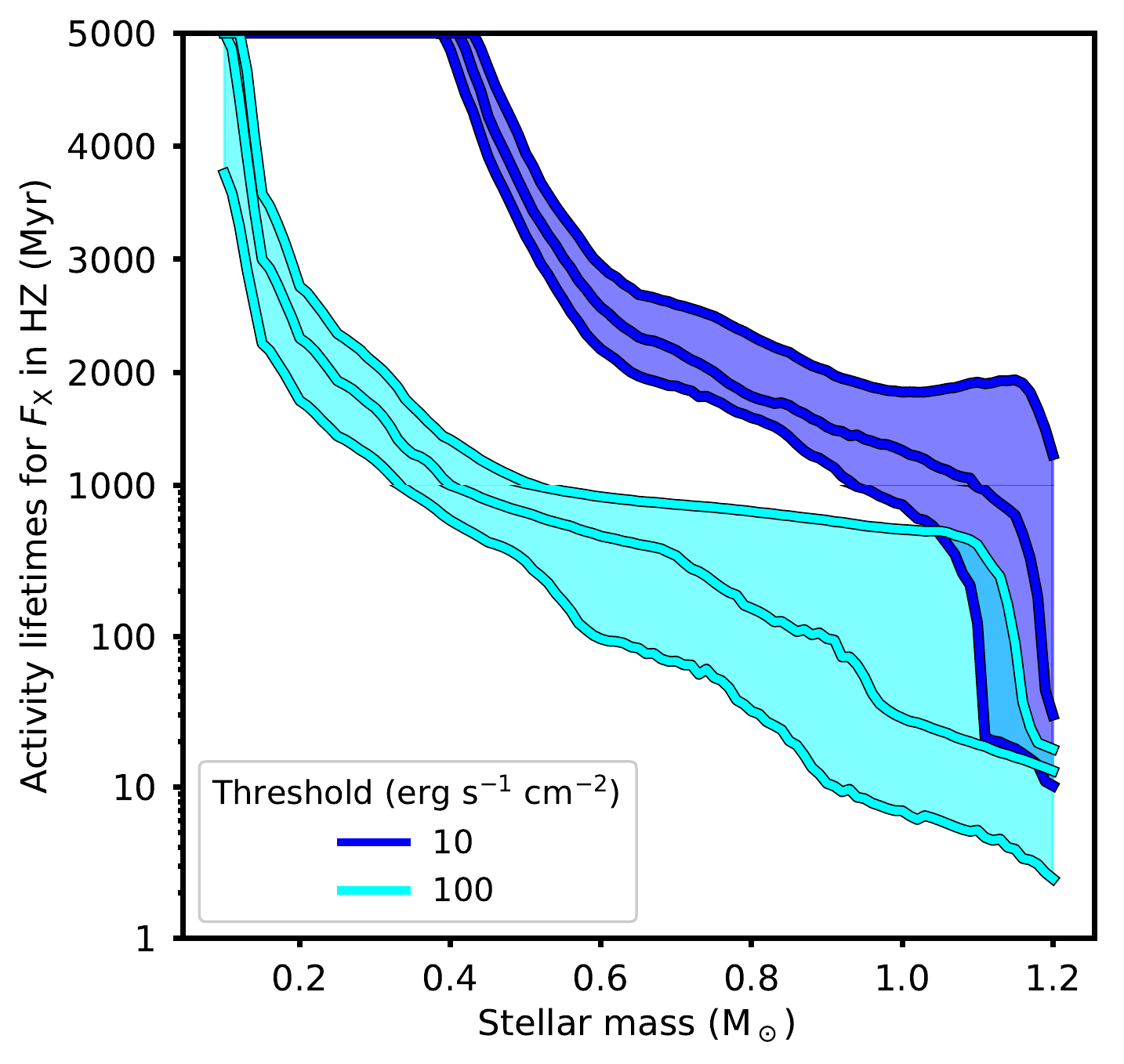}
\caption{ 
Ages of stars when they fall below threshold values for their X-ray luminosity (\emph{left-panel}) and X-ray flux in the habitable zone (\emph{right-panel}) as functions of stellar mass.
In both panels, each color represents a specific choice for the threshold value and the lines from bottom to top for each color shows the values for slow, medium, and fast rotator cases.
Note that in both panels, the y-axes are logarithmic for ages up to 1000~Myr and linear at later ages. 
}
\label{fig:activelifetime}
\end{figure*}

\subsection{Observations of X-ray evolution in young clusters}

To verify the description of X-ray evolution presented in the next section, we use measured X-ray distributions from 10 young clusters with known ages collected from the literature.
These are Taurus ($\sim$2~Myr), h~Per ($\sim$12~Myr), NGC~2547 ($\sim$40~Myr), $\alpha$~Per ($\sim$50~Myr), Blanco~I ($\sim$50~Myr), Pleiades ($\sim$125~Myr), NGC~2516 ($\sim$150~Myr), NGC~6475 ($\sim$300~Myr), M37 ($\sim$550~Myr), and Hyades ($\sim$650~Myr).
Combining $\alpha$~Per and Blanco~I into one bin at 50~Myr and NGC~2516 and Pleiades into one bin at 150~Myr gives us eight age bins with measured X-ray distributions.
These age bins and the references for our sources of the X-ray measurements are given in Table.~\ref{table:xraycluster} and the distributions are shown in Fig.~\ref{fig:XrayClusters}.
For six of these clusters, we use the X-ray luminosities determined by \citet{Nunez16} instead of the values from the original studies: these clusters and the original studies are NGC~2547 (\citealt{Jeffries06}), NGC~2516 (\citealt{Pillitteri06}), Pleiades (\citealt{Stauffer94}; \citealt{Micela99}; \citealt{Stelzer00}), NGC~6475 (\citealt{Prosser95}; \citealt{JamesJeffries97}), and Hyades (\citealt{Stern95}; \citealt{Stelzer00}).
For an additional comparison, we include also X-ray measurements for nearby field stars from ROSAT collected from the NEXXUS database (\citealt{SchmittLiefke04}), converting between (B-V)$_0$ and mass using the data from \citet{PecautMamajek13}.

\subsection{Results: X-ray evolution}

Combining our rotation models with the relation for X-ray emission in the previous section allows us to understand the evolution of stellar X-ray emission.
In Section~\ref{sect:rotresults}, we used our rotational evolution model to evolve the observed 150~Myr rotation distribution from 1~Myr to 5~Gyr, which allows us to derive a model distribution at the ages of the clusters with observed X-ray luminosity distributions.
For each star in the model rotation distribution, we predict an average $\log R_\mathrm{X}$ using Eqn.~\ref{eqn:RoRx} and add to that a random value selected using a normal distribution that represents the spread around the best fit relation.
In Fig.~\ref{fig:XrayClusters}, blue circles show our model distributions for each observed cluster, and red circles and green triangles show the measured distributions for the young clusters described above.
We find very good agreement, providing important validation of our approach. 
The comparison is difficult however due to observational limitations, especially due to detection thresholds for many clusters being at high X-ray luminosities.
This is especially a problem for low mass stars since they are typically less X-ray luminous. 

While measurements exist for a large number of older field stars, using these to test the evolution of X-ray emission is difficult given the lack of known ages for these stars.
As a demonstration of the later X-ray evolution, we show our model $L_\mathrm{X}$ distribution at 1 and 5~Gyr in the lower right panel of Fig.~\ref{fig:XrayClusters} and compare these to ROSAT measurements of nearby field stars.
Most of the field star measurements are contained well within the 1 and 5~Gyr distributions and the trend of higher $L_\mathrm{X}$ for higher mass stars is clearly visible, though as expected, some outliers are present given the large number of stars, the fact that not all stars in the sample are indeed within this age range, and the likely presence of tight binary systems. 

In Fig.~\ref{fig:XrayEvo}, we summarise the evolution of rotation (left-panels) and X-ray emission (middle panels) for stellar masses between 0.1 and 1.2~M$_\odot$. 
In each panel, the red, green, and blue lines show the slow, medium, and fast rotator tracks, defined as the 5$^\mathrm{th}$, 50$^\mathrm{th}$, and 95$^\mathrm{th}$ percentiles of the 150~Myr rotation distribution, and the dashed line shows the saturation threshold.
At the start of the evolutionary sequence, there is a large spread in rotation rates at all masses, but this does not translate into a large spread in emission since the saturation threshold is at such low rotation rates that there are no unsaturated stars.
There is a strong positive dependence of $L_\mathrm{X}$ on stellar mass due to the fact that higher mass stars have higher bolometric luminosities.
At 10~Myr, the situation is similar, but the $L_\mathrm{X}$ values of all stars have decreased due to the decrease in bolometric luminosity on the pre-main-sequence.
Some of the most slowly rotating stars with masses similar to that of the Sun have fallen out of saturation and have $L_\mathrm{X}$ values slightly below the saturation threshold.
This is due to the increase in the saturation threshold rotation rate and not to rotational evolution.
By 100~Myr, the $L_\mathrm{X}$ of the entire distribution has decreased and most solar mass stars have become unsaturated, leading to an increased spread in $L_\mathrm{X}$ values for these masses.
These changes in the $L_\mathrm{X}$ distribution happen despite the fact that stars spin up on the pre-main-sequence and are due to the evolution of the decrease in the bolometric luminosities of stars and an increase in the rotation rate of the saturation threshold (driven by decreasing convective turnover times).
After 100~Myr, the stellar bolometric luminosities and saturation threshold rotation rates remain approximately constant and most changes are a result of rotational evolution. 

As stars spin down, a larger fraction of them enter the unsaturated regime, with higher mass stars tending to become unsaturated earlier than lower mass stars due to a combination of two effects. 
Firstly, due primarily to the lower rotation rates for the saturation threshold, rapidly rotating lower mass stars feel a smaller spin-down torque than higher mass stars with similar rotation rates, and therefore remain rapidly rotating longer. 
Secondly, the lower X-ray saturation threshold rotation rate for lower mass stars means that they must spin down more in order to become unsaturated.  
By 1~Gyr, most stars with masses $>$0.4~M$_\odot$ and almost all stars with masses $>$0.6~M$_\odot$ are unsaturated.
By 5~Gyr, almost all stars are unsaturated, except those with masses of $\sim$0.1~M$_\odot$, and the convergence of the rotation distribution means that the initial rotation rate is no longer a factor influencing the emission.   
The dependence of the time that stars remain in the saturated regime on stellar mass and initial rotation rate is shown in Fig.~\ref{fig:satage}.

In Fig.~\ref{fig:xraytracks}, we show slow, medium, and fast rotator evolutionary tracks for $L_\mathrm{X}$ for several stellar masses. 
The shaded areas around each line represent the spread around the best fit relation between Rossby number and X-ray emission and extends one standard deviation of this spread above and below the average lines.
Assuming that this spread is a result of activity variability only, stars spend approximately 90\% of the time within these shaded regions.
In the solar mass case, the X-ray emission of the slow and fast rotator tracks start to diverge in the first 10~Myr and are different by approximately a factor of $\sim$50 for most of the first $\sim$500~Myr. 
For the lower mass cases, this difference is smaller due to the saturation threshold being at a much lower rotation rate.
For the 0.5~M$_\odot$ and 0.25~M$_\odot$ cases, the slow and fast rotator tracks are almost identical until 200~Myr and 2~Gyr respectively. 

In Fig.~\ref{fig:activelifetime}, we show as functions of mass the ages at which the average $L_\mathrm{X}$ decays below values of $10^{28}$, $10^{29}$, and $10^{30}$~erg~s$^{-1}$.
These can be seen as measures of the activity lifetimes for stars with different masses and initial rotation rates.
Two separate regimes can be seen in the figure: at higher stellar masses, slow rotators cross the thresholds earlier than rapid rotators, and at lower stellar masses, all three rotation tracks cross the thresholds at approximately the same age.
The reason for this difference is that higher mass stars decay across these thresholds due to rotational spin-down whereas the less luminous lower mass stars cross the thresholds due to the decay in their bolometric luminosities while still in the saturated regime.
Likely the most unexpected feature of these results is the fact that when using $L_\mathrm{X}$ to measure activity lifetimes, higher mass stars remain active longer than lower mass stars.

\begin{figure}
\centering
\includegraphics[width=0.49\textwidth]{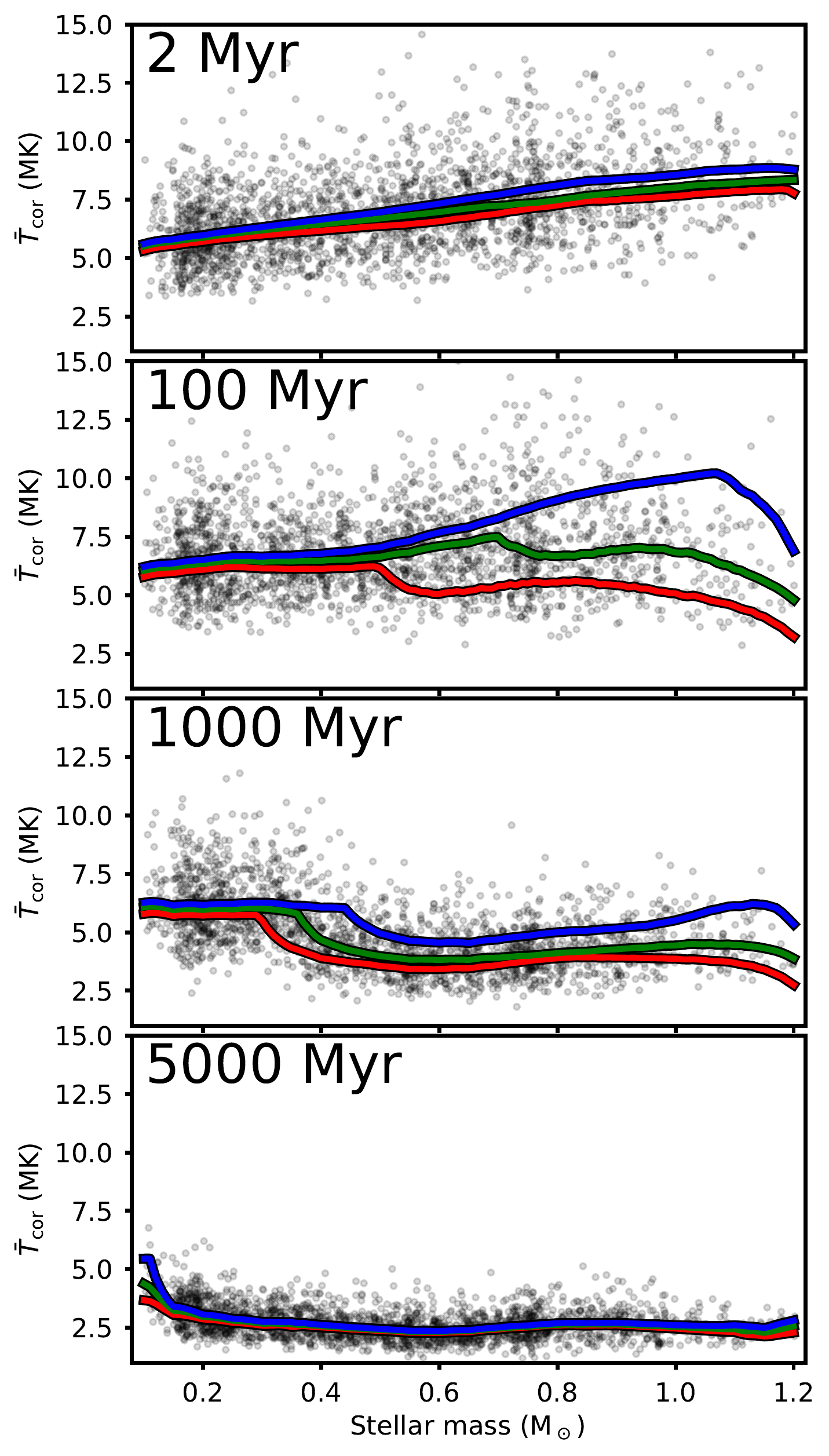}
\caption{ 
{
The evolution of average coronal temperature as a function of stellar mass.
Grey circles show our model distribution and blue, green, and red lines show our slow, medium, and fast rotator tracks.
}
}
\label{fig:tcor}
\end{figure}

\begin{figure}
\centering
\includegraphics[width=0.49\textwidth]{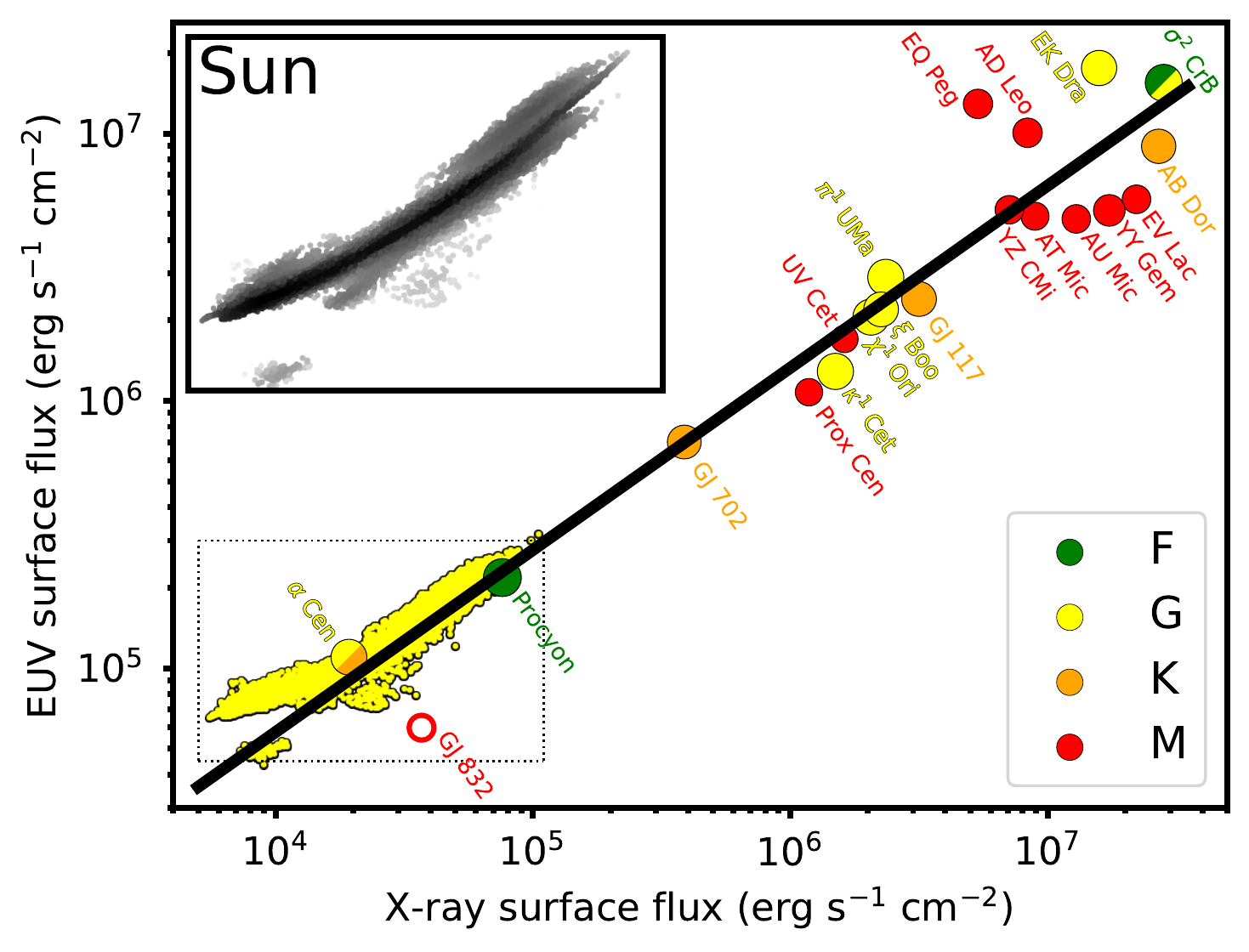}
\caption{ 
{
Relation between X-ray (0.517--12.4~nm) and EUV (10--36~nm) surface fluxes, $F_\mathrm{X}$ and $F_\mathrm{EUV}$, for the sample of stars given in Table~\ref{table:EUVXray}. 
The colors and sizes of the circles represent spectral type and mass and the yellow region and the inset shows the Sun at different times during the solar cycle.
The open red circle shows the M~dwarf GJ~832, with fluxes derived from the stellar parameters and semi-empirical model spectrum of \citet{Fontenla16}.
}
}
\label{fig:xrayeuv}
\end{figure}


\section{Stellar EUV and Ly-$\alpha$ emission} \label{sect:xuvplanet}

A primary motivation for this study is the need to understand the influences of stellar activity on the formation and evolution of planetary atmospheres.
In previous sections, we concentrate only on X-ray emission since it has clear and abundant observational constraints, but in most cases the ultraviolet part of the spectrum contributes more to the heating and expansion of planetary upper atmospheres. 
A complication is that the range of wavelengths of the stellar X-ray and ultraviolet spectrum relevant for atmospheric escape depends on the chemical composition of the atmosphere (see Section~3 of \citealt{Johnstone19a}) since different chemical species have different wavelength-dependent absorption cross-sections.
For example, much of the heating in the Earth's thermosphere and upper mesosphere is due to absorption by O$_2$ and O$_3$, which are effective at absorbing radiation at wavelengths up to $\sim$200~nm for O$_2$ and beyond for O$_3$.
This is in stark contrast to primordial atmospheres composed primarily of hydrogen since H$_2$ does not absorb radiation longward of 112~nm.
For highly irradiated atmospheres undergoing strong hydrodynamic escape in the form of a transonic Parker wind, such as short-period planets with hydrogen dominated atmospheres, the escape can be driven mostly by X-rays if EUV does not penetrate below the sonic point (\citealt{OwenJackson12}).
In this section, we discuss the evolution of EUV and Ly-$\alpha$ emission.
For discussions of emission at far-ultraviolet and longer wavelengths and its relation to X-ray and EUV emission, see \citet{France13,France18}, \citet{Linsky14}, \citet{Shkolnik14}, and \citet{Peacock20}.

\subsection{EUV emission}

We define the term `XUV' here to refer to the wavelength range 0.1 and 92~nm
We define EUV as the wavelength range 10 to 92~nm and {to be consistent with the literature, we define X-rays to be 2.4 to 0.1~keV (0.517--12.4~nm).
Although there is slight overlap, the additional 2.4~nm contributes only a few percent of the emission at EUV wavelengths.
The 2.4 keV boundary is largely arbitrary and as shown in Table~1 of \citet{Telleschi05}, increasing it to 10~keV (0.124~nm) leads to negligible increases in $L_\mathrm{X}$. 
} 
Despite interstellar absorption making EUV observable for only very nearby stars, some useful observational constraints on EUV emission are available (e.g. \citealt{Bowyer00}; \citealt{SanzForcada11}; \citealt{Drake20}).
The EUV emission of stars follow similar trends with rotation and spectral type as X-rays (\citealt{Mathioudakis95}), but since more active stars have hotter coronae (\citealt{Schmitt97}; \citealt{Telleschi05}), {a larger fraction of the emitted energy is at shorter wavelengths}, meaning that the X-ray to EUV luminosity ratios are higher and EUV emission decays slower with rotational spin-down (\citealt{Ribas05}).

An important question is which quantities should be related: possible choices include the luminosities ($L_\mathrm{X}$ and $L_\mathrm{EUV}$), the surface fluxes ($F_\mathrm{X}$ and $F_\mathrm{EUV}$), and the luminosities normalised to the bolometric luminosities ($R_\mathrm{X}$ and $R_\mathrm{EUV}$).
We want to know which of these best correlates with the physical nature of the emitting plasma since this determines the spectral energy distribution.
This question was addressed by \citet{JohnstoneGuedel15} who used measurements of the temperatures of coronal plasma for stars with different spectral types and found a single tight mass-independent dependence on X-ray surface flux, given by 
\begin{equation} \label{eqn:Tcor}
\bar{T}_\mathrm{cor} = 0.11 F_\mathrm{X}^{0.26}.
\end{equation}
where $\bar{T}_\mathrm{cor}$ is the emission measure weighted average temperature of the X-ray emitting plasma.  
A similar relation was found by \citet{Wood18} and no such mass-independent relation exists for $L_\mathrm{X}$ and $R_\mathrm{X}$.
It is therefore reasonable to assume a single relation between X-ray and EUV surface fluxes for all spectral types. 
{
Although we do not use Eqn.~\ref{eqn:Tcor} to derive the X-ray--EUV relation, it provides useful physical understanding of the evolutionary trends.
Stars with higher X-ray surface fluxes have hotter coronae, and since coronae dominate emission at wavelengths below $\sim$40~nm (as shown in Fig.~10 of \citealt{Fontenla09}), a larger fraction of their XUV emission is at shorter wavelengths.
The evolution of coronal temperature is shown in Fig.~\ref{fig:tcor}.
}

\begin{table*}
\caption{Sample of stars with EUV constraints}
\label{table:EUVXray}
\centering
\begin{tabular}{c c c c c c c c c}
\hline\hline
Star & Spec. & $M_\star$ & $R_\star$ & Dist. & $\log N_\mathrm{H}$ & $\log L_{\mathrm{Ly}\alpha}$ & $\log L_\mathrm{EUV,1}$ & $\log L_\mathrm{X}$ \\
name & type& (M$_\odot$) & (R$_\odot$) & (pc) & (cm$^{-2}$) & (erg s$^{-1}$) & (erg s$^{-1}$) & (erg s$^{-1}$) \\
\hline
Procyon A & 			F5IV$^{(1)}$ & 		1.43$^{(2)}$ & 			2.03$^{(3)}$ & 			3.51 & 		17.88$^{(4)}$ & 	29.34 & 	28.72 & 	28.28 \\
$\sigma^2$ CrB A/B & 	F6V/G0V$^{(5)}$ & 	1.14/1.09$^{(6)}$ & 	1.24/1.24$^{(6)}$ &		23.30 & 	18.40$^{(7)}$ &  	(30.04) & 	30.43 & 	30.73 \\
$\chi^1$ Ori & 			G0V$^{(8)}$ & 		1.03$^{(9)}$ & 			0.98$^{(9)}$ & 			8.66 & 		17.80$^{(4)}$ & 	29.06 & 	29.05 & 	29.08 \\
$\pi^1$ UMa & 			G0.5V$^{(10)}$ & 	1.03$^{(11)}$ & 		0.95$^{(11)}$ & 		14.30 & 	18.00$^{(11)}$ & 	29.07 & 	29.18 & 	29.11 \\
EK Dra & 				G1.5V$^{(12)}$ & 	0.95$^{(13)}$ & 		0.94$^{(13)}$ & 		33.94 & 	18.18$^{(11)}$ & 	(29.46) & 	29.95 & 	29.93 \\
$\alpha$ Cen A/B & 		G2V/K1V$^{(14)}$ & 	1.10/0.91$^{(15)}$ & 	1.22/0.86$^{(16)}$ & 	1.35 & 		17.60$^{(4)}$ & 	28.73 & 	28.16 & 	27.42 \\
$\kappa^1$ Cet & 		G5V$^{(17)}$ & 		1.04$^{(18)}$ & 		0.92$^{(18)}$ & 		9.14 & 		17.50$^{(4)}$ & 	28.93 & 	28.79 & 	28.89 \\
$\xi$ Boo A & 			G7V$^{(19)}$ & 		0.90$^{(20)}$ & 		0.80$^{(21)}$ &	 		6.71 & 		17.90$^{(4)}$ & 	29.00 & 	28.91 & 	28.94 \\
AB Dor A & 				K0V$^{(22)}$ & 		0.86$^{(23)}$ & 		0.96$^{(24)}$ & 		14.90 & 	18.38$^{(25)}$ & 	(29.51) & 	29.66 & 	30.18 \\
GJ 702 A/B & 			K0V/K4V$^{(8)}$ & 	0.88/0.73$^{(26)}$ & 	0.86/0.88$^{(26)}$ & 	5.08 & 		18.06$^{(27)}$ & 	28.82 & 	28.70 & 	28.46 \\
GJ 117 & 				K1V$^{(28)}$ & 		0.90$^{(29)}$ & 		0.79$^{\mathrm{c}}$ & 	10.40 & 	18.00$^{(4)}$ & 	28.89 & 	28.93 & 	29.08 \\
AU Mic & 				M1V$^{(30)}$ & 		0.31$^{(31)}$ & 		0.84$^{(32)}$ & 		9.90 & 		18.20$^{(4)}$ & 	29.09 & 	29.25 & 	29.74 \\
YY Gem A/B & 			M0.5V$^{(30)}$ & 	0.60/0.60$^{(33)}$ & 	0.62/0.62$^{(33)}$ & 	15.60 & 	18.00$^{(34)}$ & 	(29.14) & 	29.34 & 	29.91 \\
EQ Peg A & 				M3.5V$^{(35)}$ & 	0.39$^{(36)}$ & 		0.35$^{(36)}$ & 		6.26 & 		18.13$^{(37)}$ & 	(28.37) & 	28.95 & 	28.60 \\
AD Leo & 				M4V$^{(8)}$ & 		0.40$^{(38)}$ & 		0.38$^{(38)}$ & 		4.90 & 		18.50$^{(4)}$ & 	28.42 & 	28.92 & 	28.86 \\
YZ CMi & 				M4V$^{(39)}$ & 		0.31$^{(40)}$ & 		0.32$^{(40)}$ & 		5.99 & 		17.80$^{(34)}$ & 	28.28 & 	28.48 & 	28.66 \\
EV Lac & 				M4V$^{(41)}$ & 		0.32$^{(36)}$ & 		0.30$^{(36)}$ & 		5.02 & 		18.00$^{(4)}$ & 	27.92 & 	28.46 & 	29.08 \\
AT Mic A/B & 			M4.5V$^{(42)}$ & 	0.27/0.25$^{(43)}$ & 	0.61/0.59$^{(44)}$ & 	10.70 & 	18.20$^{(4)}$ & 	(29.22) & 	29.30 & 	29.59 \\
Prox Cen & 				M5.5V$^{(45)}$ & 	0.12$^{(46)}$ & 		0.15$^{(46)}$ & 		1.30 & 		17.61$^{(27)}$ & 	26.93 & 	27.16 & 	27.23 \\
BL/UV Cet & 			M6V$^{(47)}$ & 		0.10/0.10$^{(48)}$ & 	0.14/0.14$^{(49)}$ & 	2.68 & 		17.78$^{(49)}$ & 	(27.68) & 	27.58 & 	27.59 \\
\hline
\end{tabular}
\tablefoot{ 
\tablefoottext{a}{
Both components in YY Gem A/B, AT Mic A/B, and BL/UV Cet are expected to have the same spectral type.
}
\tablefoottext{b}{
Distances derived by Hipparcos and Gaia (\citealt{}; \citealt{}).
}
\tablefoottext{c}{
The EUV and X-ray values give luminosities in the 10--36~nm and 0.517--12.4~nm (2.4--0.1~keV) range.
Ly-$\alpha$ luminosities in parentheses are estimated using Eqn.~\ref{eqn:FxLya} and measured values were reported by \citet{Wood14} for $\pi^1$ UMa, \citet{Linsky14} for YZ~CMi, and \citet{Wood05} for the rest.
}
\tablefoottext{c}{
The radius of GJ~117 is estimated from the stellar models of \citet{Spada13} assuming the mass and age (150~Myr) given by \citet{Vigan17}.
}
}
\tablebib{ 
(1) \citet{Kervella04};
(2) \citet{Gatewood06};
(3) \citet{Aufdenberg05};
(4) \citet{RedfieldLinsky08};
(5) \citet{Strassmeier94};
(6) \citet{Raghavan09};
(7) \citet{SanzForcada03};
(8) \citet{KeenanMcNeil89};
(9) \citet{Boyajian12};
(10) \citet{Gray01};
(11) \citet{Ribas05};
(12) \citet{Montes01};
(13) \citet{Waite17};
(14) \citet{Torres06};
(15) \citet{Thevenin02};
(16) \citet{Kervella17};
(17) \citet{KeenanMcNeil89};
(18) \citet{Boyajian12};
(19) \citet{LevatoAbt78}
(20) \citet{Fernandes98};
(21) \citet{Petit05};
(22) \citet{Torres06};
(23) \citet{Guirado06}; 
(24) \citet{Guirado11}; 
(25) \citet{Rucinski95}; 
(26) \citet{Eggenberger08}; 
(27) \citet{Wood05}; 
(28) \citet{Houk88};
(29) \citet{Vigan17}; 
(30) \citet{KeenanMcNeil89};
(31) \citet{Vigan17}; 
(32) \citet{Plavchan09}; 
(33) \citet{TorresRibas02}; 
(34) \citet{Linsky14}; 
(35) \citet{Mason01};
(36) \citet{Morin08}; 
(37) \citet{MonsignoriFossi95}; 
(38) \citet{Favata00}; 
(39) \citet{Davison15};
(40) \citet{Newton17}; 
(41) \citet{Lepine13};
(42) \citet{JoyAbt74};
(43) \citet{Caballero09}; 
(44) \citet{Messina17}; 
(45) \citet{Bessell91};
(46) \citet{Mann15}; 
(47) \citet{Kirkpatrick91}z
(48) \citet{Delfosse00}; 
(49) \citet{Audard03}.
}
\end{table*}

To constrain the X-ray--EUV relation, we compare measurements of X-ray and EUV emission for a sample of nearby stars observed by the Extreme Ultraviolet Explorer (\emph{EUVE}) spacecraft\footnotemark.
We include most F, G, K, and M dwarfs in the sample of \citet{Craig97}, including $\sigma^2$~CrB, which is an RS CVn binary composed of main-sequence F and G~dwarfs. 
We include additionally EK~Dra, $\pi^1$ UMa, and BL/UV~Cet.
For each star, we collect masses, radii, and interstellar absorption from the literature, as summarised in Table.~\ref{table:EUVXray}.
For binaries, we have two options: firstly, when both stars are expected to contribute to the observed emission, we calculate their surface fluxes by summing the surface areas of the two stars and list both components in Table.~\ref{table:EUVXray}, and secondly when one component is expected to dominate, we consider only that component.
Due to ISM extinction, the sample cannot be used to reliably estimate the X-ray--EUV relation at wavelengths beyond 36~nm, so we break the EUV into segment 1 with the range 10-36~nm and segment 2 with the range 36--92~nm.

\footnotetext{
EUVE spectra for can be obtained from the MAST archive at \url{https://archive.stsci.edu/euve/} and particularly for the \citet{Craig97} sample from \url{https://archive.stsci.edu/prepds/atlaseuve/datalist.html}.
}

We first correct the EUVE spectra for ISM absorption, which for many of our stars is significant at longer wavelengths within the EUV wavelength range considered. 
We compute the interstellar absorption (the factor \mbox{$\exp \left(-\tau \right)$} by which the unabsorbed spectrum is reduced, where $\tau$ is the optical depth) in XSPEC (\citealt{Arnaud96}).
We use the tbabs absorption model for a standard composition of the interstellar gas with average admixtures of dust, based on \citet{Wilms00}.
We then integrate the unabsorbed spectrum between 10 and 36~nm to get $L_\mathrm{EUV,1}$.
To derive $L_\mathrm{X}$, we use the results of the \emph{ROSAT} All-Sky Survey (RASS) since that provides a consistent and complete set of X-ray measurements for all stars and all RASS measurements are taken from the NEXXUS database (\citealt{SchmittLiefke04})\footnotemark, except EK~Dra which is not included and instead we use $L_\mathrm{X}$ determined by \citet{Guedel97}, correcting for the slightly different distance estimate that they used.
The RASS $L_\mathrm{X}$ values listed in the NEXXUS database are determined assuming a single count rate to energy flux conversion factor, but since this factor should be dependent on the spectral shape, we recalculate $L_\mathrm{X}$ using the count rates and hardness-ratios (listed in NEXXUS as `HR1'), the distances listed in Table.~\ref{table:EUVXray}, and the hardness-ratio dependent conversion factor given by \citet{Fleming95} and \citet{Schmitt95}.

\footnotetext{
RASS X-ray measurements can be obtained from \url{https://hsweb.hs.uni-hamburg.de/projects/nexxus/index.html}.
}

The correlation between X-ray and EUV surface fluxes is shown in Fig.~\ref{fig:xrayeuv} and confirms that a single mass independent scaling law between $F_\mathrm{X}$ and $F_\mathrm{EUV,1}$ is appropriate.
Since AU~Mic and AT~Mic are likely on the pre-main-sequence, Fig.~\ref{fig:xrayeuv} suggests that this is also valid for pre-main-sequence stars.
We include also daily average solar spectra over several solar cycles calculated from the Flare Irradiance Spectrum Model (\citealt{Chamberlin07}).
At medium and high solar activity, the slope of this distribution is consistent with the stellar sample, but there is a turnoff at low activity, which could indicate this relation changes for very low activity stars, though \citet{King18} did not find such a dramatic change in the $F_\mathrm{X}$--$F_\mathrm{EUV,1}$ relation based on solar spectra derived from the \emph{TIMED/SEE} mission.
To include the Sun, we split the daily averages into low, medium, and high activity states based on $F_\mathrm{X}$ assuming three equal-sized bins spaced between the minimum and maximum values and then calculate the average X-ray and EUV fluxes for each bin.
Using these three points in our fit ensures the Sun is weighted higher than other stars but still does not dominate the fit. 
Using the OLS Bisector method, our fit to all stars in the sample gives
\begin{equation} \label{eqn:FxFeuv}
\log F_\mathrm{EUV,1} = 2.04 + 0.681 \log F_\mathrm{X} ,
\end{equation}
which gives
\begin{equation} \label{eqn:FxFeuv2}
\frac{F_\mathrm{EUV,1}}{F_\mathrm{X}} = 110 F_\mathrm{X}^{-0.319} ,
\end{equation}
where both fluxes are given in erg~s$^{-1}$~cm$^{-2}$.
We find a steeper dependence of EUV emission on X-ray emission that those derived by \citet{Chadney15} and \citet{King18} largely because we do not fit our relation to the Sun only.
Our dependence is shallower than the relation \citet{SanzForcada11} because we relate surface fluxes instead of luminosities.

There are several sources of uncertainty in empirical X-ray--EUV scaling relations.
Stellar radii are rarely accurately determined causing uncertainties in the surface fluxes. 
The lack of simultaneous X-ray and EUV measurements, which combined with activity variability adds significant scatter to Fig.~\ref{fig:xrayeuv}, is another source of uncertainty and could explain why there is more scatter at the high flux part of Fig.~\ref{fig:xrayeuv} since M~dwarfs are typically more variable on short timescales (Section~\ref{sect:variability}). 
For example, EQ~Peg is the main outlier and it has a large number of $L_\mathrm{X}$ determinations listed on the NEXXUS database, most of which are \mbox{$\sim 10^{28.9}$~erg~s$^{-1}$} which would put it very close to our best fit line, whereas the value from RASS that we use is significantly lower. 
Another issue is that M~dwarfs with very low surface fluxes are not included in the EUVE sample, likely because they necessarily have low EUV luminosities and were not detectable due to ISM absorption. 
We instead show in Fig.~\ref{fig:xrayeuv} (though do not include in our fit) the M~dwarf GJ~832, which has an XUV spectrum derived from the semi-empirical models of \citet{Fontenla16} as part of the MUSCLES Treasury Survey.
This star sits close to out best fit relation, which is reassuring, but it is more luminous in X-rays than expected given its EUV flux, possibly suggesting a small mass dependence in the slope of the $F_\mathrm{X}$--$F_\mathrm{EUV,1}$ relation.
This is however difficult to test since the only stars with surface fluxes similar to that of the Sun are $\alpha$~Cen and Procyon and the M~dwarf with the smallest surface fluxes is Proxima Centauri which lies two orders of magnitude above the Sun in Fig.~\ref{fig:xrayeuv}.
Our sample does not contain many low activity stars because ISM abosrption makes such stars undetectable at EUV wavelengths for all but the nearest stars.
As can be seen in Fig.~7 of \citet{Schmitt97}, there are many nearby stars with X-ray surface fluxes similar to that of the Sun ($10^4$--$10^5$~erg~s$^{-1}$~cm$^{-2}$), including also K and M~dwarfs, but these stars were not detected by EUVE and are therefore not included in our sample (except $\alpha$~Cen and Procyon).

\begin{figure}
\centering
\includegraphics[width=0.49\textwidth]{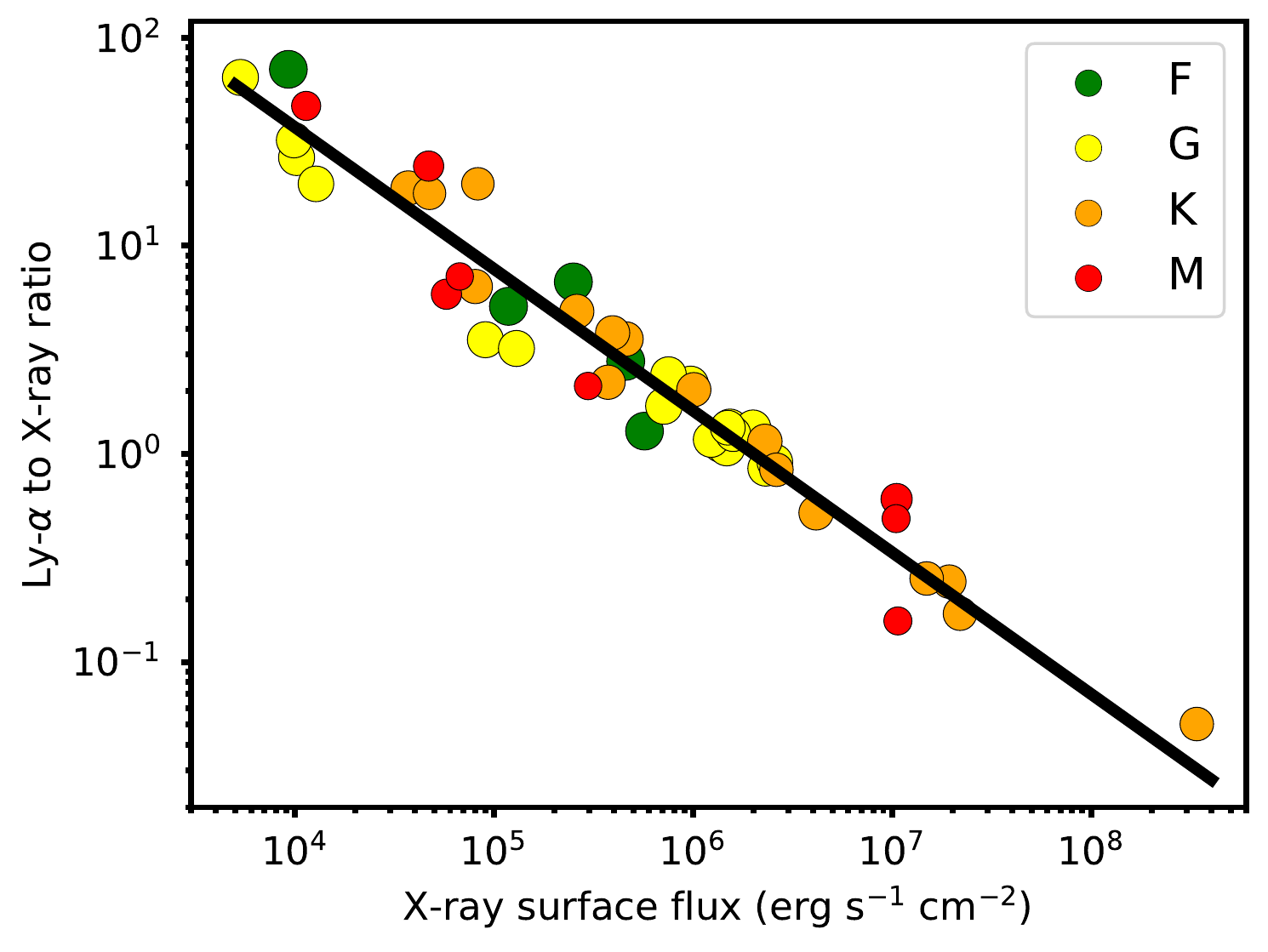}
\caption{ 
{
The Ly-$\alpha$ to X-ray ratio as a function of X-ray surface flux, $F_\mathrm{X}$, for F, G, K, and M stars derived from the data given in Table~1 of \citet{Linsky13}.
}
}
\label{fig:xraylya}
\end{figure}

To constrain the relation between $F_\mathrm{EUV,1}$ (10--36~nm) and $F_\mathrm{EUV,2}$ (36--92~nm), we use the Sun only. 
As shown in Fig.~\ref{fig:xrayeuv}, the X-ray--EUV relation derived from our sample of stars is consistent with the relation that we get considering only the active Sun.
We therefore consider only solar values with $L_\mathrm{X}$ above $10^{27}$~erg~s$^{-1}$ since with this threshold value, we get a $F_\mathrm{X}$--$F_\mathrm{EUV,1}$ relation from the Sun only that is consistent with Eqn.~\ref{eqn:FxFeuv}.
We find
\begin{equation} \label{eqn:FeuvFeuv}
\log F_\mathrm{EUV,2} = -0.341 + 0.920 \log F_\mathrm{EUV,1} ,
\end{equation}
which gives
\begin{equation} \label{eqn:FeuvFeuv2}
\frac{F_\mathrm{EUV,2}}{F_\mathrm{EUV,1}} = 0.924 F_\mathrm{EUV,1}^{-0.0798}.
\end{equation}
where both fluxes are given in erg~s$^{-1}$~cm$^{-2}$.
This is less reliable than the $F_\mathrm{X}$--$F_\mathrm{EUV,1}$ relation derived above since it is derived considering only the Sun, which varies over only a small fraction of the parameter space.
This part of the EUV spectrum is also less important since for the Sun, it contributes typically between 30 and 45\% of the total EUV (10--92~nm) emission, and this contribution is likely much less for very active stars.

\begin{figure*}
\centering
\includegraphics[width=0.98\textwidth]{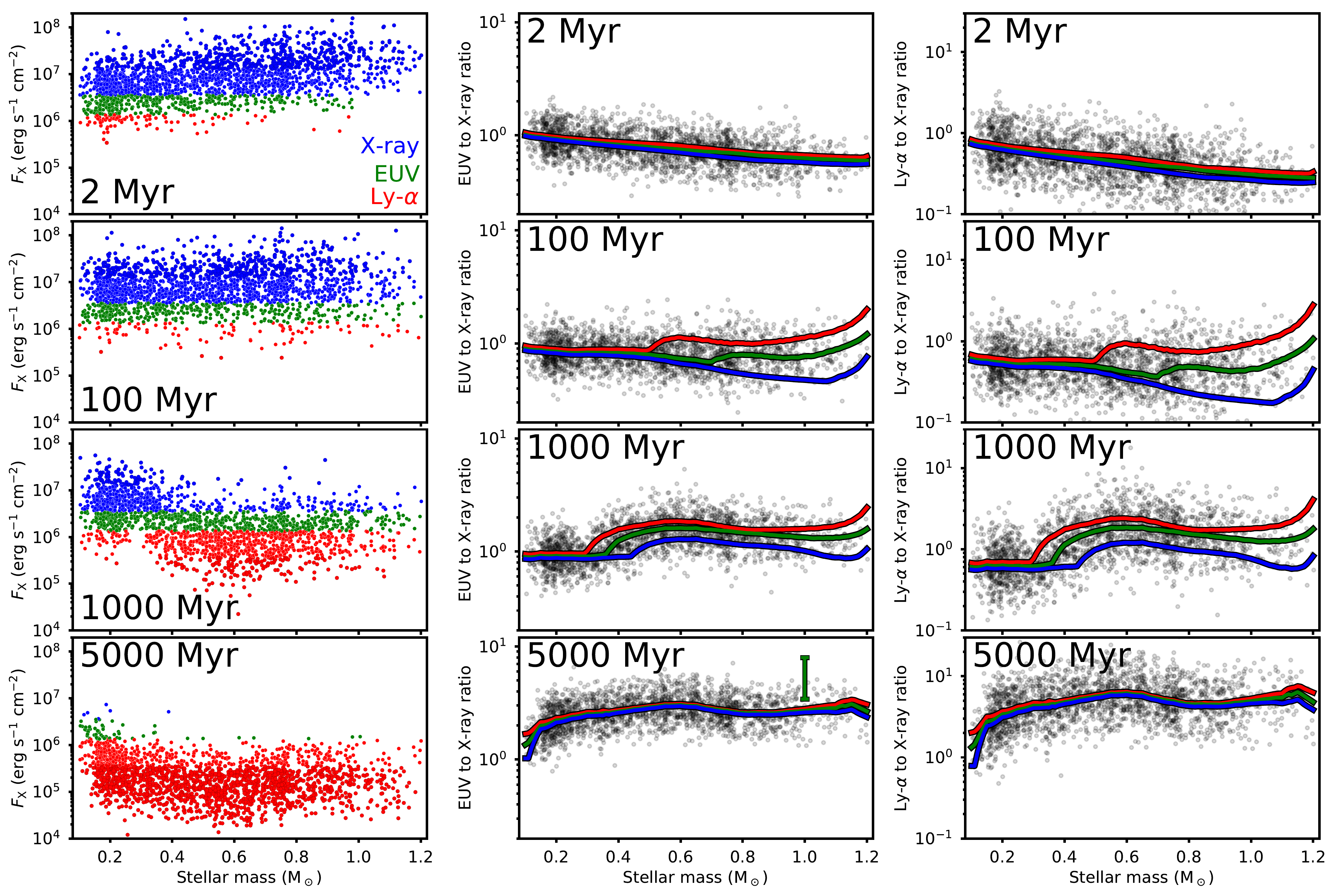}
\caption{ 
The evolution of X-ray surface flux (\emph{left column}), the EUV to X-ray ratio (\emph{middle column}), and the Ly-$\alpha$ to X-ray ratio (\emph{right column}) as functions of stellar mass.
The EUV emission is calculated considering both 10--36 and 36--92~nm.
In the left column, circle colors shows which out of X-ray (blue), EUV (green), and Ly-$\alpha$ (red) has the highest luminosity, and circles with dark outlines show that this is more luminous than the other two combined.  
For example, blue circles with white outlines are used when \mbox{$L_\mathrm{X} > L_\mathrm{EUV}$} and \mbox{$L_\mathrm{X} > L_{\mathrm{Ly}\alpha}$} but \mbox{$L_\mathrm{X} < L_\mathrm{EUV} + L_{\mathrm{Ly}\alpha}$}, and blue circles with dark outlines are used when \mbox{$L_\mathrm{X} > L_\mathrm{EUV} + L_{\mathrm{Ly}\alpha}$}. 
In the middle and right columns, grey circles show our model distribution and blue, green, and red lines show our slow, medium, and fast rotator tracks.
The vertical green line in the lower middle panel shows the range of values for the Sun at activity maximum.
}
\label{fig:xrayeuvevo}
\end{figure*}

\subsection{Ly-$\alpha$ emission}

The most important feature in a star's far ultraviolet spectrum is the \mbox{Ly-$\alpha$} emission line at 121.6~nm, which is formed in the transition region and upper chromosphere (\citealt{AvrettLoeser08}) and as another manifestation of magnetic activity, Ly-$\alpha$ correlates well with emission at shorter wavelengths (\citealt{Linsky14}).
The Ly-$\alpha$ line has been used to constrain the properties of stellar winds and planetary atmospheres (\citealt{Ehrenreich08}; \citealt{Wood14}; \citealt{Kislyakova14}) and since it often has a luminosity higher than that of the entire X-ray and EUV, it is important to understand its evolution.
Although most of the line is absorbed by the ISM, reconstructions of the intrinsic \mbox{Ly-$\alpha$} line fluxes for a large number of stars are available in the literature (\citealt{Wood05}).

The relation between X-ray and \mbox{Ly-$\alpha$} was studied by \citet{Linsky13} who showed that the ratio of the Ly-$\alpha$ to X-ray emission has a powerlaw dependence on the X-ray flux at 1~AU (which is proportional to $L_\mathrm{X}$).
They found that for F, G and K~dwarfs this dependence is very similar with only a small offset for K~dwarfs, but for M~dwarfs, the dependence is shifted to lower Ly-$\alpha$ to X-ray ratios at each X-ray flux and has a much larger scatter (see their Fig.~7).
The reason for these differences is that, similar to coronal temperature and EUV emission, Ly-$\alpha$ emission depends not on $L_\mathrm{X}$, but scales with the X-ray surface flux, $F_\mathrm{X}$, as shown by \citet{Wood05}.
In Fig.~\ref{fig:xraylya}, we show the relation between $F_{\mathrm{Ly}\alpha}/F_\mathrm{X}$ (or equivalently $L_{\mathrm{Ly}\alpha}/L_\mathrm{X}$) and $F_\mathrm{X}$ using the fluxes given in Table~1 of \citet{Linsky13}.
For this, we convert their 1~AU fluxes into $F_\mathrm{X}$ using stellar radii determined from the spectral types listed in their Table~1 and the data given by \citet{PecautMamajek13}.
When $F_\mathrm{X}$ is used instead of the flux at 1~AU, the relation becomes mass independent and we no longer see the large scatter in the values for M~dwarfs.
This scatter is a result of M~dwarfs having a large range of radii (the surface area of an M0V star is 30 times larger than that of an M9.5V star).
The best fit relation between $F_{\mathrm{Ly}\alpha}$ and $F_\mathrm{X}$ is 
\begin{equation} \label{eqn:FxLya}
\log F_{\mathrm{Ly}\alpha} = 3.97 + 0.375 \log F_\mathrm{X} ,
\end{equation}
which gives
\begin{equation} \label{eqn:FxLya2}
\frac{F_{\mathrm{Ly}\alpha}}{F_\mathrm{X}} = 1.96 \times 10^{4} F_\mathrm{X}^{-0.681} ,
\end{equation}
where both fluxes are given in erg~s$^{-1}$~cm$^{-2}$.

\subsection{EUV and Ly-$\alpha$ evolution}

We show in Fig.~\ref{fig:xrayeuvevo} the evolution of X-ray surface flux, \mbox{$F_\mathrm{EUV}/F_\mathrm{X}$}, and \mbox{$F_{\mathrm{Ly}\alpha}/F_\mathrm{X}$}.
In the left column of Fig.~\ref{fig:xrayeuvevo}, circle colours represent which of the three parts of the spectrum considered (X-ray, EUV, and Ly-$\alpha$) is the most luminous for each star in the model distribution and dark outlines indicate that this part of the spectrum is more luminous than the other two combined.
We note that the EUV luminosity is always either the first or second most luminous of these three. 
It is also important that the spread in activity levels shown in Fig.~\ref{fig:xrayeuvevo} is likely a result of short term variability, and at each mass and age, we expect stars to spend some time at each location within this spread.

At 2~Myr, most stars are more luminous in X-rays than in EUV and Ly-$\alpha$, though the distribution does contain some stars that are more EUV luminous, and even some low activity outliers that are more Ly-$\alpha$ luminous, especially at lower masses. 
At later ages, the decline in activity leads to cooler coronal temperatures and a decay in emission that is more rapid in X-ray than in EUV and Ly-$\alpha$, which leads to an increase in \mbox{$F_\mathrm{EUV}/F_\mathrm{X}$} and an even more rapid increase in \mbox{$F_{\mathrm{Ly}\alpha}/F_\mathrm{X}$}.
By 5~Gyr, the EUV to X-ray ratios are mostly distributed between 1.5 and 4.
This is lower than the ratio for the Sun, which at activity maximum is typically between 4 and 7, possibly because the Sun appears to be less active than other stars with similar masses and rotation rates (\citealt{Reinhold2020}). 
At this age, some low-activity stars are more than an order of magnitude more luminous in Ly-$\alpha$ than in X-rays.

\begin{figure}
\centering
\includegraphics[width=0.49\textwidth]{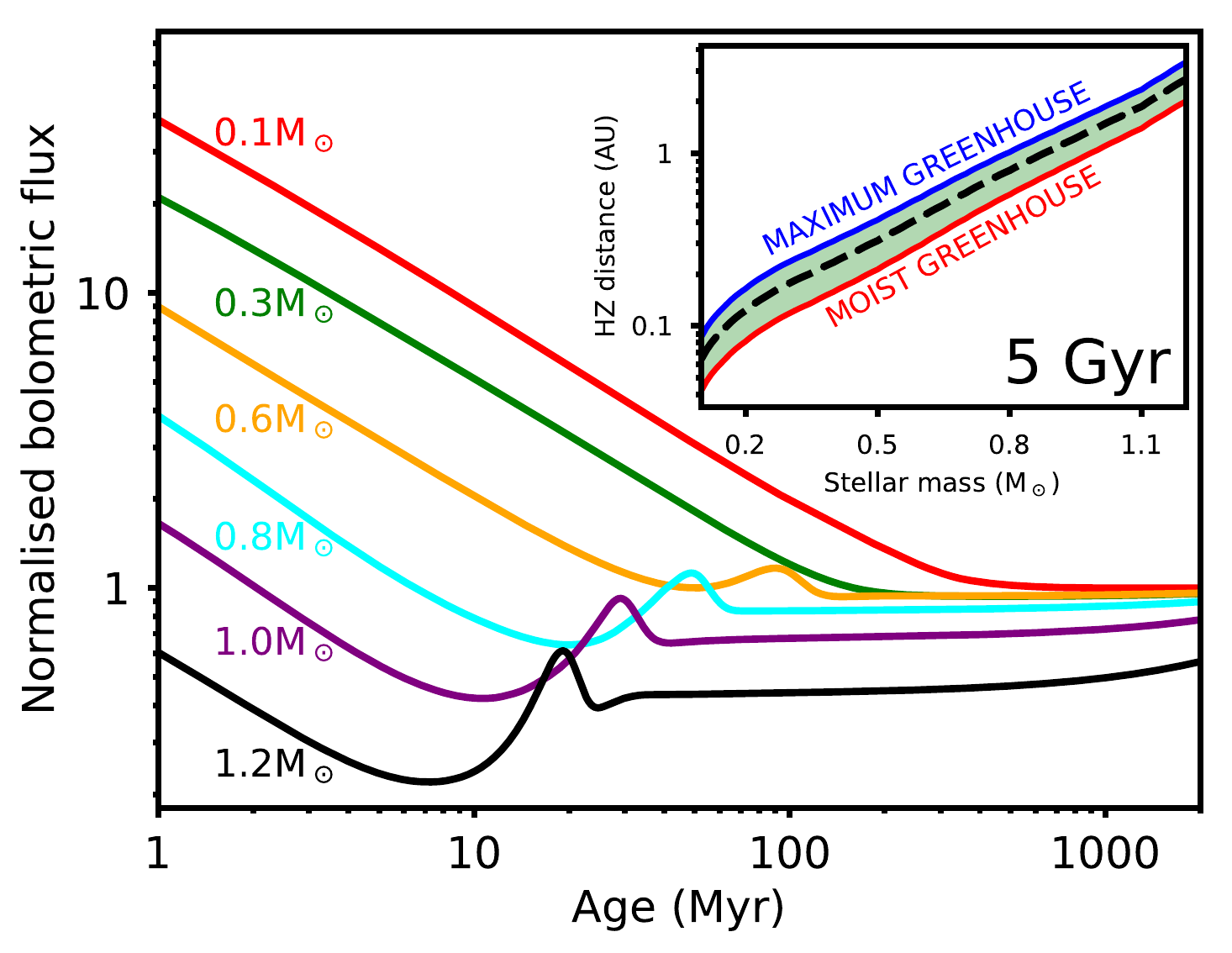}
\caption{
Evolution of stellar bolometric emission as a fraction of the value at 5~Gyr for different stellar masses from the stellar evolution models of \citet{Spada13}.
The values on the y-axis can be interpreted both as the normalised bolometric luminosities and as the normalised bolometric fluxes in the habitable zone.
The habitable zone orbital distances that we use are based on the 5~Gyr stellar properties and are shown in the inset as the dashed black line. 
}
\label{fig:HZflux}
\end{figure}

\section{Fluxes in the habitable zone} \label{sect:hz}

\begin{figure*}
\centering
\includegraphics[width=0.45\textwidth]{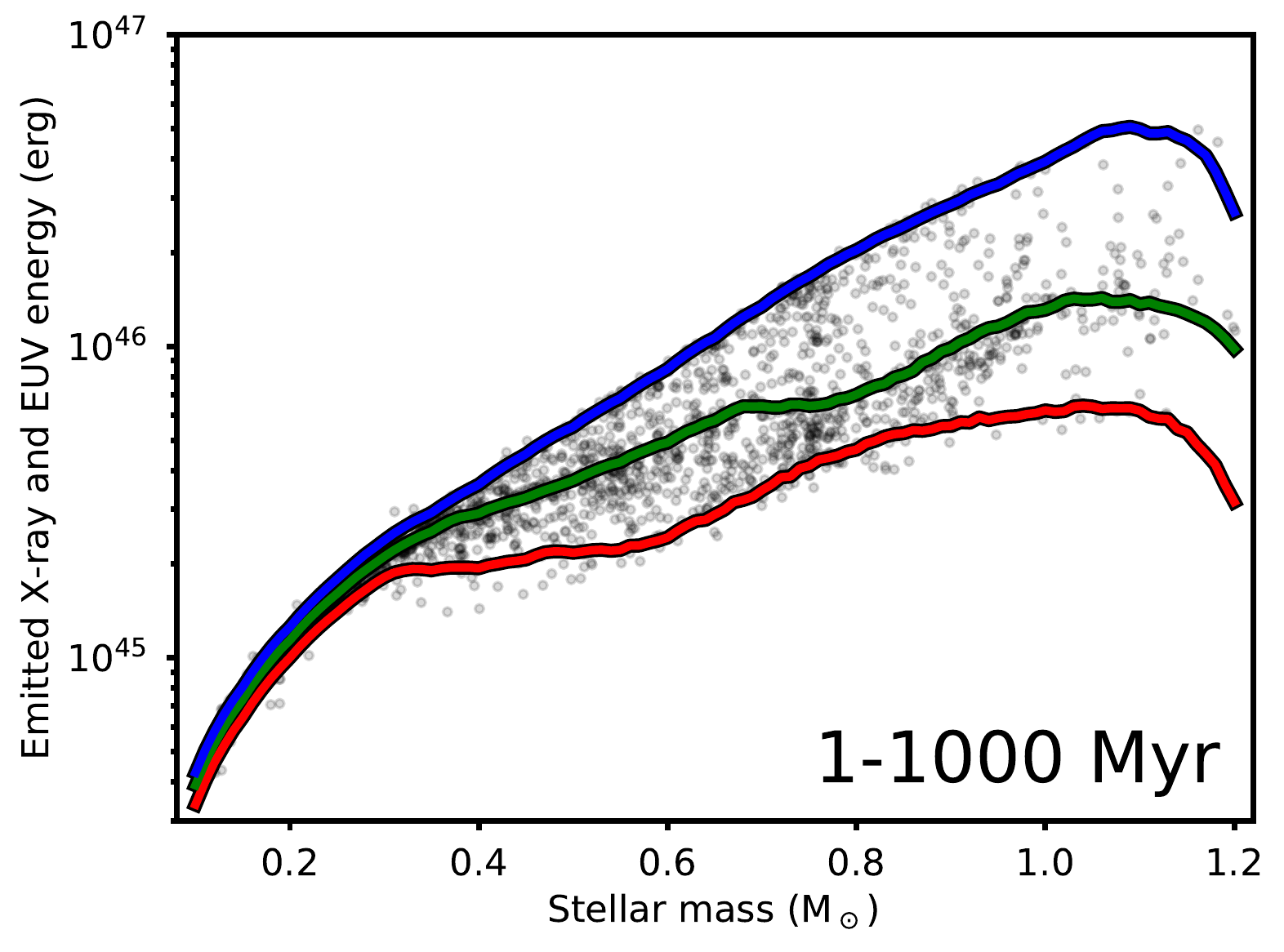}
\includegraphics[width=0.45\textwidth]{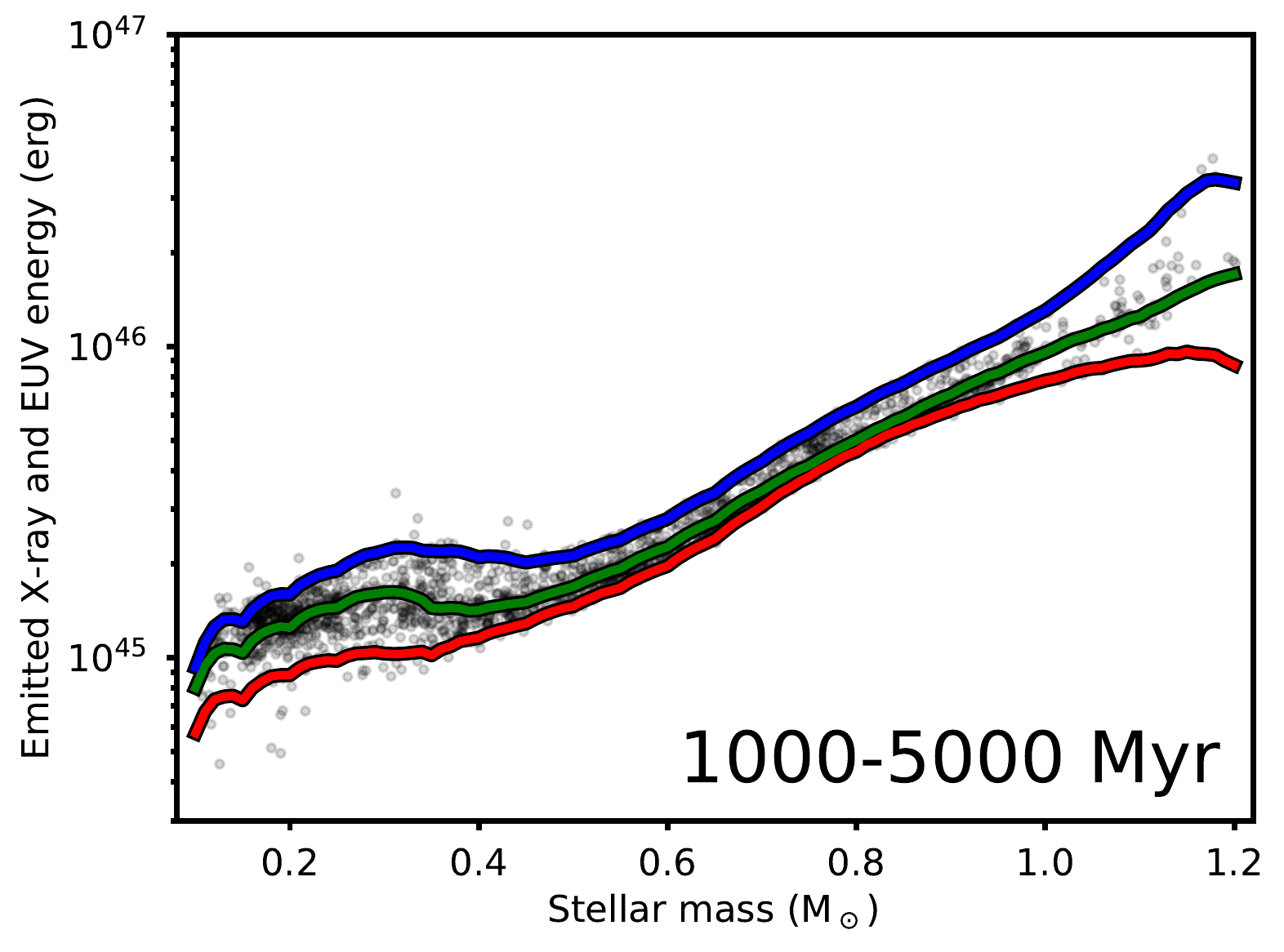}
\includegraphics[width=0.45\textwidth]{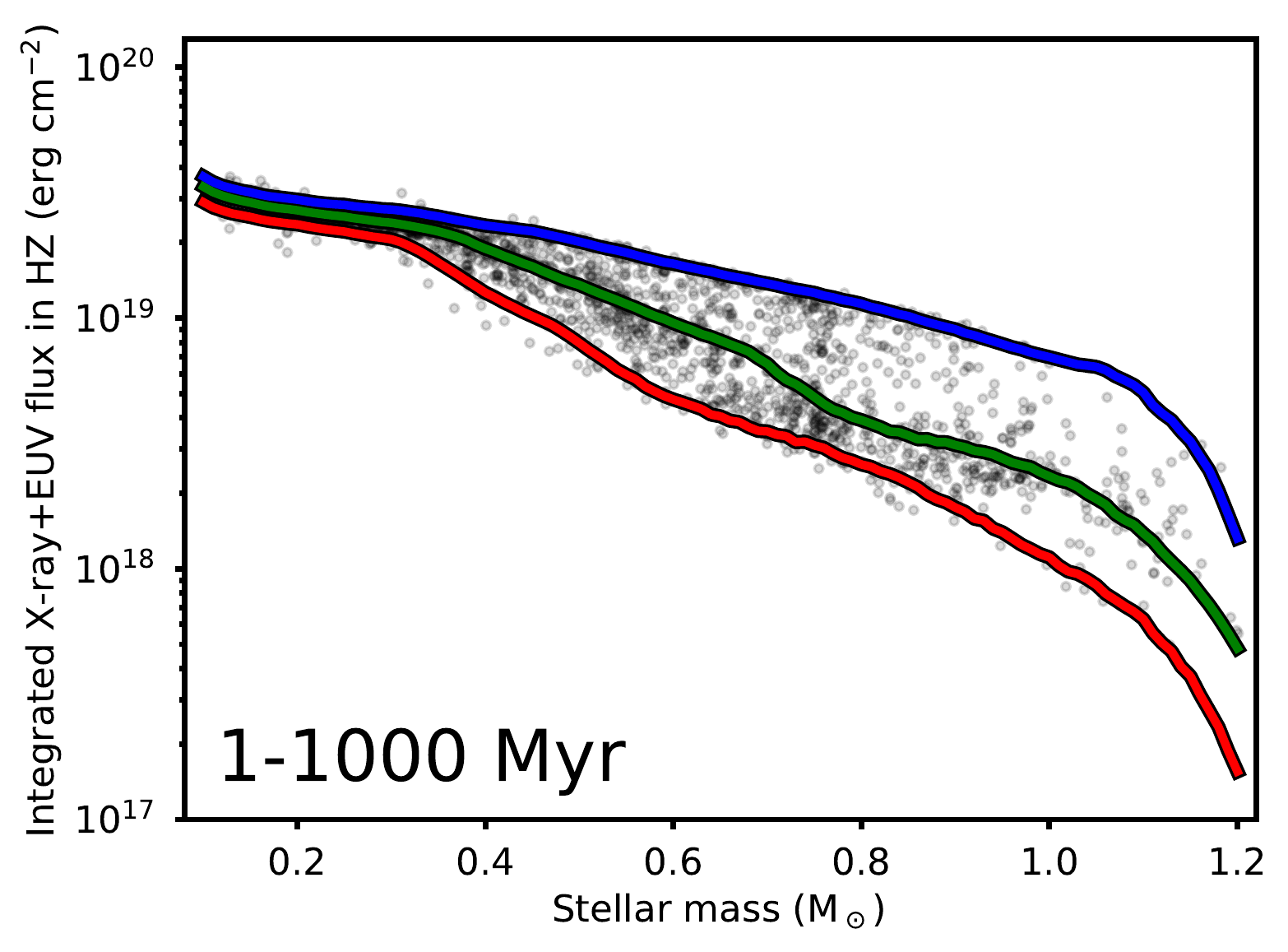}
\includegraphics[width=0.45\textwidth]{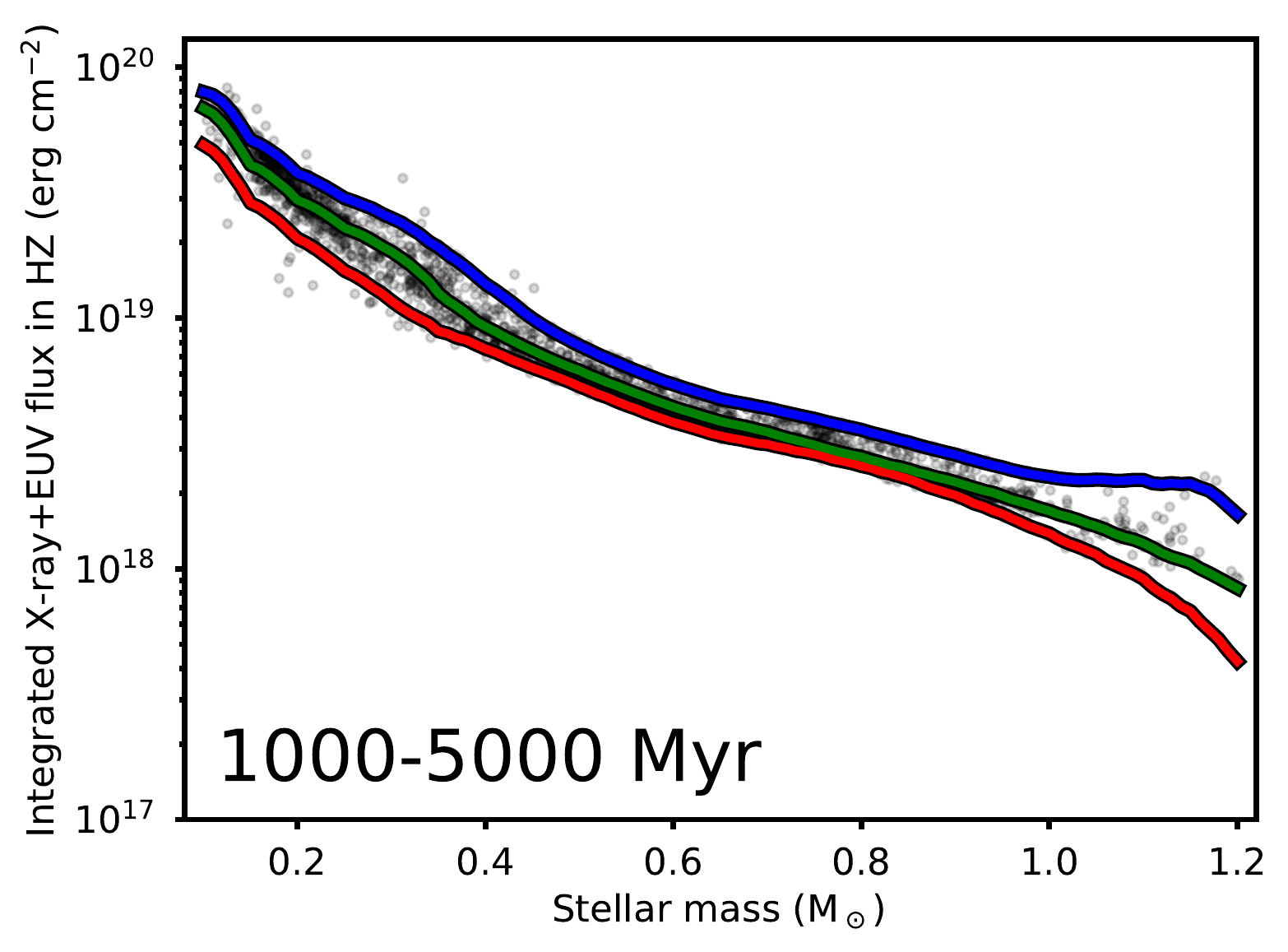}
\caption{ 
Total stellar XUV ($<100$~nm) emission integrated between 1 and 1000~Myr (\emph{upper-left}) and between 1000 and 5000~Myr (\emph{upper-right}) and XUV fluence (time integrated XUV flux) in the habitable zone between 1 and 1000~Myr (\emph{lower-left}) and between 1000 and 5000~Myr (\emph{lower-right}) as functions of stellar mass for the slow, medium, and fast rotator cases.
The background circles show the values for each star in our model distribution.
}
\label{fig:totalenergy}
\end{figure*}

It is interesting to consider also the mass dependence and evolution of X-ray and EUV fluxes in the habitable zone.
We define the habitable zone orbital distance as being half way between the moist and maximum greenhouse limits calculated using the relations derived by \citet{Kopparapu13}.
Although stellar properties evolve significantly on the pre-main-sequence, we are interested in planets that spend billions of years in the habitable zone and therefore calculate time-independent habitable zone boundaries assuming stellar properties at 5~Gyr, shown in Fig.~\ref{fig:HZflux}.
While the bolometric luminosity is the main factor determining these boundaries, the effective temperature, $T_\mathrm{eff}$ also has an effect.
Cooler stars have photospheric spectra that are shifted to longer wavelengths relative to the Sun's making them more effective at heating the surfaces of planets, meaning that habitable zone planets orbiting stars with lower $T_\mathrm{eff}$ values generally receive lower bolometric fluxes from their host stars.
This could cause habitable zone planets orbiting lower mass stars to receive a smaller X-ray flux than those orbiting higher mass stars, so long as both stars are in the saturated regime and on the main-sequence, but we find that this effect is not significant. 

In Fig.~\ref{fig:XrayEvo} and Fig.~\ref{fig:xraytracks}, we show the evolution of the X-ray flux in the habitable zone, $F_\mathrm{X,HZ}$, for stars of different masses and we see that $L_\mathrm{X}$ and $F_\mathrm{X,HZ}$ have very different mass dependences at all ages.
At 2~Myr, habitable zone planets orbiting lower mass stars receive a much higher X-ray flux than those orbiting higher mass stars despite higher mass stars being more luminous.
This might be initially surprising since all stars at this age are saturated and the saturation $L_\mathrm{X}$ is proportional to the bolometric luminosity, meaning we would expect similar X-ray fluxes in the habitable zone.
The reason can be understood if we consider the ratio of the bolometric luminosity at 2~Myr to the value on the main-sequence, remembering that we define the HZ by the main-sequence stellar properties.
This is shown in Fig.~\ref{fig:HZflux}.
Due to their slower pre-main-sequence evolution, this ratio is larger for lower mass stars, meaning that HZ planets receive a larger bolometric flux and therefore a larger X-ray flux. 

During the first 100~Myr, $F_\mathrm{X,HZ}$ decreases by more than an order of magnitude for very low mass stars due to the decreases in their bolometric luminosities. 
For solar mass stars, a similarly large decrease in $F_\mathrm{X,HZ}$ can be seen for slow and medium rotators, but the effect is caused by the decrease in convective turnover times causing these stars to enter the unsaturated regime.
Planets orbiting rapidly rotating solar mass stars receive approximately constant X-ray fluxes.
The subsequent spin-down at later ages leads to rapid declines in $F_\mathrm{X,HZ}$ for almost all stellar masses and this is especially the case for the higher mass stars, leading to a strong relation between stellar mass and $F_\mathrm{X,HZ}$ at later ages.
At 5~Gyr, HZ planets orbiting low mass M dwarfs receive X-ray fluxes that are two orders of magnitude higher those received by HZ planets orbiting G dwarfs. 

In Fig.~\ref{fig:activelifetime}, we show as functions of mass the ages at which the X-ray fluxes in the HZ decay below values of 10, and 100~erg~s$^{-1}$~cm$^{-2}$.
For comparison, the X-ray flux received by the modern Earth typically varies between 0.15 and 1.15~erg~s$^{-1}$~cm$^{-2}$.
This is also a measure of the activity lifetimes of stars, though it is very different to the activity lifetimes for $L_\mathrm{X}$ discussed in the previous section and these two different ways to measure activity lifetimes have very different mass dependences.
For solar mass stars, the HZ X-ray flux decays below 100~erg~s$^{-1}$~cm$^{-2}$ at ages of 10~Myr for initilly slow rotators and 700~Myr for initially fast rotators. 
For lower mass stars, the activity HZ X-ray fluxes remain above this threshold for much longer and 0.1~M$_\odot$ stars can take up to 5~Gyr to cross the threshold.
These timescales are much longer for the 10~erg~s$^{-1}$~cm$^{-2}$ threshold and stars with masses below 0.4~M$_\odot$ will likely never cross the threshold.

We consider also the total XUV energy emitted by the star and the total energy absorbed by HZ planets integrated over evolutionary timescales. 
These quantities influence the amount of atmospheric gas planets can lose over their lifetimes, though other factors such as the masses of the planets, the compositions of the atmospheres, which escape mechanisms dominate, and the amount of gas available in the atmospheres at different times also influence total losses.
In the upper panels of Fig.~\ref{fig:totalenergy}, we show the stellar XUV luminosity integrated between 1 and 1000~Myr and between 1000 and 5000~Myr as functions of stellar mass for the slow, medium, and fast rotator cases.
The integrated luminosity is useful for understanding how stars with different masses and initial rotation rates influence planets with the same orbital distances.
For stars with masses above $\sim$0.5~M$_\odot$, there is a significant difference in the total energy emitted by initially slow and initially fast rotating stars.
For solar mass stars, this difference is approximately an order of magnitude in the first 1000~Myr and a factor of 2 between 1000 and 5000~Myr.  
In the first 1000~Myr, there is a very strong dependence on stellar mass, with more massive stars emitting significantly more XUV radiation than less massive stars, and the magnitude of this difference depends on the initial rotation rates of the stars. 
A 0.2~M$_\odot$ star emits two orders of magnitude less energy than a rapidly rotating solar mass star and one order of magnitude less than a slowly rotating solar mass star.
Between 1000 and 5000~Myr, the dependence on initial rotation is much smaller, but the mass dependence is approximately the same.

In the lower panels of Fig.~\ref{fig:totalenergy}, we show the stellar XUV fluence, defined as the time integrated XUV flux, in the habitable zone in these two time periods. 
The fluence is useful for understanding how stars with different masses and initial rotation rates influence the atmospheres of habitable zone planets.
The dependence on stellar mass for the HZ fluence is very different to the dependence for the integrated luminosity. 
A planet in the habitable zone of a lower mass star receives significantly more XUV energy over both time intervals.
In the first 1000~Myr, a HZ planet orbiting a 0.2~M$_\odot$ star receives two orders of magnitude more energy than a HZ planet orbiting an initially slowly rotating solar mass star but only a factor of a few more than a HZ planet orbiting an initially rapidly rotating solar mass star. 
Between 1000 and 5000~Myr, this mass dependence is even larger. 


\section{The contribution of flares} \label{sect:flares}

\begin{figure}
\centering
\includegraphics[width=0.49\textwidth]{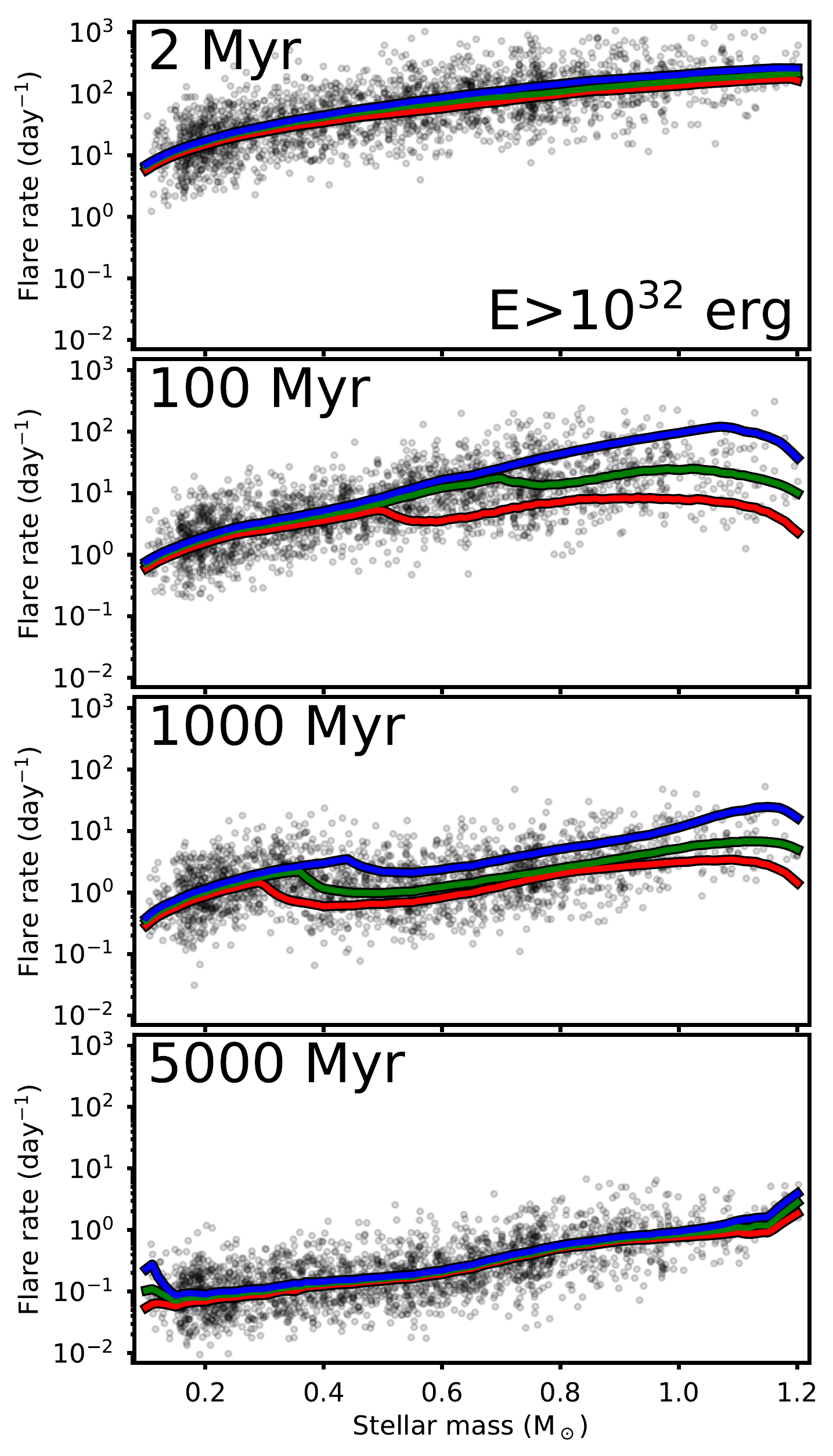}
\caption{ 
{
The evolution of the rate of flares with total emitted X-ray and EUV energies above $10^{32}$~erg for different stellar masses. 
Red, green and blue lines refer to fast, medium, and slow rotators and grey circles show our model distribution.
}
}
\label{fig:flarerates}
\end{figure}

\begin{figure*}
\centering
\includegraphics[width=0.9\textwidth]{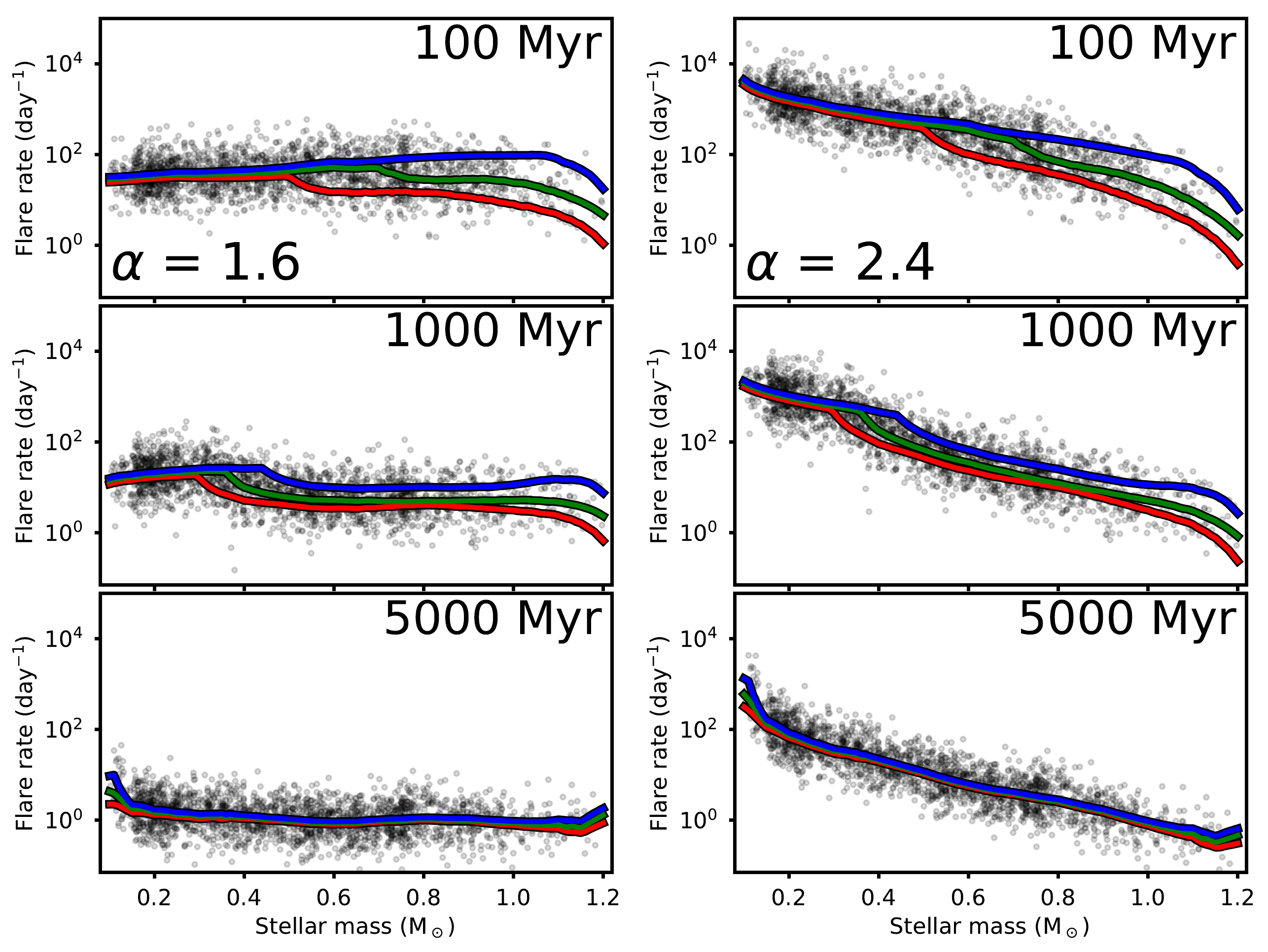}
\caption{ 
{
The evolution of the rate of flares with total emitted X-ray and EUV fluences in the habitable zone above \mbox{$1.8 \times 10^4$~erg~cm$^{-2}$} for different stellar masses.
The left and right columns show results for $\alpha$ (Eqn.~\ref{eqn:flarerate}) of 1.6 and 2.4 respectively showing that our uncertainties in the flare energy distribution can significantly influence our understanding of the effects of flares on habitable zone planets.  
Red, green and blue lines refer to fast, medium, and slow rotators and grey circles show our model distribution.
}
}
\label{fig:flareratesHZ}
\end{figure*}

The influence of flares is implicitly considered throughout this study since we do not distinguish between quiescent and flare states in any of the observational constraints on XUV emission.
Our description of variability describes how often stars spend at each activity level for each evolutionary stage taking into account both quiescent and flaring emission.
In fact, the distinction between quiescent and flaring states is largely arbitrary since much of the quiescent XUV emission is from flares that are not energetic enough to be distinguished from the background emission and it is even possible that all XUV emission from active stars is from flares happening at such high rates that the superposition of each flare lightcurve creates the appearance of steady non-flaring emission (\citealt{Kashyap02}; \citealt{Guedel03}; \citealt{Telleschi05}), which could explain the correlation between stellar activity and coronal temperature (\citealt{Guedel04}).
This could even be true for inactive stars like the modern Sun since it is possible that the background heating that maintains the large temperatures in the Sun's outer atmosphere is provided by very small scale nanoflares (\citealt{Parker88}; \citealt{Jess19}). 
The distinction between truly quiescent emission and flares is in this picture not possible although part of the apparently steady radiation could be from non-flaring active regions. 
XUV flares with a given radiated energy produce higher contrast in stars with smaller surface areas, such as M~dwarfs, and in lower activity stars for a given spectral type because less unrelated active region area contributes to the flaring and non-flaring background emission (see for example \citealt{Reale04} for an M dwarf and \citealt{Telleschi05} for solar analogs at different activity levels) biasing the detectability of individual flares in light curves.

Although flares do not cause additional XUV emission not considered in preceding sections, it is interesting to consider flare activity more explicitly to understand the evolution and spectral type dependence of flare rates. 
In recent years, large samples of optical flares have been observed by the \emph{Kepler} spacecraft (\citealt{Shibayama13}; \citealt{Davenport16}).
These measurements show that rapid rotators flare more often than slow rotators, leading to a decrease in flare rates on the main-sequence over evolutionary timescales as stars spin-down (\citealt{Davenport19}).
They also find that higher mass stars have much more energetic flares: for example, in the catalogue of \emph{Kepler} flares compiled by \citet{YangLiu19}, the upper bound (99$^\mathrm{th}$ percentile) of flare energies for 0.1--0.2~M$_\odot$ stars is $\sim10^{33}$~erg, whereas this upper bound for solar mass stars is $\sim10^{36}$~erg, where these values refer to the total energies emitted by the flares in the \emph{Kepler} bandpass.

The fraction of stars determined to be flaring and the measured optical flare rates were higher for lower mass stars, which has been interpreted to mean M and K~dwarfs flare more often than G and F~dwarfs.
This interpretation is likely incorrect and the higher flaring rates seen on low-mass stars is likely caused by detection biases related to the contrast between photospheric and flare emission (\citealt{Balona15}).
Although flares are ubiquitous among main-sequence F, G, K, and M dwarfs, flares are typically only seen on a few percent of \emph{Kepler} targets.
It is more difficult to detect flares on higher mass stars since the stars have much higher bolometric luminosities and photospheric temperatures closer to the temperatures of flaring plasma in the photospheric and chromospheric footpoints.
The lower limit for the energies of flares that can be detected is therefore much higher for higher mass stars, biasing our flare rate statistics.
For example, in the flare catalogue of \citet{YangLiu19}, the lower bound (1$^\mathrm{st}$ percentile) for flares on solar mass stars is $\sim10^{33}$~erg, which is similar to the upper bound (99$^\mathrm{th}$ percentile) for flares detected in that catalogue on late M~dwarfs and is above the white light energies of even the most energetic solar flares (\citealt{Woods06}; \citealt{Schrijver12}).

An unbiased comparison of flare rates between different types of stars requires a common lower energy threshold for all stars in a sample above which all flares are counted.
\citet{Audard00} studied the EUV lightcurves and flare statistics of F, G, K, and M dwarfs and found that the frequencies of flares is almost linearly proportional to X-ray luminosity with stars of all spectral types following the same relation. 
They found the frequencies of flares with total emitted X-ray and EUV energies above $10^{32}$~erg to be given by
\begin{equation} \label{eqn:flarerate3}
N(>10^{32}~\mathrm{erg}) = 1.9 \times 10^{-27} L_\mathrm{X}^{0.95}, 
\end{equation}
where $N$ is in day$^{-1}$ and $L_\mathrm{X}$ is in erg~s$^{-1}$.
\citet{Audard00} also showed that among the stars considered, flares with energies above $10^{32}$~erg typically contribute $\sim$10\% of the energy emitted at X-ray wavelengths.
In Fig.~\ref{fig:flarerates}, we combine Eqn.~\ref{eqn:flarerate3} with our estimates for the evolution of $L_\mathrm{X}$ described in previous sections to show how the rates of flares depend on stellar mass and age. 
As expected, long term activity decay leads to decreases in flare rates at all masses.
Since flare rates scale with X-ray luminosity, low mass stars flare less often at energies above $10^{32}$~erg than high mass stars and this trend is visible at all ages. 

The rate of flares with energy $E$ is given by
\begin{equation} \label{eqn:flarerate}
\frac{dN}{dE} \propto E^{-\alpha} ,
\end{equation}
where estimates of $\alpha$ for solar flares range from 1.35 to 2.90 (\citealt{Jess19}).
For stellar flares at XUV wavelengths, \citet{Audard00} found values typically between 1.6 and 2.4 with no clear dependence on spectral type or activity level.
The rates of flares above energy thresholds of $E_1$ and $E_2$ are related by
\begin{equation} \label{eqn:flarerate2}
\frac{N\left(>E_1\right)}{N\left(>E_2\right)} = \left( \frac{E_1}{E_2} \right)^{1-\alpha} .
\end{equation}
We should expect therefore that the rates of flares above any threshold energy follow similar dependences on mass and age as those shown in Fig.~\ref{fig:flarerates}, though our poor constraints on $\alpha$ make conclusions of this sort uncertain.

To understand the likely effects of flares on habitable zone planets, it is also interesting to consider flares with XUV fluences in the habitable zone above a given threshold.
This fluence is the XUV flux from the flare in the habitable zone integrated over the flare duration, given by \mbox{$E/4 \pi a_\mathrm{HZ}^2$}, and we assume here our definition of the habitable zone orbital distance, $a_\mathrm{HZ}$, discussed in Section~\ref{sect:hz}.
We assume a threshold fluence of \mbox{$1.8 \times 10^4$~erg~cm$^{-2}$}, equivalent to the fluence received by a habitable zone planet orbiting a G~dwarf from a flare with an XUV energy of $10^{32}$~erg.
For each stellar mass, we calculate the corresponding threshold energy needed to give a fluence in the habitable zone equal to this value, which for 0.75, 0.5, and 0.25~M$_\odot$ stars is $10^{31.4}$, $10^{30.7}$, and $10^{30.1}$~erg respectively.
Using Eqn.~\ref{eqn:flarerate2}, we then calculate the rates of flares above these threshold energies for each stellar mass and age and the results are shown in Fig.~\ref{fig:flareratesHZ}.
Since Eqn.~\ref{eqn:flarerate2} requires an assumption for $\alpha$, we calculate separately cases for values of 1.6 and 2.4 to demonstrate how this influences the results.
From X-ray emission of pre-main-sequence stars in the Taurus molecular cloud, \citet{Stelzer07} found \mbox{$\alpha = 2.4 \pm 0.5$} and for solar analogues at different activity levels, \citet{Telleschi05} estimated $\alpha$ values between 2.2 and 2.8,  so it is reasonable to expect our \mbox{$\alpha = 2.4$} case to be closer to reality for XUV flares, but it is also uncertain how far below $10^{32}$~erg we can extrapolate the flare rate distribution without a significant decrease in $\alpha$.

As we go to lower mass stars, the threshold energy goes down so we count a larger fraction of flares in our flare rate statistic.
This is true for both $\alpha$ values, and in both cases this effect compensates for the fact that lower mass stars flare less often overall. 
For \mbox{$\alpha = 1.6$}, these two effects approximately cancel out and we get a mass independent distribution, meaning that habitable zone planets receive the same amount of XUV energy from flares above our fluence threshold regardless of their host star mass. 
For \mbox{$\alpha = 2.4$}, the first effect is stronger and so we get higher flare rates for lower mass stars, meaning that habitable zone planets orbiting lower mass stars are exposed to significantly more flares above our fluence threshold. 
If the latter case is closer to reality, we can expect that the atmospheres of habitable zone planets are more strongly influenced by the XUV emission of large flares in M~dwarf systems than in systems with higher mass stars despite M~dwarfs flaring less often.


\section{Discussion and conclusions} \label{sect:discuss}

In this paper, we develop a comprehensive and empirical description of the rotation and XUV evolution of F, G, K, and M~dwarfs.
Our model is constrained and validated by an extensive catalogue of stellar rotation and XUV emission measurements from the literature, especially in young stellar clusters.
The evolution of stellar rotation and XUV emission can be summarised as follows:-
\begin{itemize}

\item
At ages of $\sim$1~Myr, the rotation rates of stars are distributed between approximately 1 and 50~$\Omega_\odot$ and this distribution is mass-independent, at least for masses above 0.4~M$_\odot$.

\item
This wide distribution gets wider during the pre-main-sequence spin-up phase and then converges on the main-sequence due to wind driven spin-down. 
Lower mass stars remain rapidly rotating longer and at later ages spin down to slower rotation rates.	

\item
There is a mass-independent relation between stellar Rossby number and $R_\mathrm{X}$ described by separate power-laws in the saturated (low $Ro$) and unsaturated (high $Ro$) regimes.
This relation has a large scatter due likely to stellar variability.
At each evolutionary stage, stars likely spend almost 20\% of their time with X-ray luminosties that are a factor of three above or below their long-term averages.  

\item
In the first few Myr, the high convective turnover times mean that the saturation threshold rotation rates, $\Omega_\mathrm{sat}$, are very low and stellar XUV emission depends not on rotation, but on mass only, with higher mass stars being more XUV luminous.
During the pre-main-sequence, $\Omega_\mathrm{sat}$ increases which causes slowly rotating stars to fall out of saturation despite spinning up. 

\item
Lower mass stars have lower $\Omega_\mathrm{sat}$ and therefore must spin down more before entering the unsaturated regime.
Combined with the longer early period of rapid rotation, this means that lower mass stars remain saturated at their peak activity levels for longer.
Late M~dwarfs remain saturated for billions of years.

\item
At young ages, the early spread in rotation causes an additional wide spread in XUV emission, which is especially large for higher mass stars such as G and F~dwarfs due to their lower $\Omega_\mathrm{sat}$.
At all masses, initially rapid rotators remain active longer and emit more XUV energy over their lifetimes than initially slow rotators.
A star's mass and initial rotation rate are the two main parameters for determining its XUV evolution. 

\item
Stellar XUV emission decays over evolutionary timescales due to the decreasing $L_\mathrm{bol}$ and increasing $\Omega_\mathrm{sat}$ on the pre-main-sequence and rotational spin-down on the main-sequence. 
At all ages, higher mass stars tend to be more XUV luminous than lower mass stars.

\item  
At all ages, the XUV fluxes in the habitable zone (assuming HZ boundaries at 5~Gyr) are higher for lower mass stars due to their closer habitable zones and longer evolutionary timescales. 
Important are both the longer phases of decreasing $L_\mathrm{bol}$ on the pre-main-sequence and the longer saturation times on the main-sequence.

\item
{
As activity decays, XUV spectra become more shifted to longer wavelengths causing emission at shorter wavelengths to decay more rapidly. 
For most inactive stars, the Ly-$\alpha$ emission line is more luminous than both X-ray and EUV.
}

\item
{
At all ages, higher mass stars flare more often at all energies than lower mass stars, but habitable zone planets likely receive more XUV energy from flares when orbiting lower mass stars. 
Flare rates at all energies decrease as activity decays.
}

\end{itemize}

As we demonstrate in this paper, a realistic description of the evolution of stellar activity, and especially of X-ray and EUV emission, must be based on a description of rotational evolution and an understanding of how the rotation distribution as a function of stellar mass evolves with time. 
Single unique decay laws for stellar activity are unable to describe the range of possible evolutionary tracks that stars with different initial rotation rates can follow.
Also important is the fact that stars are variable on short timescales and likely spend much of their lives significantly more and less active than the long-term average. 
We show that it is important to be clear about which measure of activity is being used when describing the mass dependence of activity: for example, XUV luminosity and XUV flux in the habitable zone lead to very different descriptions.

It is commonly believed, especially in the exoplanet community, that M dwarfs are more X-ray and EUV active than G dwarfs, implying that planets orbiting lower mass stars will be subject to higher rates of atmospheric escape. 
As we show, M~dwarfs are in fact less XUV luminous than higher mass stars at all ages and emit much less XUV radiation over their lifetimes.
If we consider planets with similar orbital distances, a planet orbiting an M dwarf receives less radiation over its lifetime and therefore likely experiences less overall atmospheric erosion than a similar planet orbiting a G dwarf.
However, if we instead consider planets with similar effective temperatures, such as those in the habitable zone, the idea that M dwarfs are more active is justified due to their much longer evolutionary timescales.
{
These conclusions extend to stellar flares: while the rates of flares with a given total energy are higher for G~dwarfs at all ages, the rates of flares with a given habitable zone fluence are likely higher for M~dwarfs, at least for the most energetic flares. 
}

{
In this paper, we do not concentrate on evolution beyond the age of the Sun, largely because of this phase is less interesting for planetary atmosphere formation. 
Until recently, we have lacked observational constraints on the later evolution of rotation and activity and it has usually been assumed that stars continue to follow the expected Skumanich spin-down (\mbox{$P_\mathrm{rot} \propto t^{0.5}$}) until the end of the main-sequence.
This assumption has been challenged by recent astroseimological age determinations for older field stars with \citet{vanSaders16} finding evidence for older field stars that rotate faster than expected, suggesting that at approximately the age (or Rossby number) of the modern Sun, changes in magnetic activity cause a reduction in the spin-down rate.
If correct, this could indicate that stars in the middle of their main-sequence evolution transition into lower activity states (\citealt{Metcalfe16}; \citealt{MetcalfeEgeland19}).
Evidence for the Sun being less active than expected given its mass and rotation (\citealt{Reinhold2020}) could suggest that it has already undergone such a transition, and could explain our need for an additional factor in Eqn.~\ref{eqn:modifyMatttorque} to reproduce Skumanich spin-down using solar properties.
}

Fully understanding stellar activity and its evolution is an important step to interpreting the results of recent and upcoming exoplanetary missions.
We make available with this study a large number of evolutionary tracks for rotation and XUV emission for the range of stellar masses that we consider which are intended to be used as essential input in studies into the long term evolution of atmospheres and their surface conditions.
In future studies, we will combine our results with state-of-the-art planetary atmosphere evolution models to explore the different ways that stars with different masses and rotation rates influence the atmospheres of terrestrial planets.

\section{Acknowledgments} 

This study was carried out with the support by the Austrian Science Fund (FWF) project S11601-N16 ``Pathways to Habitability: From Disk to Active Stars, Planets and Life'' and the related subproject S11604-N16.
This work was also scientifically supported by the ExoplANETS-A project (Exoplanet Atmosphere New Emission Transmission Spectra Analysis; \url{http://exoplanet-atmosphere.eu}) funded from the EU's Horizon-2020 programme (grant agreement no. 776403).


\begin{appendix}

\section{Grid of rotation and XUV evolution tracks} \label{appendix:package}

With this paper, we provide a comprehensive set of evolutionary tracks for stellar rotation and XUV emission.
For each case, we provide two files: files labeled `basic' include only the evolution of the surface rotation rate and the X-ray, EUV, and Ly-$\alpha$ luminosities, and files labeled `extended' include more detailed information about rotation and XUV emission including envelope and core rotation rates, Rossby number, the various torques calculated in the model, wind mass loss rates, dipole field strengths, and the break-up rotation rates.
The XUV emission quantities included in the extended data files are X-ray and EUV luminosities, X-ray surface fluxes, $R_\mathrm{X}$, emission measure weighted average coronal temperature, and the X-ray and EUV fluxes in the habitable zones.
We also include X-ray and EUV luminosities at one and two standard deviations of the spread in the $Ro$--$R_\mathrm{X}$ relation above and below the averages, representing the likely variability of stars on non-evolutionary timescales.
For each case, we also include the star's percentile in the rotation distribution at 150~Myr for its mass and the habitable zone boundaries for that mass calculated assuming the 5~Gyr stellar properties. 

We provide a basic grid of models for stars with masses between 0.1 and 1.2~$M_\odot$ and with initial (1~Myr) rotation rates between 0.1~$\Omega_\odot$ and the break-up rotation rate at 1~Myr.
Although we there is no observational support for stars at 1~Myr rotating slower than $\sim$1~$\Omega_\odot$, we include slower rotating tracks for completeness. 
The tracks start at an age of 1~Myr and end at 12~Gyr for stars with masses below 0.95~M$_\odot$ and near the end of the main-sequence for higher mass stars. 
In addition, we include also a second grid of models where the initial rotation rates have been binned instead by the percentile of our model rotation distribution based on the rotation distribution at 150~Myr.
We include all integer percentiles between the 2$^\mathrm{nd}$ and the 98$^\mathrm{th}$. 
As described above, this distribution is based on the rotation distribution of the 150~Myr combined cluster with stars that we find to be rotating unrealistically slow or fast removed. 
Finally, we also include also tracks for each of the stars in our model distribution and the distributions at each age from 1~Myr to 10~Gyr.
We do not include in these distributions stars that have evolved off the main-sequence.

\section{Fitting procedure for rotation model} \label{appendix:fitting}

The rotational evolution model described in Section~\ref{sect:rotmodel} contains five unconstrained parameters: these are $a_\mathrm{w}$ and $b_\mathrm{w}$ from Eqn.~\ref{eqn:MdotRossby}, and $a_\mathrm{ce}$, $b_\mathrm{ce}$, and $c_\mathrm{ce}$ from Eqn.~\ref{eqn:tCE}. 
We constrain these parameters empirically using the observational constraints described in Section~\ref{sect:rotobs} and a Markov-Chain Monte-Carlo (MCMC) method described here. 
While our results could reveal useful information about the physical properties of stellar winds and the angular momentum transport in stellar interiors, this is not our aim in this paper.

The goodness-of-fit parameter that we use in our fitting procedure is the likelihood, given by
\begin{equation} \label{eqn:likelihood1}
\log L = \frac{1}{N_\mathrm{obs}} \sum_{i=1}^{N_\mathrm{mass}} \gamma_i \sum_{j=1}^{N_\mathrm{age,i}} \log_{10} L_{ij}
\end{equation}
where $N_\mathrm{obs}$ is the total number of observed stars used for all mass bins and ages, $N_\mathrm{mass}$ is the number of mass bins, $\gamma_i$ is a parameter used to adjust the importance of individual mass bins in the fitting procedure, $N_\mathrm{age,i}$ is the number of age bins considered in the $i$th mass bin, and $\log L_{ij}$ is the log (base 10) of the likelihood for the $i$th mass and $j$th age bin.
This is given by
\begin{equation} \label{eqn:likelihood2}
\log L_{ij} =  \sum_{k=1}^{N_{\star ij}} \log f_{ij} \left( \Omega_{\star ijk} \right) 
\end{equation}
where $N_{\star ij}$ is the number of stars observed at this mass and age bin, $f_{ij}$ is the probability density function (PDF) describing how likely it is for a given star at this mass and age to have a rotation rate given by $\Omega_{\star ijk}$.
The first and second sums in Eqn.~\ref{eqn:likelihood1} are over all mass and age bins and the sum in Eqn.~\ref{eqn:likelihood2} is over all observed stars at this mass and age.
The values of $\Omega_{\star ijk}$ are taken from the observational constraints described in Section~\ref{sect:rotobs}.

Calculating $f_{ij}$ for each mass and age requires the initial PDF describing the distribution of rotation rates at 1~Myr and the specification of the fit parameters in our rotational evolution model. 
Our main assumption is that at each mass and age, the underlying PDF for the rotation rates has a double normal distribution dependence on \mbox{$\log_{10} \Omega_\star$}, which is reasonable given the forms of the observational constraints. 
For a given mass and age, the PDF is given by
\begin{equation} \label{eqn:doublegaussianpdf}
f (\Omega_\star) = \sum_1^2 A_{i} \frac{1}{\sigma_i \sqrt{2 \pi}} \exp \left[ \frac{\left( \log_{10} \Omega_\star - \mu_i \right)^2}{2 \sigma_i^2} \right]
\end{equation}
where $\mu_i$ and $\sigma_i$ are the means and standard deviations of the two normal distributions and $A_1$ and $A_2$ controls how important the two normal distributions are relative to each other and are related by \mbox{$A_1+A_2=1$}.
For the starting distribution, we use \mbox{$A_1 = 0.618$}, \mbox{$\mu_1 = 0.714$}, \mbox{$\sigma_1 = 0.255$}, \mbox{$A_2 = 0.382$}, \mbox{$\mu_2 = 1.342$}, and \mbox{$\sigma_2 = 0.180$}.
These values are derived from the rotation distributions measured in the Orion Nebula Cluster (\citealt{RodriguezLedesma09}) and NGC~6540 (\citealt{HendersonStassun12}) assuming that $\Omega_\star$ is given in units of \mbox{$\Omega_\odot = 2.67 \times 10^{-6}$~rad~s$^{-1}$}.
To evolve this function forward in time for a given stellar mass, we start with a distribution of 500 stars at 1~Myr with rotation rates chosen randomly based on this starting PDF.
We then evolve each star forward in time to the observed ages using our physical model and at each of the observed ages we fit a double normal distribution (the parameters in Eqn.~\ref{eqn:doublegaussianpdf}) using an MCMC method. 
Doing this separately for each mass bin gives us \mbox{$f_{ij} \left( \Omega_{\star ijk} \right)$} from Eqn.~\ref{eqn:likelihood2} needed to calculate the likelihood.

The method described above allows us to calculate our goodness-of-fit parameter for a given set of our unconstrained model parameters, which we describe collectively as $\mathbf{X}$.
Our aim is to find the value of $\mathbf{X}$ that corresponds to the maximum value of our goodness-of-fit parameter, which we do using an MCMC method based on the Metropolis-Hastings algorithm.
Starting from an initial $\mathbf{X}$ chosen randomly between reasonable limits, we iteratively evolve $\mathbf{X}$ by making small changes to each of the parameters.
The sizes and directions of the changes to the parameters in each step are determined randomly, making our exploration of the parameter space into a random walk. 
To best explore the parameter space, we perform 1600 separate fit attempts and our final set of parameters are those that give the largest likelihood from all 1600. 


\section{Fitting X-ray relation} \label{appendix:RoRx}

To use the relation between $Ro$ and $R_\mathrm{X}$ given in Eqn.~\ref{eqn:RoRx}, we need to empirically determine the parameters $\beta_1$, $\beta_2$, $Ro_\mathrm{sat}$, and $R_\mathrm{X,sat}$.
To fit this relation, we use the OLS($Y|X$) method, where $Y$ is $\log R_\mathrm{X}$ and $X$ is $\log Ro$.  
This method involves finding the set of parameters that minimised the squares of the vertical distances in Fig.~\ref{fig:RoRx} between the observed data points and the power law relation.
This is the method recommended by \citet{Isobe90} when $Y$ has a causal dependence on $X$ or when we want to use the relation to predict $Y$ from a measured $X$, which most likely best represents our situation. 
In this case, the cause is the star's magnetic dynamo, the strength of which is characterised by $Ro$, and the effect is the resulting magnetic surface field and X-ray emission, measured by $R_\mathrm{X}$.
Similarly, our aim in this paper is to use our empirical relation to predict evolutionary tracks for X-ray emission based on observationally constrained tracks for rotation. 
It therefore seems that OLS($Y|X$) is the most appropriate for our purposes, but we cannot rule out that other methods such as the OLS Bisector method could be more appropriate. 
Given the amount of spread in the relation between $Ro$ and $R_\mathrm{X}$, the OLS Bisector method would lead to a steeper power-law fit in the unsaturated regime (\citealt{Reiners14}).

Our fitting procedure involves minimising $S$, given by
\begin{equation}
S = \sum_i \left[ \log R_{\mathrm{X},i} - \log R_\mathrm{X} (Ro_i) \right]^2 ,
\end{equation}
where the sum is over all stars in our observed sample, $Ro_i$ and $R_{\mathrm{X},i}$ are the measured values for the $i$th star in the sample, and \mbox{$R_\mathrm{X} (Ro_i)$} is the corresponding $R_\mathrm{X}$ calculated from Eqn.~\ref{eqn:RoRx} with a Rossby number $Ro_i$.
For a given set of $Ro_\mathrm{sat}$, and $R_\mathrm{X,sat}$, we can easily minimise $S$ separately for all stars with \mbox{$Ro_i < Ro_\mathrm{sat}$} to get $\beta_1$ and \mbox{$Ro_i > Ro_\mathrm{sat}$} to get $\beta_2$. 
The only constraint on these fits is the necessity that the lines pass through the saturation point, meaning that the fit must be done numerically, which is trivial given that \mbox{$dS/d\beta$} is a linear function of $\beta$ and the fit can therefore be performed by finding where \mbox{$dS/d\beta = 0$}.
We find the values of $Ro_\mathrm{sat}$ and $R_\mathrm{X,sat}$ that minimise $S$ using a simple gradient descent method and uncertainties on these parameters are estimated using the bootstrapping method.


\end{appendix}

\bibliographystyle{aa}
\bibliography{mybib}

\end{document}